\documentclass[twocolumn,showpacs, nofootinbib]{revtex4-1}

\usepackage{amssymb}
\usepackage{appendix}
\usepackage{ctable}
\usepackage{enumitem}
\usepackage{epsfig}
\usepackage{epstopdf}
\usepackage{graphicx}
\usepackage{lipsum}
\usepackage{multirow}
\usepackage{subfigure}
\usepackage{hyphenat}
\usepackage{hyperref}
\usepackage{tikz}
\usepackage{amsmath}

\definecolor{cof}{RGB}{219,144,71}
\definecolor{pur}{RGB}{186,146,162}
\definecolor{greeo}{RGB}{91,173,69}
\definecolor{greet}{RGB}{52,111,72}

\newcommand{\specialcell}[2][c]{\footnotesize \begin{tabular}[#1]{@{}c@{}}#2\end{tabular}}

\begin{document}

\title{Statistical Topology of Three-Dimensional \\Poisson-Voronoi Cells and Cell Boundary Networks}
\author{Emanuel A. Lazar$^{1,5}$, Jeremy K. Mason$^{2,3}$, Robert D. MacPherson$^4$, David J. Srolovitz$^5$}
\affiliation
{$^1$Applied Physics and Applied Mathematics, Columbia University, New York, New York 10027.\\
 $^2$Lawrence Livermore National Laboratory, Livermore, California 94550.\\
 $^3$Bo\u{g}azi\c{c}i University, Bebek, Istanbul 34342 T\"urkiye.\\
 $^4$School of Mathematics, Institute for Advanced Study, Princeton, New Jersey 08540.\\ 
 $^5$Materials Science and Engineering, University of Pennsylvania, Philadelphia, PA 19104.} 
\date{\today}

\begin{abstract}  
Voronoi tessellations of Poisson point processes are widely used for modeling many types of physical and biological systems.  In this paper, we analyze simulated Poisson-Voronoi structures containing a total of 250,000,000 cells to provide topological and geometrical  statistics of this important class of networks.  We also report correlations between some of these topological and geometrical measures.  Using these results, we are able to corroborate several conjectures regarding the properties of three-dimensional Poisson-Voronoi networks and refute others.  In many cases, we provide accurate fits to these data to aid further analysis.  We also demonstrate that topological measures represent powerful tools for describing cellular networks and for distinguishing among different types of networks.  
\end{abstract}
\maketitle

\section{Introduction}
Poisson-Voronoi tessellations are random subdivisions of space that have found applications as models for many physical systems \cite{1992okabe, stoyan1995stochastic}.  They have been used to study how galaxies are distributed throughout space \cite{icke1988voronoi, yoshioka1989large} and have aided in discovering new galaxies \cite{ramella2001finding}.  They have been used to study how animals establish territories \cite{1980tanemura}, how crops can be planted to minimize weed growth \cite{1973fischer}, and how atoms are arranged in crystals \cite{mackay2011stereological}, liquids \cite{finney1970random}, and glasses \cite{hentschel2007statistical, luchnikov2000voronoi}.  A more complete list of applications can be found in standard references on the subject \cite{1992okabe, stoyan1995stochastic}.

A Poisson-Voronoi tessellation is constructed as follows.  Points called {\it seeds} are obtained as the realization of a uniform Poisson point process (e.g.~see \cite{daley2003introduction, daley2007introduction, stoyan1995stochastic, kingman1992poisson, cox1980point}) in a fixed region.  {\it Cells} are the sets of all points in the region that are closer to a particular seed than to any other.  If a point is equidistant to multiple nearest seeds, then it lies on the boundary of the associated cells. In three dimensions, there is a zero probability that a point will be equidistant to five or more seeds.  All cells are convex, and this network of cells partitions the entire region.    

Many exact results have been obtained in connection with three-dimensional Poisson-Voronoi structures.  Meijering \cite{1953meijering} proved that the average number of faces per cell is $48\pi^2/35 + 2 \approx 15.535$, the average number of edges per face is $144\pi^2/(35+24\pi^2) \approx 5.228$, the average surface area per cell is $\left(256\pi/3\right)^{1/3}\Gamma(\frac{5}{3})\rho^{-2/3}$ and the average edge length per cell is $(3072\pi^5/125)^{1/3}\Gamma(\frac{4}{3})\rho^{-1/3}$, where $\rho$ is the number of seeds or cells per unit volume.  Gilbert \cite{1962gilbert} expressed the variance of the cell volume distribution as a double integral.  Using a more general approach, Brakke \cite{1985brakke} obtained integral expressions for the variances of number of cell faces, volumes,  surface areas, number of face edges, face areas, and perimeters, as well as  variances and covariances of several  other quantities of interest and the distribution of edge lengths.  In all of these cases, Brakke also solved these integrals numerically.  Much more is understood about Poisson-Voronoi structures than can be detailed here, and the interested reader is referred to standard references \cite{1992okabe, 1994moller, stoyan1995stochastic} and the more recent surveys of M{\o}ller and Stoyan \cite{moller2007stochastic} and of Calka in \cite{kendall2010new}. 

Additional properties of Poisson-Voronoi structures have been investigated through simulation.  Using a data set with 358,000 cells, Kumar {\it et al.}~\cite{1992kumar} reported the distributions of faces with fixed numbers of edges and cells with fixed numbers of faces, volumes, face areas and cell surface areas.   They also reported distributions of volumes and surface areas restricted to cells with fixed numbers of faces, and distributions of areas and perimeters restricted to faces with fixed numbers of edges.  Although their data set was relatively small by current standards, their results are the most complete set of three-dimensional Poisson-Voronoi cell  statistics available in the literature.  

The Kumar {\it et al.}~data set has since been augmented by additional results.  Marthinsen \cite{1996marthinsen} used a set of 100,000 cells to compute the distribution of cell volumes and surface areas.  Tanemura \cite{2003tanemura} later used a substantially larger data set of 5,000,000 cells to obtain more precise data for the distributions of volumes, surfaces areas, and faces, as well as volumes for cells with fixed numbers of faces.  Ferenc and N{\'e}da \cite{2007ferenc} later used a data set with 18,000,000 cells to calculate the distribution of cell volumes.  

Thorvaldsen \cite{thorvaldsen1992simulation} and Reis {\it et al.} \cite{ferro2006geometry} used the ratio between the surface area of a cell and the surface area of a sphere of equal volume to describe the ``shape isotropy'' of a cell.  Using a system of 250,000 cells, Thorvaldsen reported the distribution of this parameter among Poisson-Voronoi cells, and observed that this parameter decreases with increasing cell volume.  Using a smaller set of 10,000 cells, Reis {\it et al.} considered how this parameter depends on the number of faces of a cell.  A more sophisticated, higher-order method of measuring shape isotropy using Minkowski tensors has been recently introduced \cite{schroder2010minkowski} and used to characterize a number of natural structures \cite{kapfer2010local, schroder2010disordered}.  In particular, Kapfer {\it et al.} \cite{kapfer2010local} have used this method to characterize a data set of 160,000 Poisson-Voronoi cells \cite{kapfer2010local}.

In prior studies, the topology of individual cells has been described by counting their numbers of faces.  As we discuss below, this is a simplistic and incomplete description of the topology of a cell.\footnote{When referring to the topology of a cell we have in mind the topology of the cell and its immediate neighborhood, which includes the network of edges and faces which intersect it.}  In this report, we present distributions of many important topological features of Poisson-Voronoi structures based on a dataset of a combined total of 250,000,000 cells.  This is the largest data set available today and provides the most precise characterization of topological properties of the Poisson-Voronoi network. This resolution allows us to examine the validity of conjectures made on the basis of smaller data sets, some of which we now show are qualitatively incorrect.  We supplement the discussion of topological properties with analysis of some purely geometrical descriptions and the interrelationship between some topological and geometrical features.  We leave many results in the Supplemental Material and make the entire data set available online \cite{suppmat}.

\section{Method}
We employ the computer code {\tt vor3dsim}, developed by Ken Brakke \cite{vor3dsim}, to generate 250 Poisson-Voronoi tessellations, each of which contains 1,000,000 cells; periodic boundary conditions are used to eliminate boundary effects.  Because the statistics we consider measure neighborhoods of the structure which are small compared to the total size of the system, we expect that statistics observed in this set of smaller systems will be consistent with what we would observe in a single system with an identical number of total cells.  Details of the algorithms used to perform some of the more complex topological analyses were reported previously \cite{2012lazar, 2012mason}.

\section{Topological characteristics}

\subsection{Distribution of faces}
\begin{figure}[b]
\centering
\includegraphics[width=1.\columnwidth]{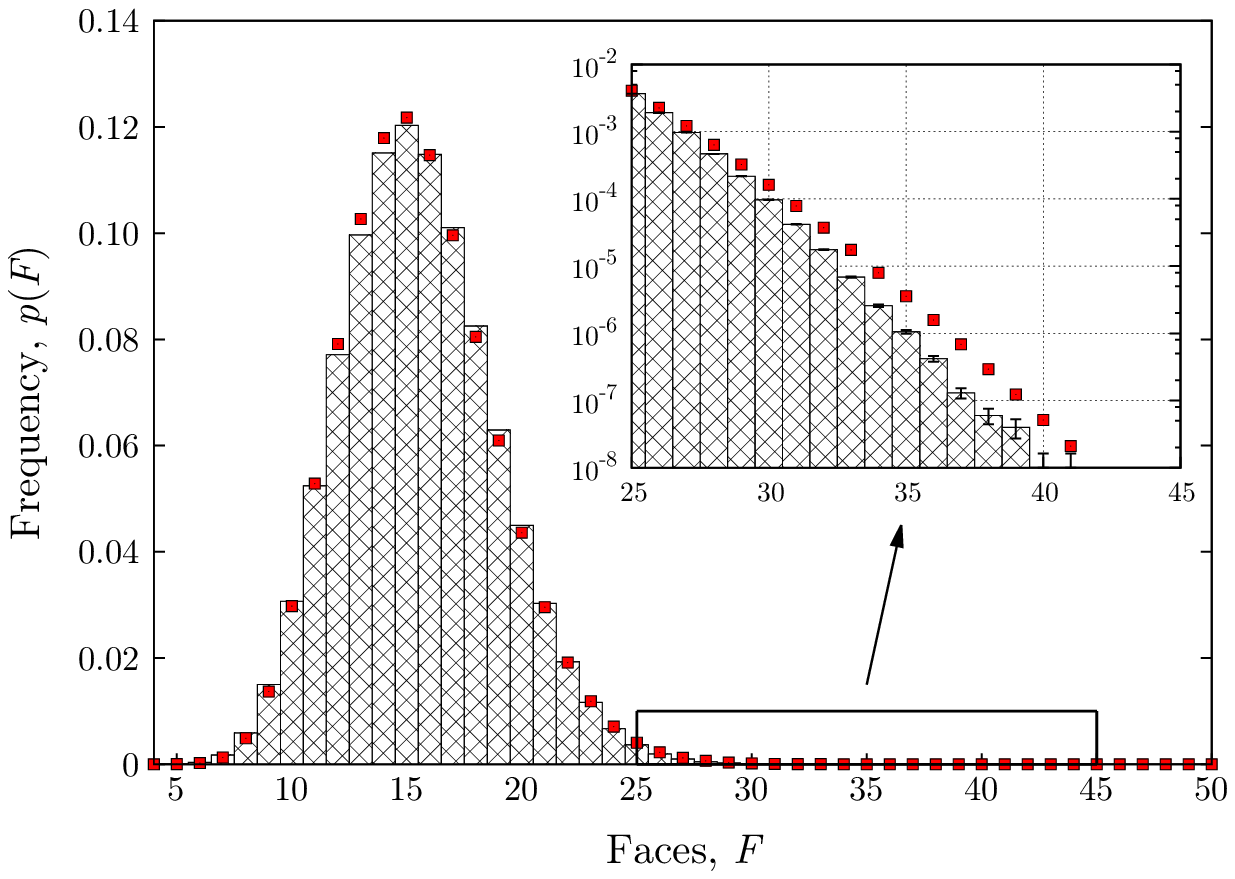}
\caption{Distribution of the number of faces per cell; squares show the discrete probability distribution of Eq.~(\ref{kumar_faces}).  The mean and standard deviation are 15.535 and 3.335, respectively, to within the accuracy of the data. The inset shows a subset of the data on a semilogarithmic plot; error bars show the standard error from the mean. }
\label{faces}
\end{figure}

The simplest way to classify the topology of a Poisson-Voronoi cell involves counting its number of faces.  This is the topological characterization most commonly quoted in the literature \cite{1952smith, 1974rhines}.  Figure \ref{faces} shows this distribution of faces per cell; these data  are consistent with those reported in \cite{1992kumar} and \cite{2003tanemura}.  
As noted earlier, Meijering \cite{1953meijering} proved that the mean of this distribution is $48\pi^2/35 + 2$.  Brakke \cite{1985brakke} obtained an integral form of the variance, which he numerically evaluated to be 11.1246.  Our data reproduce these exact results to within 0.001\% and 0.004\%, respectively.  While the distribution of faces is approximately symmetric about 15, there are no cells with fewer than 4 faces.  We note that while Kumar {\it et al.}~\cite{1992kumar} and Tanemura \cite{2003tanemura} reported no cells with more than 36 faces based on their more limited data set, we find cells with up to 41 faces.  



Kumar {\it et al.}~\cite{1992kumar} suggested that the distribution of the number of faces $F$ per cell can be described by the discretized two-parameter $\Gamma$ function:
\begin{equation}
p(F) = \int_{F-1/2}^{F+1/2}\frac{x^{a-1}}{b^a\Gamma(a)}e^{-x/b}dx.
\label{kumar_faces}
\end{equation}
The best fit to their data yielded $a=21.6292$ and $b=0.7199$.   The form of $p(F)$ in  Eq.~(\ref{kumar_faces}) gives $p(F)>0$ for all positive integers $F$, including 1, 2 and 3. Of course, this  is incorrect since there can be no  polyhedra with fewer than four faces in a Voronoi tessellation.  Moreover, since we know exactly both the mean of the distribution as well as its variance, we are left with no free parameters.  Regardless of whether we choose to include $p(1)$, $p(2)$, and $p(3)$ in normalizing the distribution,  these parameters must be $a=21.85892$ and $b=0.710714$ to  match the exact results.  
Careful inspection of the data in the inset to Fig.~\ref{faces} reveals that Eq.~(\ref{kumar_faces}) does not accurately describe the decay in $p(F)$ for large $F$.  This further demonstrates that this conjectured equation is not an exact representation of $p(F)$; we know of no such exact relation.

\subsection{Distribution of edges}
We next consider the distribution $p(n)$ of faces with $n$ edges.  Meijering \cite{1953meijering} proved that the mean of this distribution is $144\pi^2/(35+24\pi^2)$, and Brakke \cite{1985brakke} obtained an integral form of the variance, which he evaluated numerically to be 2.4846.  Our data reproduce the exact result of the mean to within 0.0002\%, and the exact result for the variance to within 0.00001\%.
\begin{figure}
\centering
\includegraphics[width=1.\columnwidth]{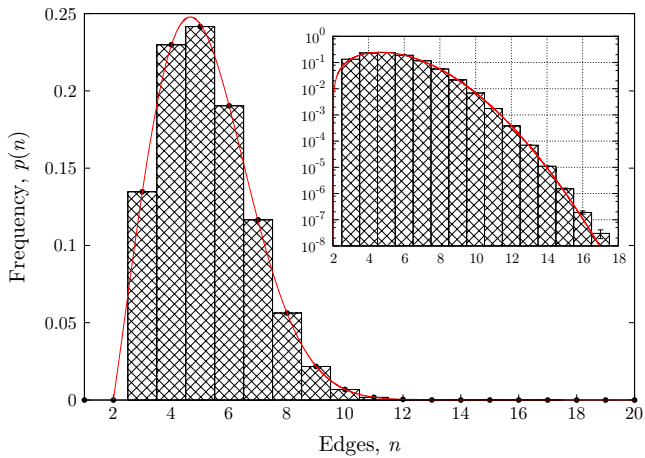}
\caption{Distribution of number of edges per face; squares show the discrete probability distribution of Eq.~(\ref{newfit}).  The mean and standard deviation for this data set is 5.228  and 1.576, respectively, to within the accuracy of the data.  The inset shows the same data on a semilogarithmic plot; error bars show the standard error from the mean.}
\label{multiedges}
\end{figure}
Figure \ref{multiedges} shows this distribution for our Poisson-Voronoi data set.  This distribution is similar to that reported earlier \cite{1992kumar}, albeit with more accurate statistics and over a greater range of $n$.  The distribution has a maximum at 5 edges per face, which is close to the mean; there are no faces with fewer than 3 edges. While Kumar {\it et al.}~\cite{1992kumar} reported no faces with more than 15 edges, our data set shows faces with up to 18 edges.

While the  mean and variance of this distribution are known exactly, little else is known and, to our knowledge, there are no proposed forms for this distribution.  We suggest the following empirical form
\begin{equation}
p(n) =
\begin{cases}
A(n-2)e^{-B(n-\hat{n})^2} & n\geq 2 \\
0 & n<2,
\end{cases}
\label{newfit}
\end{equation}
where $n$ is the number of edges of a face.  Although this empirical relation fits the data remarkably well, its origin is unclear.  By requiring that the distribution is properly normalized and that the mean and variance reproduce the exact results, all three parameters are determined: $A=0.13608070$, $B=0.093483172$, and $\hat{n}=2.64631320$.  Overall, this empirical functional form provides an excellent fit to the Poisson-Voronoi data set, including the large $n$ tail.

\subsection{Aboav-Weaire relation}
In studying two-dimensional cross-sections of polycrystalline magnesium oxide, Aboav \cite{1970aboav} and Weaire \cite{1974weaire} explored the relationship between the number of edges $n$ of a cell and the expected number of edges $m(n)$ of its $n$ neighbors.  They observed that this relationship can be described by $m(n) \approx A + B/n$, which can be understood as follows.  In two dimensions, the average number of edges per cell is $\langle n \rangle = 6$.  If this average is approximately maintained among every cluster of cells, then a cluster with an $n$-sided cell in its center should have on average  $(n+nm(n))/(n+1) \approx \langle n \rangle$ edges per cell.  This gives us an expression for $m(n)$ in terms of $n$ and $\langle n \rangle$ of the above form: $m(n) \approx 5 + 6/n$.  This equivalence is only approximate because the $\langle n \rangle = 6$ average is {\it not} maintained among every cluster of cells.  This leads to a correction term that in part depends on the variance of the distribution.  We note, for later, that this form of $m(n)$ decreases monotonically with increasing $n$.  This relationship has been used to analyze biological tissue \cite{jeune1998interactions, mombach1993mitosis}, soap foams \cite{weaire1999physics, mejia2000evolution}, and other cellular structures \cite{elias1997two, earnshaw1994topological, moriarty2002nanostructured}.  

In two dimensions, it was originally believed that this relationship also describes Poisson-Voronoi structures \cite{boots1982arrangement, kumar1993properties}.  However, Hilhorst \cite{hilhorst2005planar, hilhorst2006planar} has shown that the correct form of the relationship is $m(n) = 4+3(\pi/n)^{1/2} + O(1/n)$, in the limit of large $n$.

\begin{figure}[h!]
\centering
\vspace{3mm}
\includegraphics[width=1.\columnwidth]{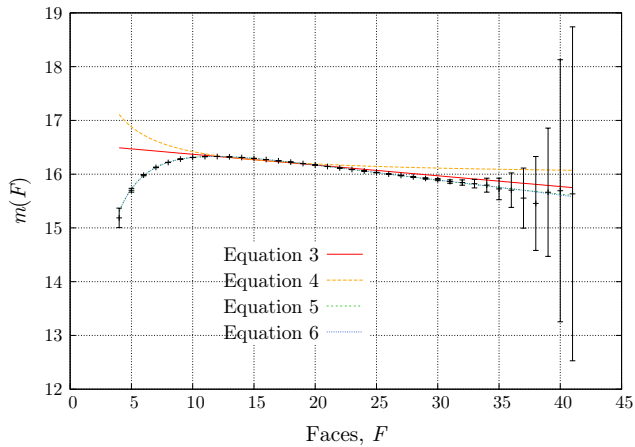}
\caption{(Color online) Expected number of faces $m(F)$ of neighbors of cells with $F$ faces; error bars indicate standard error from the mean.  The red, yellow, green, and blue curves show the forms suggested by Kumar {\it et al.}, Fortes, Hillhorst (truncated at $i=3$), and Mason {\it et al.}~(truncated at $i=4$), respectively.}
\label{weaireaboav}
\end{figure}

We now investigate the extension of this relationship to three-dimensional Poisson-Voronoi structures, i.e., the relationship between the number of faces $F$ of a cell and the expected number of faces $m(F)$ of its neighbors.  Figure \ref{weaireaboav} shows that $m(F)$ increases for small $F$, reaches a maximum at $F = 12$, and then decreases in a nearly linear manner for large $F$.  The existence of the increasing region of $m(F)$ has not been previously reported for the Poisson-Voronoi structure.

Based on limited three-dimensional Poisson-Voronoi data (3729 cells), Kumar {\it et al.}~\cite{1992kumar} fit their data to a linear function as follows:
\begin{equation}
m(F) = A - BF,
\end{equation}
and found $A=16.57$ and $B=0.02$.  Of course, such a fit is unreasonable because it suggests that $m(F)<0$ for sufficiently large $F$.  Using the same data set as Kumar {\it et al.}, Fortes \cite{fortes1993applicability} proposed fitting this data to an Aboav-Weaire-type of relation:
\begin{equation}
m(F) = A + B/F,
\end{equation}
where the constants $A=15.96$ and $B=4.60$ were found using a least square fit to this data set.  
The forms suggested by Kumar {\it et al.}~and Fortes do not provide even a qualitative fit to $m(F)$ at small $F$; they both decrease monotonically with increasing $F$, contrary to the data for $F < 12$.  Clearly, the general form of the Aboav-Weaire relation provides a poor representation of the topological correlations between nearest neighbor cells in three-dimensional Poisson-Voronoi structures.

Hilhorst \cite{hilhorst2009heuristic}, building on his earlier work on two-dimensional Poisson-Voronoi structures \cite{hilhorst2005planar, hilhorst2006planar}, provided strong theoretical arguments for a relationship of the form:
\begin{equation}
m(F) = k_0 + \sum_{i=1}^{\infty}k_iF^{-i/6} 
\label{hilhorsteq}
\end{equation}
with $k_0 = 8$.  A least squares fit of this expression (truncated at $i=3$) to our data yields $k_1=2.474$, $k_2=49.36$, and $k_3=-51.50$.  This form provides an excellent fit to the present three-dimensional Poisson-Voronoi data set, over the entire range of $F$.

Finally, Mason {\it et al.}~\cite{2012masonB}, also based on theoretical arguments, developed the following expression for $m(F)$:
\begin{equation}
m(F) = \langle F \rangle + \frac{\langle F \rangle+\mu_2}{F} - 1 - \frac{1}{\xi F} \sum_{i=1}^{\infty}k_i\left[(F-\langle F \rangle)^i - \mu_i \right],
\label{masoneq}
\end{equation}
where $\xi = 4\pi-6\cos^{-1}(-1/3)$ is a constant for three-dimensional structures, $\mu_i$ is the $i^{\text th}$ central moment of the distribution of faces, and $k_i$ are fitting parameters.  The equation shown in Fig.~\ref{weaireaboav} corresponds to a best fit when considering $i\leq 4$; we find that $k_1=-1.567$, $k_2=0.0478$, $k_3=-0.00109$, and $k_4=0.000022$.  Except at the tail end of $m(F)$, Eqs.~(\ref{hilhorsteq}) and (\ref{masoneq}) are indistinguishable.

\subsection{$p$-vectors}
\begin{figure}[h]
\begin{center}    
\begin{tabular}{cc}      
\resizebox{0.33\columnwidth}{!}{\includegraphics{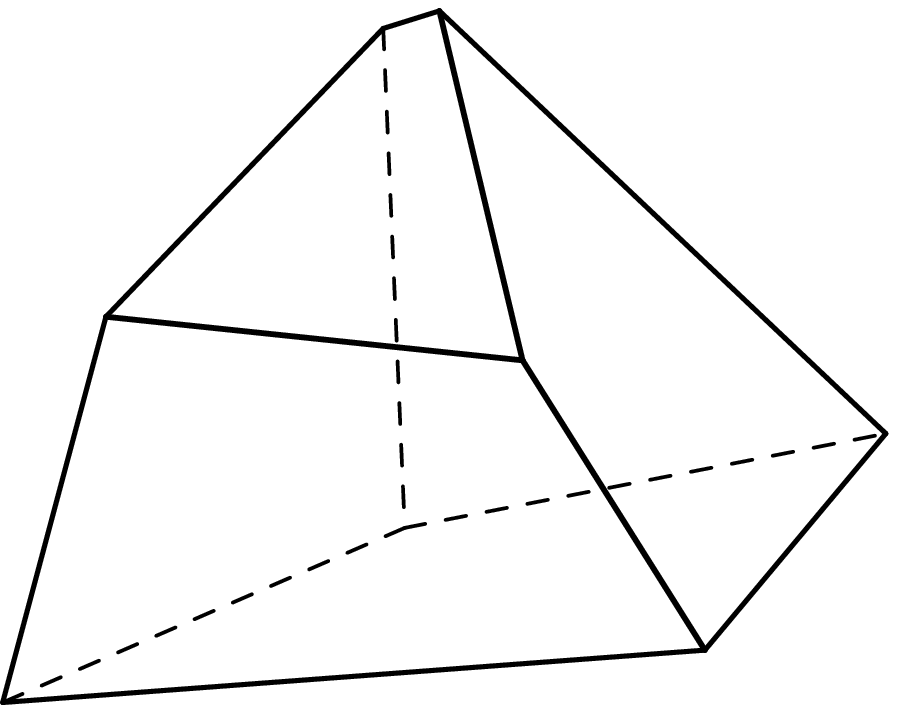}} &\quad\quad
\resizebox{0.38\columnwidth}{!}{\includegraphics{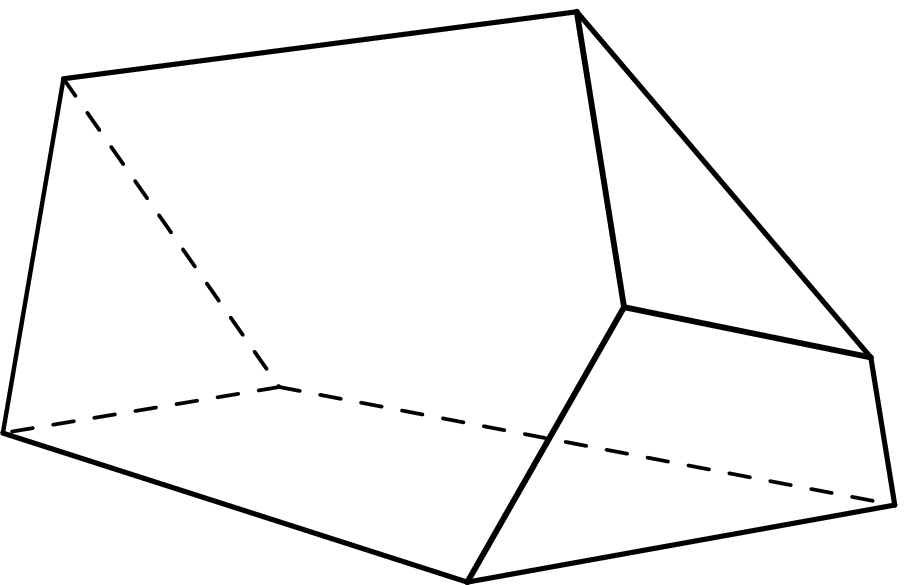}}\\
(a)&\quad\quad(b)
\end{tabular}    
\vspace{-5mm}
\end{center}
\caption{Topologically distinct cells with six faces.  Type (a) appears more than twice as frequently as type (b) in the Poisson-Voronoi structure.}
\label{sixfaces}
\end{figure}

Although counting faces can often distinguish between topologically distinct cells, it cannot do so in general.  Figure \ref{sixfaces} shows topologically distinct cells, each with six faces.  
A more refined description of the topology of a cell involves recording not only its number of faces, but also its particular types of faces.  Figure \ref{sixfaces}(a) has six four-sided faces, while Fig.~\ref{sixfaces}(b) has two three-sided faces, two four-sided faces, and two five-sided faces.  These two topological types are the only ones with six faces that appear in the Poisson-Voronoi structures.

Barnette \cite{1969barnette}, in describing the combinatorial properties of three-dimensional polytopes, defined a $p$-vector as a vector of integer entries in which the $i^{\text{th}}$ entry denotes the number of $i$-sided faces of a polyhedron.  Table \ref{pvectortable} lists the 48 most frequent $p$-vectors of the Poisson-Voronoi structure and their relative frequencies in the Poisson-Voronoi dataset.  The table shows that the Poisson-Voronoi structure is not dominated by a small set of  $p$-vectors; rather, the distribution is quite broad -- no $p$-vector occurs with a frequency greater than 0.39\%.  In the 250,000,000 Poisson-Voronoi cell data set, there are 375,410 distinct $p$-vectors; the complete distribution of observed $p$-vectors may be found in the Supplemental Material. 
\begin{table*}
\centering
\begin{tabular}{|r|c|r|c|c|}
\hline
& $p$-vector &  \multicolumn{1}{c|}{$F$} & {\it f} \\
\hline
1 & $(001343100...)$ & 12 & 0.388\% \\
2 & $(001342100...)$ & 11 & 0.342\% \\
3 & $(001433200...)$ & 13 & 0.298\% \\
4 & $(001344100...)$ & 13 & 0.289\% \\
5 & $(001423100...)$ & 11 & 0.288\% \\
6 & $(002333110...)$ & 13 & 0.284\% \\
7 & $(001332000...)$ & 9 & 0.274\% \\
8 & $(000442000...)$ & 10 & 0.265\% \\
9 & $(001352200...)$ & 13 & 0.263\% \\
10 & $(002233100...)$ & 11 & 0.261\% \\
11 & $(001432200...)$ & 12 & 0.258\% \\
12 & $(001353200...)$ & 14 & 0.258\% \\
\hline       
\end{tabular}       
\hspace{0.6mm}       
\begin{tabular}{|r|c|r|c|c|}       
\hline       
& $p$-vector & \multicolumn{1}{c|}{$F$} & {\it f} \\ 
\hline       
13 & $(002332110...)$ & 12 & 0.256\% \\
14 & $(001422100...)$ & 10 & 0.254\% \\
15 & $(002322200...)$ & 11 & 0.252\% \\
16 & $(002242200...)$ & 12 & 0.248\% \\
17 & $(002342210...)$ & 14 & 0.247\% \\
18 & $(001443110...)$ & 14 & 0.244\% \\
19 & $(000443000...)$ & 11 & 0.239\% \\
20 & $(002343210...)$ & 15 & 0.233\% \\
21 & $(001442110...)$ & 13 & 0.232\% \\
22 & $(001424100...)$ & 12 & 0.231\% \\
23 & $(001434200...)$ & 14 & 0.223\% \\
24 & $(002243200...)$ & 13 & 0.217\% \\
\hline       
\end{tabular}       
\hspace{0.6mm}       
\begin{tabular}{|r|c|r|c|c|}       
\hline       
& $p$-vector & \multicolumn{1}{c|}{$F$} & {\it f} \\ 
\hline       
25 & $(002323200...)$ & 12 & 0.213\% \\
26 & $(002232100...)$ & 10 & 0.210\% \\
27 & $(002423210...)$ & 14 & 0.203\% \\
28 & $(002334110...)$ & 14 & 0.202\% \\
29 & $(001252000...)$ & 10 & 0.201\% \\
30 & $(000533100...)$ & 12 & 0.199\% \\
31 & $(001263100...)$ & 13 & 0.198\% \\
32 & $(001341100...)$ & 10 & 0.196\% \\
33 & $(002234100...)$ & 12 & 0.193\% \\
34 & $(001354200...)$ & 15 & 0.192\% \\
35 & $(001345100...)$ & 14 & 0.191\% \\
36 & $(001334000...)$ & 11 & 0.190\% \\
\hline       
\end{tabular}       
\hspace{0.6mm}       
\begin{tabular}{|r|c|r|c|c|}       
\hline       
& $p$-vector & \multicolumn{1}{c|}{$F$} & {\it f} \\ 
\hline       
37 & $(002422210...)$ & 13 & 0.190\% \\
38 & $(002333300...)$ & 14 & 0.188\% \\
39 & $(002324200...)$ & 13 & 0.188\% \\
40 & $(001442300...)$ & 14 & 0.186\% \\
41 & $(001444110...)$ & 15 & 0.185\% \\
42 & $(001443300...)$ & 15 & 0.185\% \\
43 & $(002332300...)$ & 13 & 0.180\% \\
44 & $(001351200...)$ & 12 & 0.179\% \\
45 & $(000453100...)$ & 13 & 0.178\% \\
46 & $(001533210...)$ & 15 & 0.175\% \\
47 & $(001333000...)$ & 10 & 0.173\% \\
48 & $(001453210...)$ & 16 & 0.172\% \\
\hline
\end{tabular}
\caption{The 48 most frequent $p$-vectors in the Poisson-Voronoi structure, their number of faces $F$  and  their relative frequency ${\it f}$.  The 250,000,000 cell data set contains 375,410 distinct $p$-vectors; the complete distribution of $p$-vectors may be found in the Supplemental Material.} 
\label{pvectortable}
\end{table*}

Poisson-Voronoi cells may be contrasted with those found in other natural structures.  Matzke \cite{1946matzke} carefully recorded $p$-vector data for 1000 soap bubbles in a foam, and Williams and Smith \cite{1952williams} reported $p$-vector data for 91 individual cells in an aluminum polycrystal.  More recently, Kraynik {\it et al.}~\cite{kraynik2003structure} reported $p$-vector data to characterize over 1000 simulated monodisperse foam bubbles, and we have reported the distribution of $p$-vectors in a set of 269,555 grains in simulated grain growth microstructures \cite{2012lazar}.

Although soap foams and grain growth microstructures share much in common with Poisson-Voronoi tessellations, it is important to emphasize that they result from qualitatively different processes.  Capillarity, surface tension, and curvature play significant roles in the formation and evolution of soap foams \cite{weaire1999physics} and grain growth microstructures \cite{ratke2002growth}.  Since these forces tend to minimize interfacial areas (subject to certain constraints), we can expect that the microstructures that result from these processes somehow reflect these physics.  More specifically, we might expect to find qualitatively different microstructures than those that result from a Poisson-Voronoi tessellation, in which these forces play no role.

The data reported in \cite{1946matzke}, \cite{1952williams}, and \cite{kraynik2003structure} are insufficient to provide definitive $p$-vector distributions for either polycrystalline aluminum or soap foam structures.  However, they are sufficient to clearly distinguish those structures from the Poisson-Voronoi one.  Of the 91 aluminum cells examined by Williams and Smith, the most common $p$-vector is $(0004420...)$, and it appeared 8 times.  Seven other $p$-vectors appeared 2 or 3 times each, and the remaining 66 distinct $p$-vectors appeared only once each.  

Considering only the interior bubbles of his original sample, Matzke found that the most common $p$-vector was $(0001\,10\,2...)$.  These bubbles accounted for 20\% of the 600 interior bubbles.  Three more $p$-vectors each accounted for at least 8\% of all bubbles, five more accounted for at least 3\% each, and five more accounted for at least 1.5\% of all bubbles.  

Kraynik {\it et al.}~\cite{kraynik2003structure} found that data from simulated monodisperse foams closely resembled the experimental results of Matzke.  In particular, they found that the most common $p$-vector was $(0001\,10\,2...)$, which accounted for just under 20\% of all relaxed and annealed monodisperse foam bubbles, a result almost identical to that of Matzke.  The next most common $p$-vector in simulated monodisperse structures was $(0002840...)$, which accounted for almost 14\% of all bubbles, and then $(0001\,10\,3...)$, which accounted for just under 11\% of all bubbles.  Four other $p$-vectors each accounted for at least 5\% of bubbles.

The distribution of $p$-vectors in grain growth structures \cite{2012lazar}, which evolve through mean curvature flow, is substantially more concentrated than in the Poisson-Voronoi microstructure, but not nearly as much as in the data of Matzke, Williams and Smith, and Kraynik {\it et al.}.  In the grain growth data, the most common $p$-vector is $(0004400...)$, and it accounts for nearly 3\% of all cells; each of the 10 most common $p$-vectors accounts for at least 1\% of all cells.  

Although the exact nature of the heavy bias towards certain $p$-vectors in each of the structures is not completely understood, it can already be used to distinguish the different structures from the Poisson-Voronoi structure and from each other by standard statistical tests (e.g., a chi-squared test).  The distribution of $p$-vectors, hence,  provides a useful means  to distinguish between cellular structures of different physical origin.  Despite its early introduction, this method has not been widely adopted.

\subsection{Distribution of topological types}

\begin{figure}[h]
\begin{center}    
\begin{tabular}{ccc}      
\includegraphics[height=0.23\columnwidth]{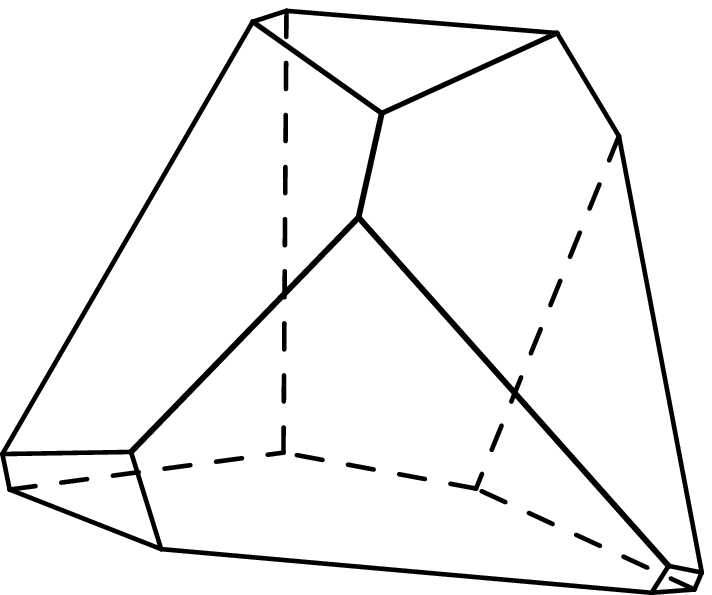}\quad\quad&
\includegraphics[height=0.25\columnwidth]{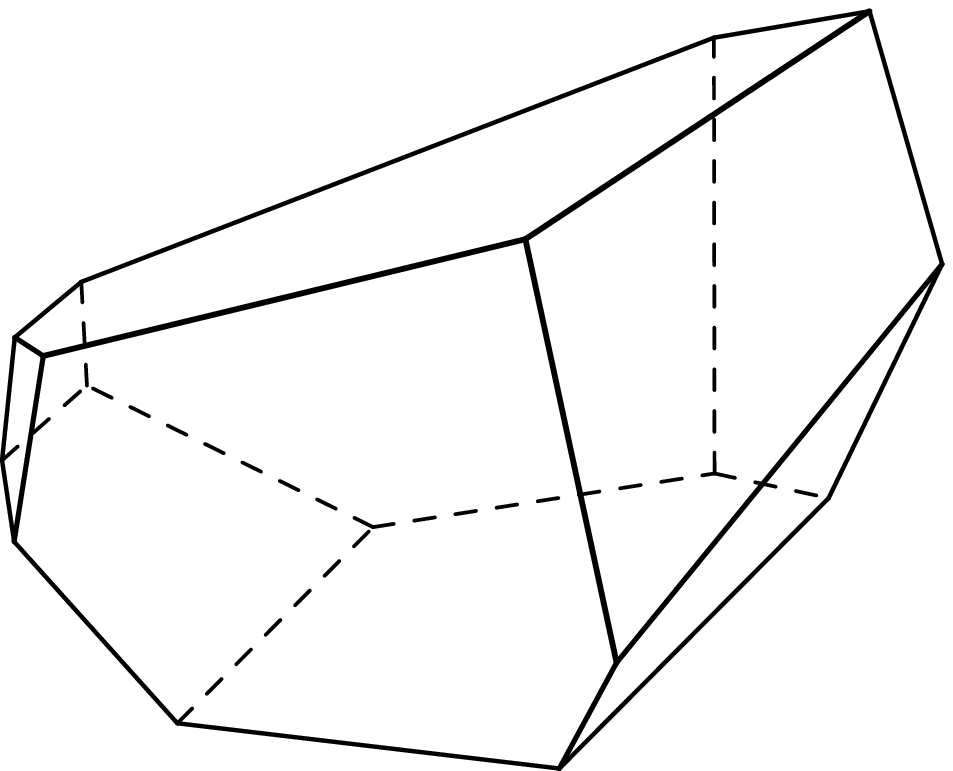}\quad\quad&
\includegraphics[height=0.25\columnwidth]{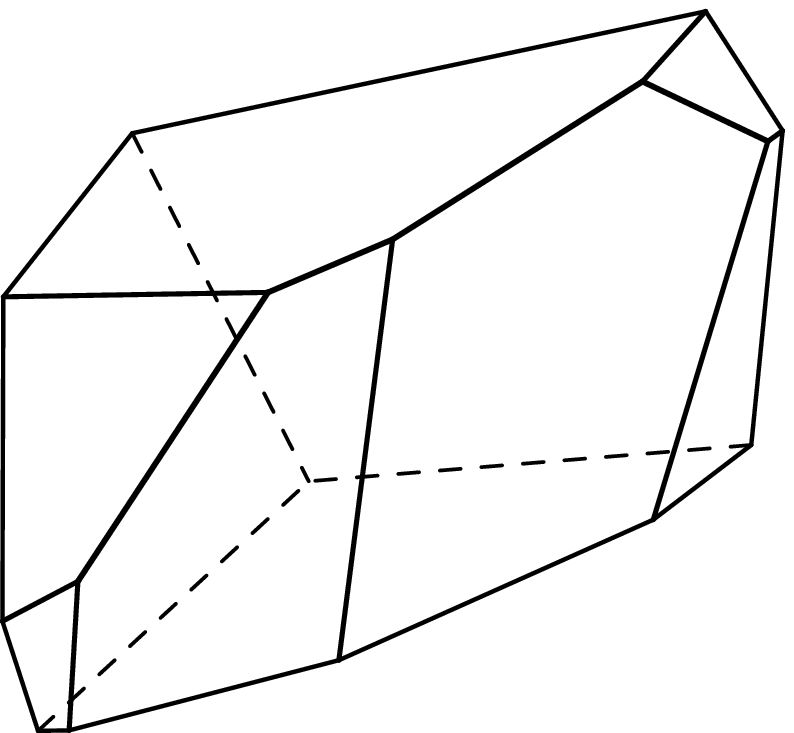}\\
(a)\quad&(b)\quad&(c)
\end{tabular}    
\end{center}
\vspace{-5mm}
\caption{Three topologically distinct cells, each with four quadrilateral, four pentagonal, and two hexagonal faces.  Type (a) appears roughly twice as frequently as type (b), which appears roughly fives times as frequently as type (c), in the Poisson-Voronoi structure.}
\label{gons8}
\end{figure}

Although the $p$-vector of a cell provides more information than a mere count of its faces, it too does not completely describe its topology.  For example, Fig.~\ref{gons8} shows three cells that share the $p$-vector $(0004420...)$ and yet are topologically distinct.  

In an earlier paper \cite{2012lazar}, we developed a method to succinctly characterize the complete topology of a cell.  That work was built on earlier work of Weinberg \cite{weinberg1965plane, weinberg1966simple}, who developed an efficient graph-theoretic algorithm to determine whether two triply-connected planar graphs are isomorphic.  We showed that the edges and vertices of a cell can be treated as a planar graph, and Weinberg's method can then be used to calculate what we call a {\it Weinberg vector} for each cell.  A Weinberg vector for a cell with $F$ faces is a vector with $6(F-2)$ integer entries that can be computed in time linear in $F^2$.  Two cells are topologically identical if and only if their Weinberg vectors are identical.  Moreover, the method by which the Weinberg vector is calculated also determines the order of the cell's associated symmetry group \cite{1966weinberg2}.  The topological type of each cell in the structures is recorded, along with its $p$-vector, symmetry order, and frequency.  We do not reproduce the algorithm for creating a Weinberg vector here, but simply refer the interested reader to \cite{2012lazar}. 

\begin{figure*}[ht]
\centering
\begin{tabular}{|c|c|c|c|c|c|c|c|}
\hline
\specialcell[t]{1. $f$=0.274\%\\\includegraphics[scale=0.275]{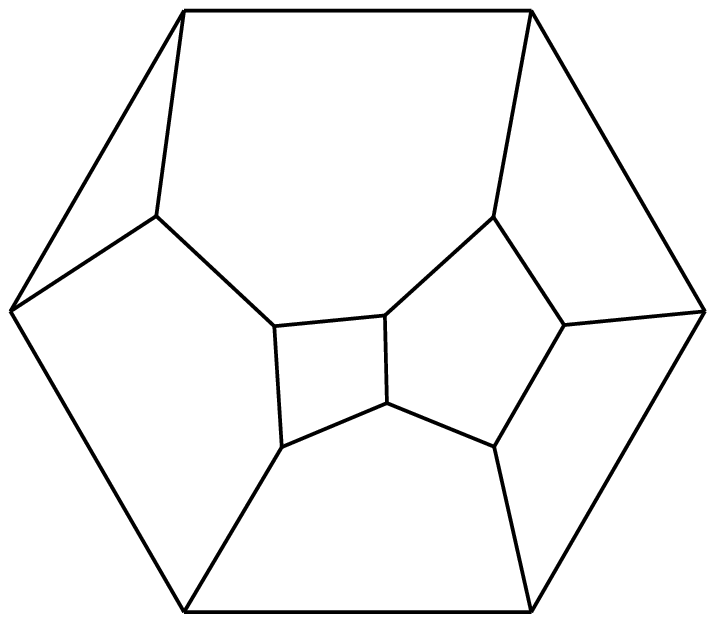}\\(0013320...)\\$F$=9, $S$=1 } & 
\specialcell[t]{2. $f$=0.166\%\\\includegraphics[scale=0.275]{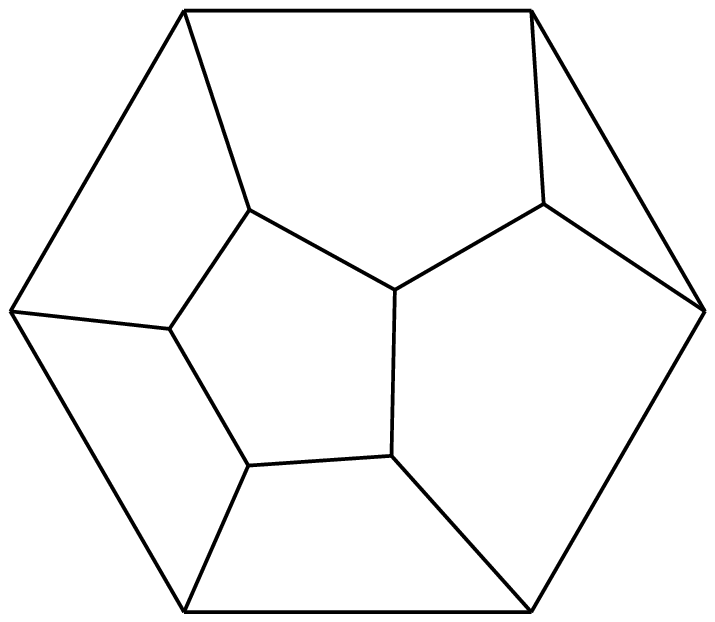}\\(0013310...)\\$F$=8, $S$=2 } & 
\specialcell[t]{3. $f$=0.158\%\\\includegraphics[scale=0.275]{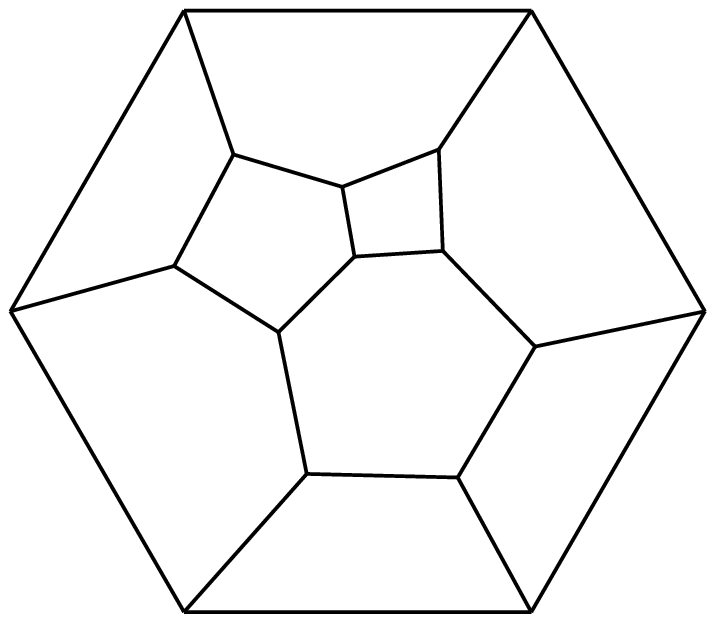}\\(0004420...)\\$F$=10, $S$=2 } & 
\specialcell[t]{4. $f$=0.120\%\\\includegraphics[scale=0.275]{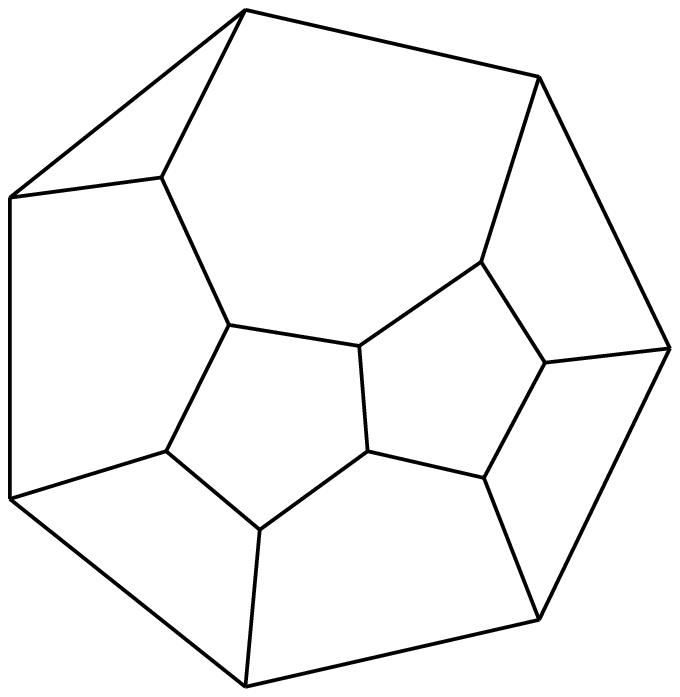}\\(0013411...)\\$F$=10, $S$=1 } & 
\specialcell[t]{5. $f$=0.117\%\\\includegraphics[scale=0.275]{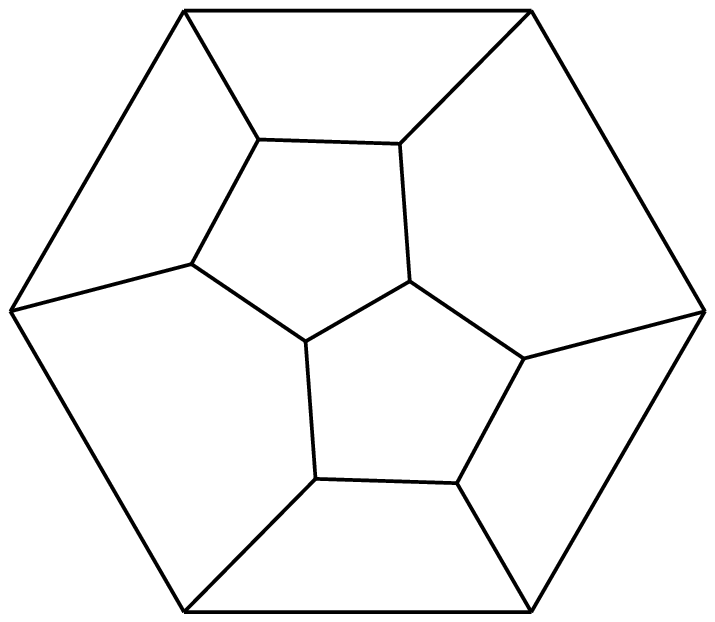}\\(0004410...)\\$F$=9, $S$=4 } & 
\specialcell[t]{6. $f$=0.101\%\\\includegraphics[scale=0.275]{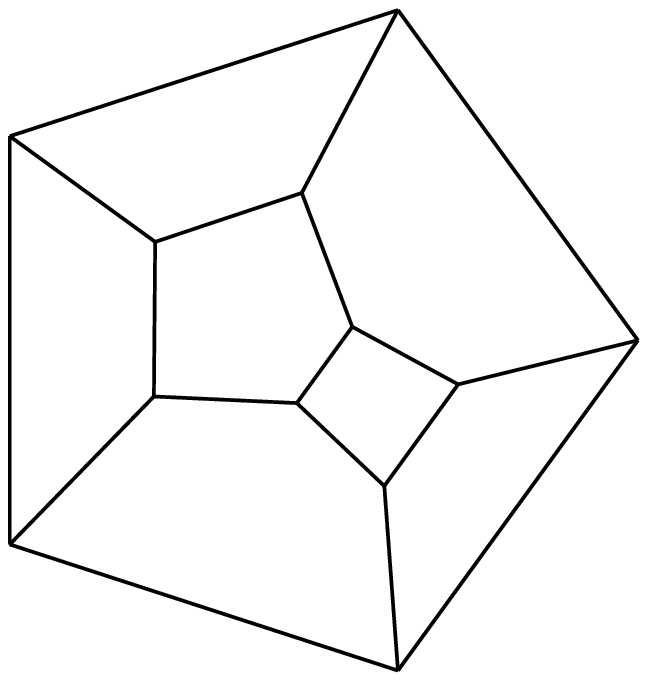}\\(0004400...)\\$F$=8, $S$=8 } & 
\specialcell[t]{7. $f$=0.096\%\\\includegraphics[scale=0.275]{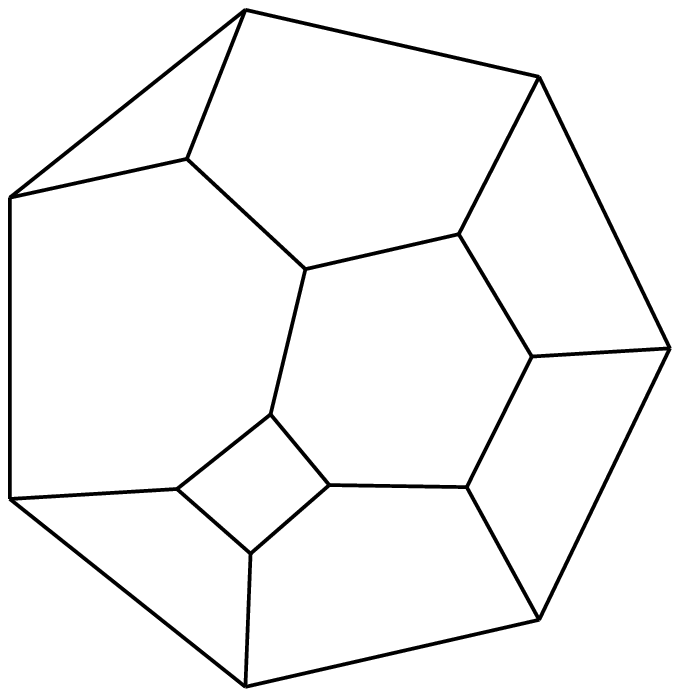}\\(0014221...)\\$F$=10, $S$=1 } &
\specialcell[t]{8. $f$=0.095\%\\\includegraphics[scale=0.275]{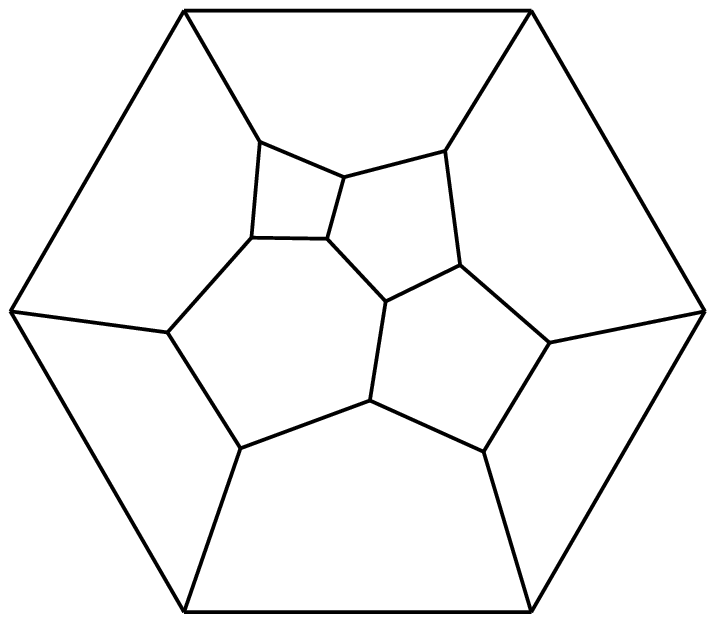}\\(0003620...)\\$F$=11, $S$=2 } \\ 
\hline
\hline
\specialcell[t]{9. $f$=0.095\%\\\includegraphics[scale=0.275]{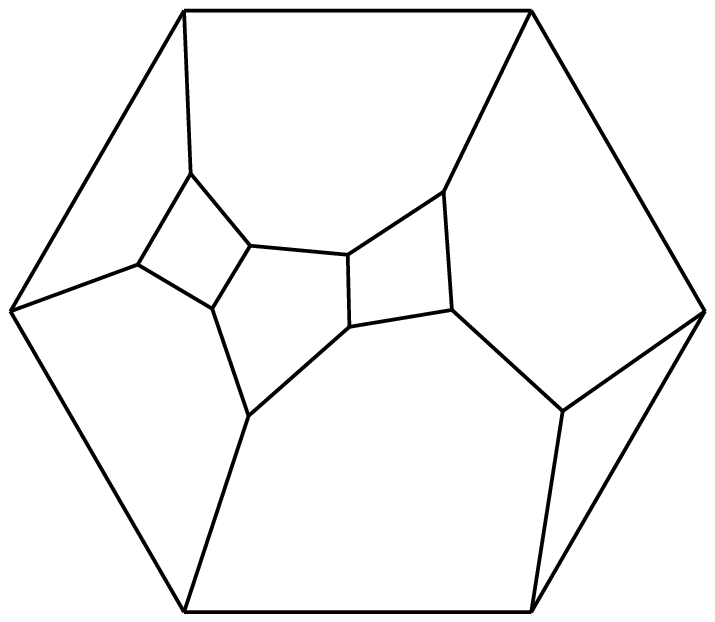}\\(0013330...)\\$F$=10, $S$=1 } & 
\specialcell[t]{10. $f$=0.094\%\\\includegraphics[scale=0.275]{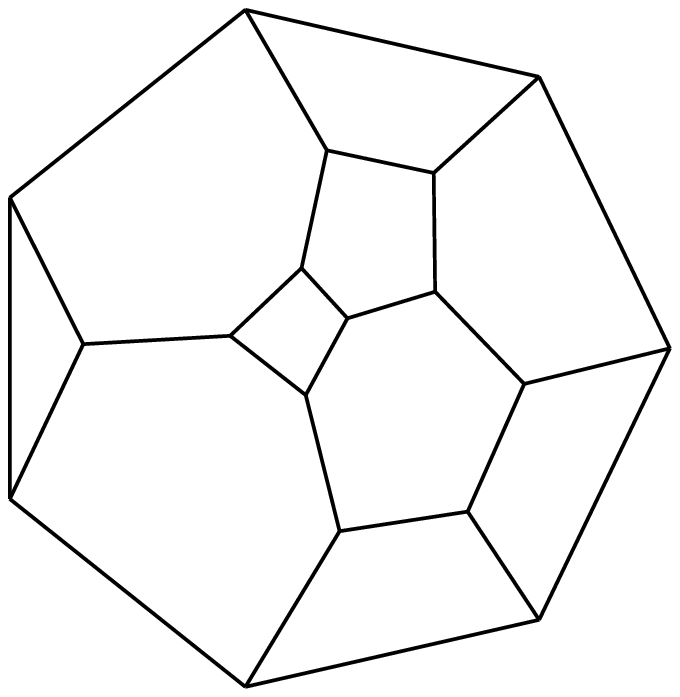}\\(0004430...)\\$F$=11, $S$=1 } & 
\specialcell[t]{11. $f$=0.094\%\\\includegraphics[scale=0.275]{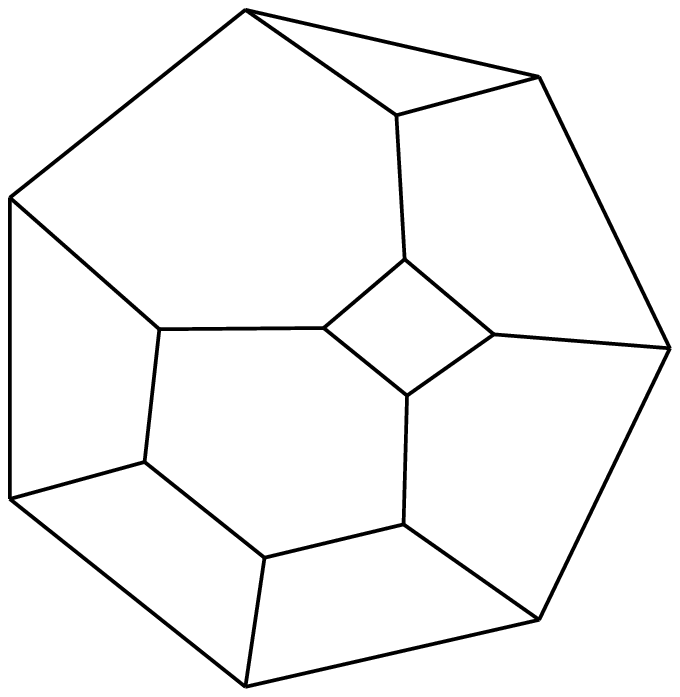}\\(0014221...)\\$F$=10, $S$=1 } & 
\specialcell[t]{12. $f$=0.093\%\\\includegraphics[scale=0.275]{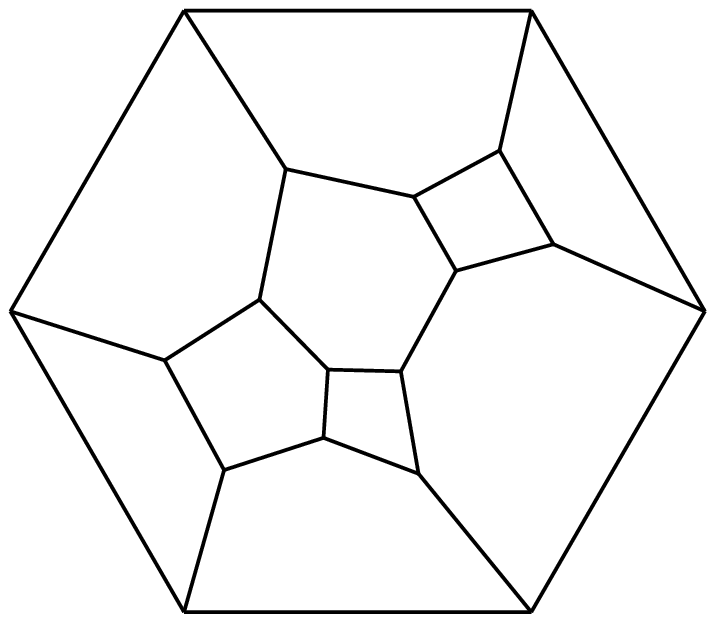}\\(0014231...)\\$F$=11, $S$=1 } & 
\specialcell[t]{13. $f$=0.091\%\\\includegraphics[scale=0.275]{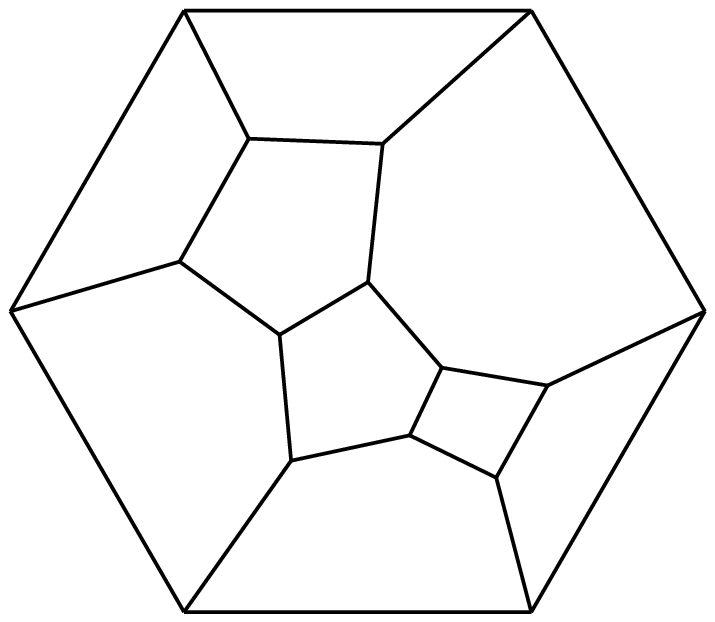}\\(0004420...)\\$F$=10, $S$=2 } & 
\specialcell[t]{14. $f$=0.091\%\\\includegraphics[scale=0.275]{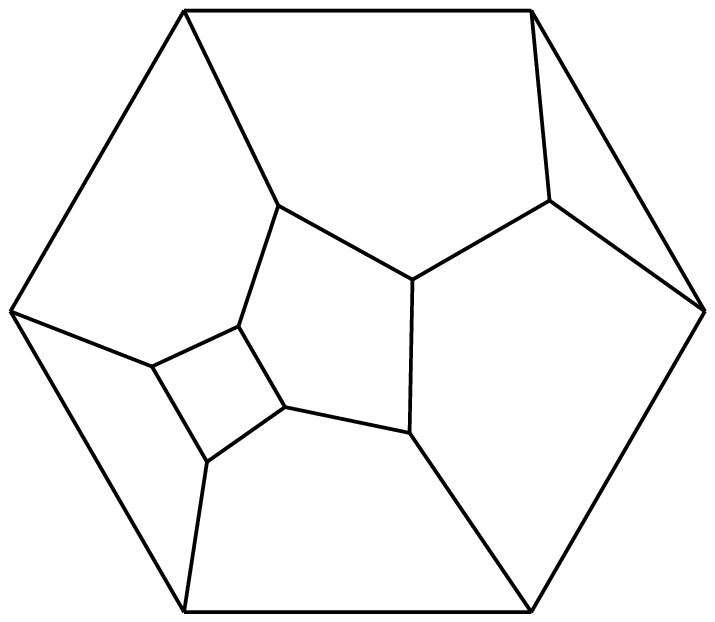}\\(0012510...)\\$F$=9, $S$=2 } & 
\specialcell[t]{15. $f$=0.090\%\\\includegraphics[scale=0.275]{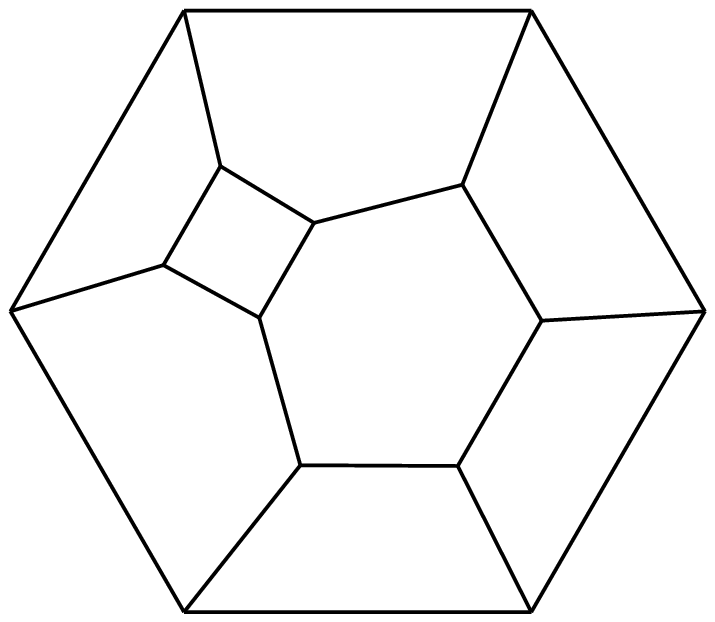}\\(0005220...)\\$F$=9, $S$=4 } & 
\specialcell[t]{16. $f$=0.088\%\\\includegraphics[scale=0.275]{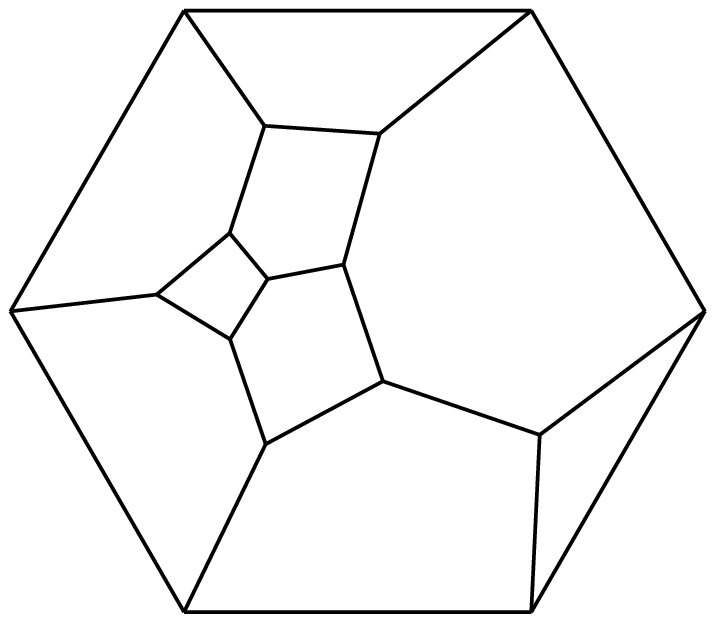}\\(0012520...)\\$F$=10, $S$=2 } \\ 
\hline
\hline
\specialcell[t]{17. $f$=0.082\%\\\includegraphics[scale=0.275]{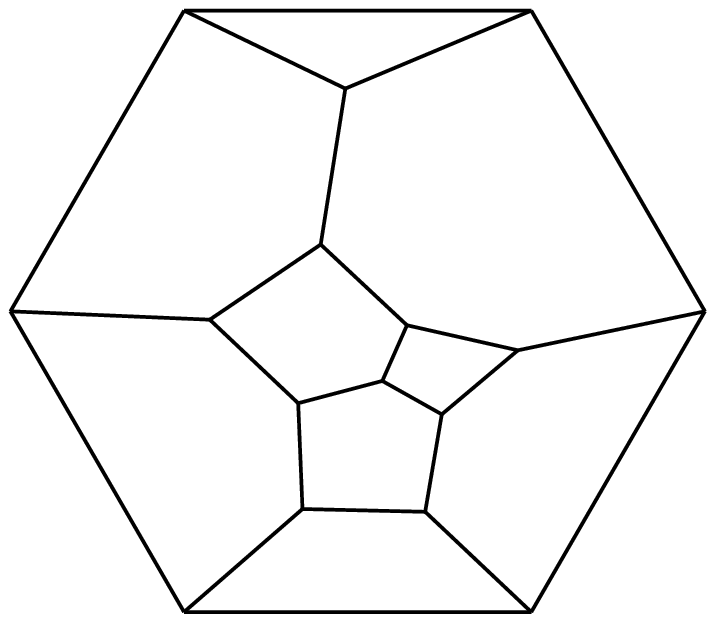}\\(0012520...)\\$F$=10, $S$=2 } & 
\specialcell[t]{18. $f$=0.082\%\\\includegraphics[scale=0.275]{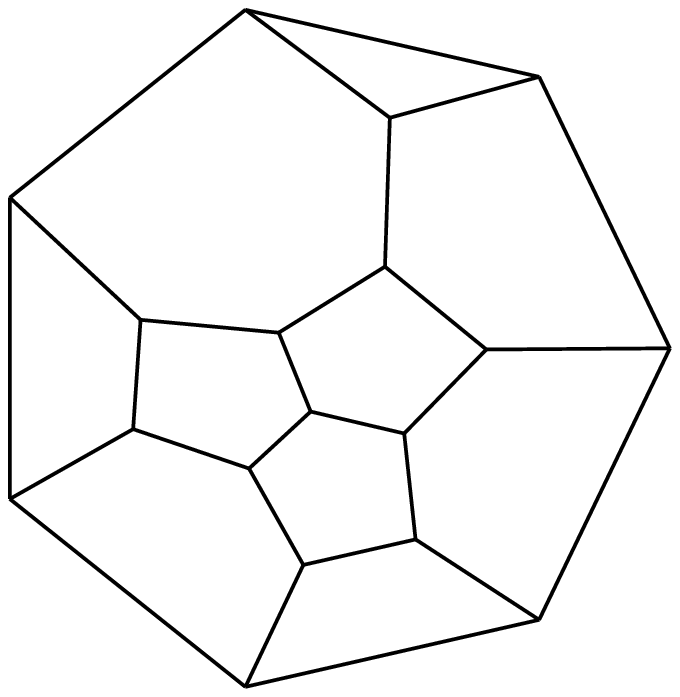}\\(0012611...)\\$F$=11, $S$=1 } & 
\specialcell[t]{19. $f$=0.081\%\\\includegraphics[scale=0.275]{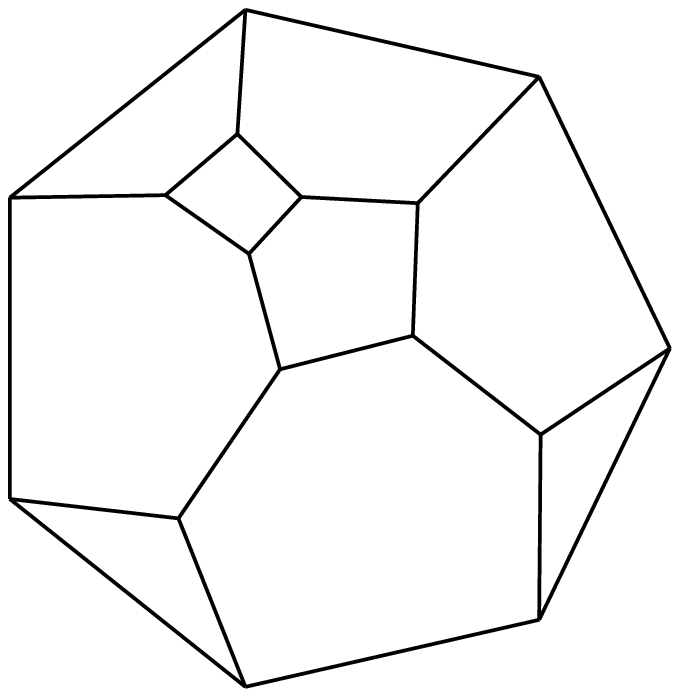}\\(0022321...)\\$F$=10, $S$=1 } & 
\specialcell[t]{20. $f$=0.080\%\\\includegraphics[scale=0.275]{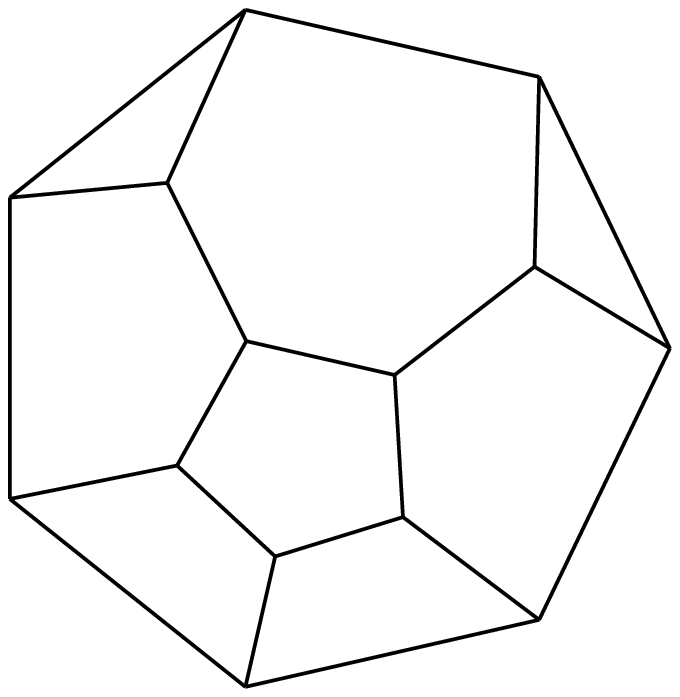}\\(0022311...)\\$F$=9, $S$=2 } & 
\specialcell[t]{21. $f$=0.079\%\\\includegraphics[scale=0.275]{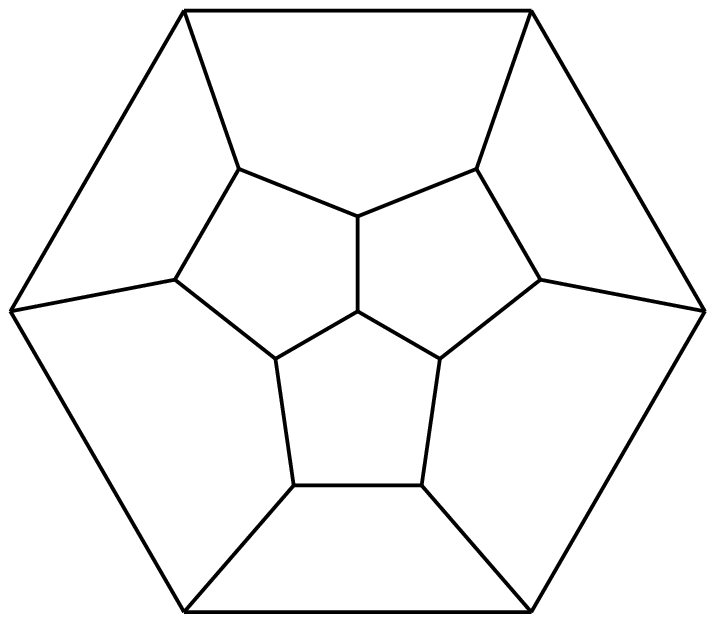}\\(0003610...)\\$F$=10, $S$=6 } & 
\specialcell[t]{22. $f$=0.077\%\\\includegraphics[scale=0.275]{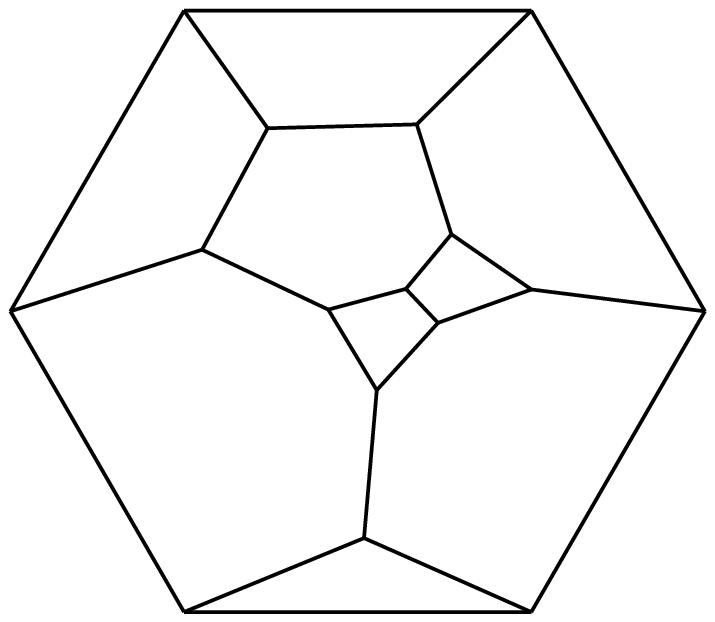}\\(0014140...)\\$F$=10, $S$=2 } & 
\specialcell[t]{23. $f$=0.075\%\\\includegraphics[scale=0.275]{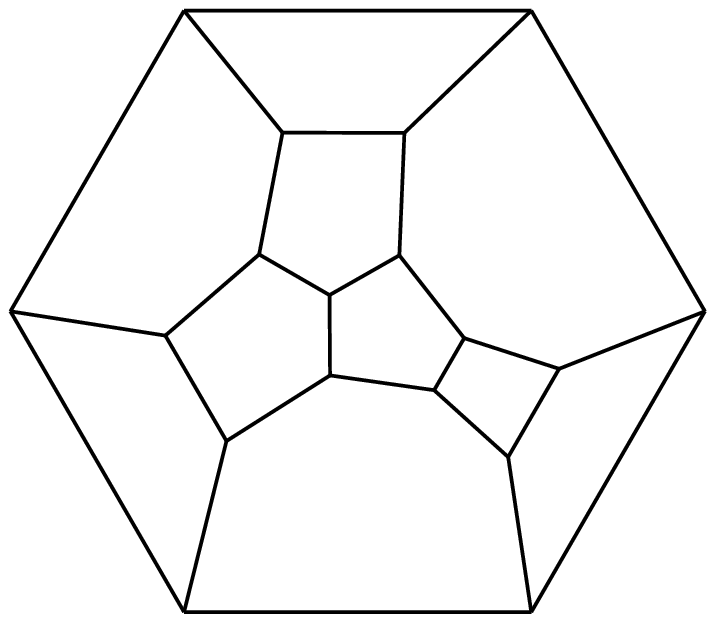}\\(0004430...)\\$F$=11, $S$=2 } & 
\specialcell[t]{24. $f$=0.073\%\\\includegraphics[scale=0.275]{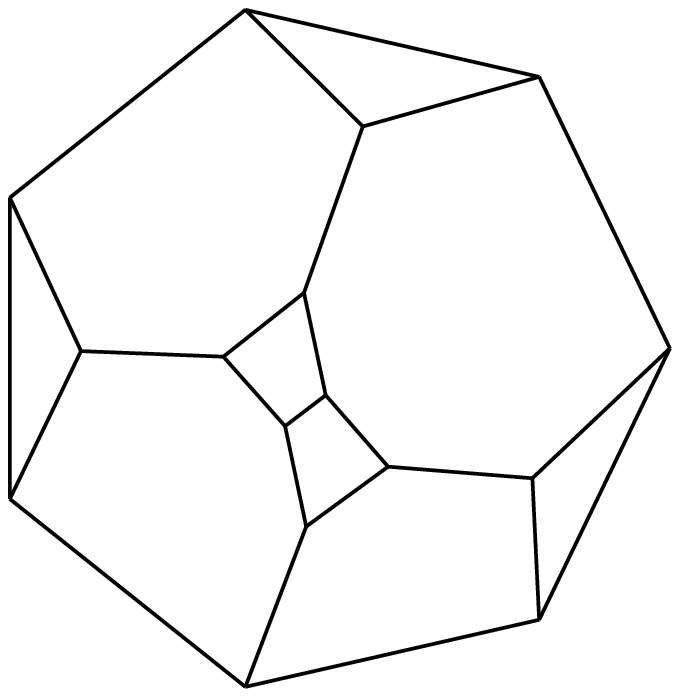}\\(0032122...)\\$F$=10, $S$=1 } \\ 
\hline
\end{tabular}
\caption{Schlegel diagrams of the 24 most common topological types among the Poisson-Voronoi cells.  Listed for each type is its frequency $f$, $p$-vector, number of faces $F$, and order $S$ of its associated symmetry group.  In these data, there are four pairs of Weinberg vectors which share $p$-vectors.}
\label{schlegel-diagrams}
\end{figure*}

In the remainder of this paper, we use Schlegel diagrams to help visualize topological types.  A {\it Schlegel diagram} is constructed by projecting the boundary of a cell onto one of its faces in a way that vertices not belonging to that face lie inside it and no edges cross \cite{schlegel1883theorie, schlegel1886uber}.  Figure \ref{schlegel-diagrams} shows Schlegel diagrams for the 24 most common topological types that appear in the Poisson-Voronoi structure.  Along with the Schlegel diagram for each, we show its frequency $f$, number of faces $F$, $p$-vector, and order $S$ of its symmetry group.

\setlength{\tabcolsep}{3pt}

Each of the six most common topological types have 10 or fewer faces.  This may be surprising in light of the fact that of the 48 most commonly occurring $p$-vectors, only one had fewer than 10 faces.  This can be understood by considering that many distinct topological types can share the same $p$-vector, as illustrated earlier in Fig.~\ref{gons8}.  This degeneracy increases with the number of faces, and so $p$-vectors of cells with many faces can appear frequently even if no single topological type with that $p$-vector appears frequently.  Conversely, $p$-vectors with few faces are typically shared by few distinct topological types.  The most frequently occurring $p$-vector (001343100...) is shared by 38 distinct topological types,\footnote{This can be extracted from data available on {\it The Manifold Page} \cite{2013lutz}; data for these 38 types are included in Fig.~20 of the Supplemental Material.} not one of which appears among the 24 most common types.  

The most common topological type in the Poisson-Voronoi structure (Fig.~\ref{schlegel-diagrams}, entry 1) has $p$-vector $(0013320...)$ and occurs with frequency 0.273\%.  Two factors appear to contribute to its relative high frequency.  First, its distribution of face types closely resembles that of the structure as a whole (Fig.~\ref{multiedges}).  Specifically, four- and five-sided faces appear most frequently, followed by six-sided faces and then three-sided faces.  Second, no other topological type shares this $p$-vector.  Despite its frequency, however, it is difficult to describe it as a ``typical'' Poisson-Voronoi cell, given how few cells are of this type.  

\begin{figure*}[ht!]
\centering
\begin{tabular}{|c|c|c|c|c|c|c|c|}
\hline
\specialcell[t]{$N=325$\\$F$=4, $S$=24 \\\includegraphics[height=1.9cm, angle=0]{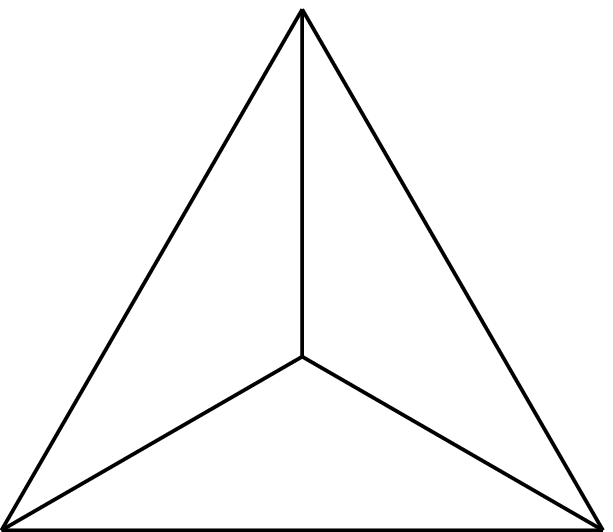}\\ \\$tetrahedron$\\} & 
\specialcell[t]{$N=22227$\\$F$=8, $S$=24 \\\includegraphics[height=1.9cm, angle=0]{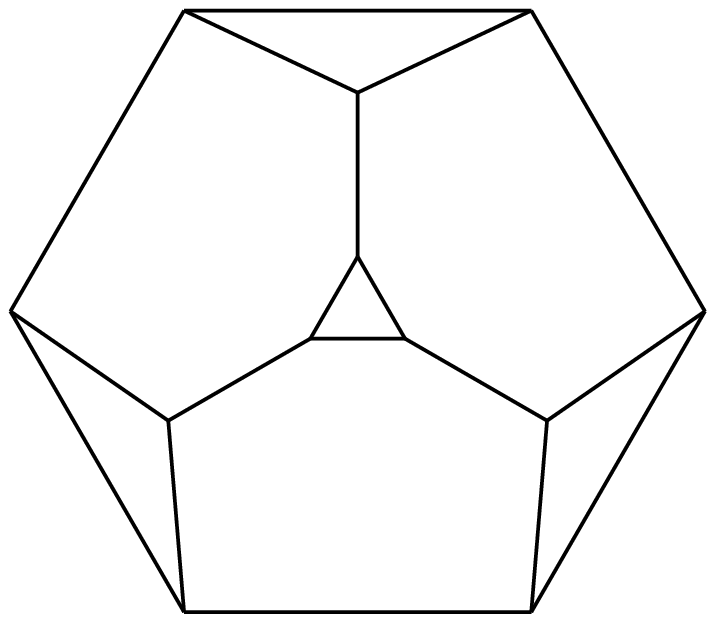}\\\\$truncated$\\$tetrahedron$} & 
\specialcell[t]{$N=$ 23744\\$F$=6, $S$=48\\\includegraphics[height=1.9cm]{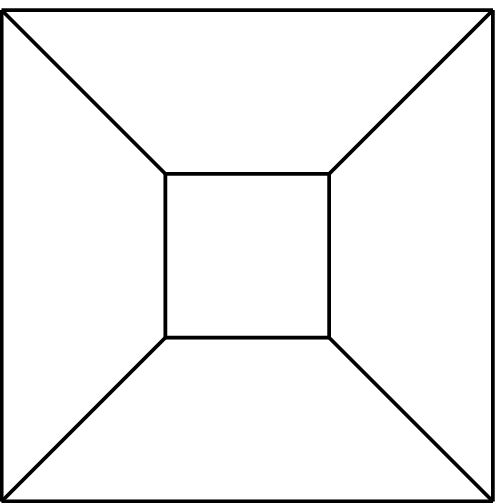}\\  \\$cube$\\} & 
\specialcell[t]{$N=41$\\$F$=14, $S$=48 \\\includegraphics[height=1.9cm, angle=0]{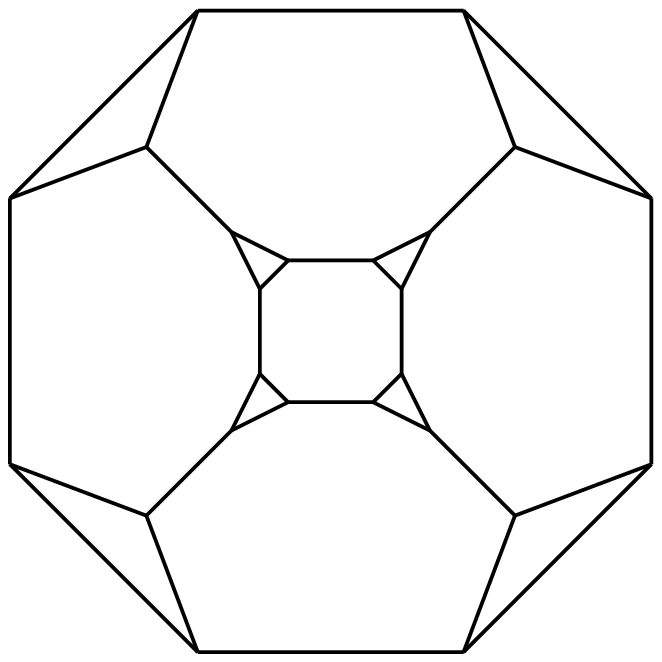}\\\\$truncated$\\$cube$} & 
\specialcell[t]{$N=3612$\\$F$=12, $S$=120\\\includegraphics[height=1.9cm]{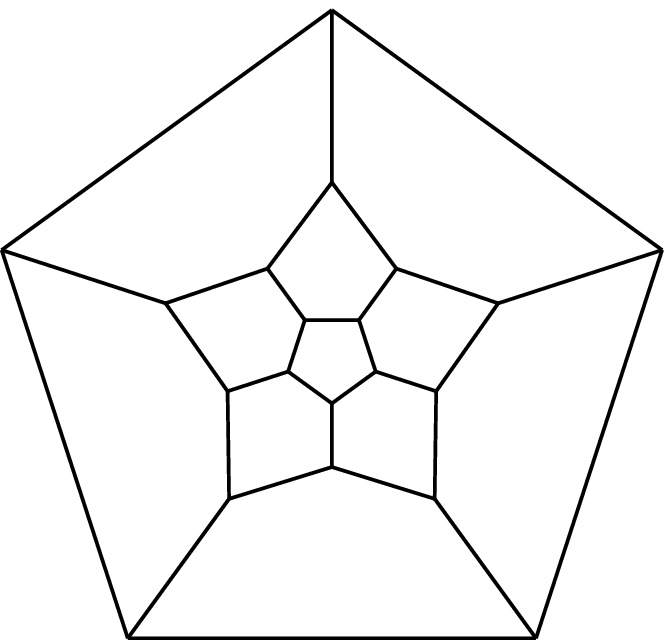}\\\\ $pentagonal$\\$dodecahedron$} & 
\specialcell[t]{$N=1$\\$F$=17, $S$=20 \\\includegraphics[height=1.9cm, angle=0]{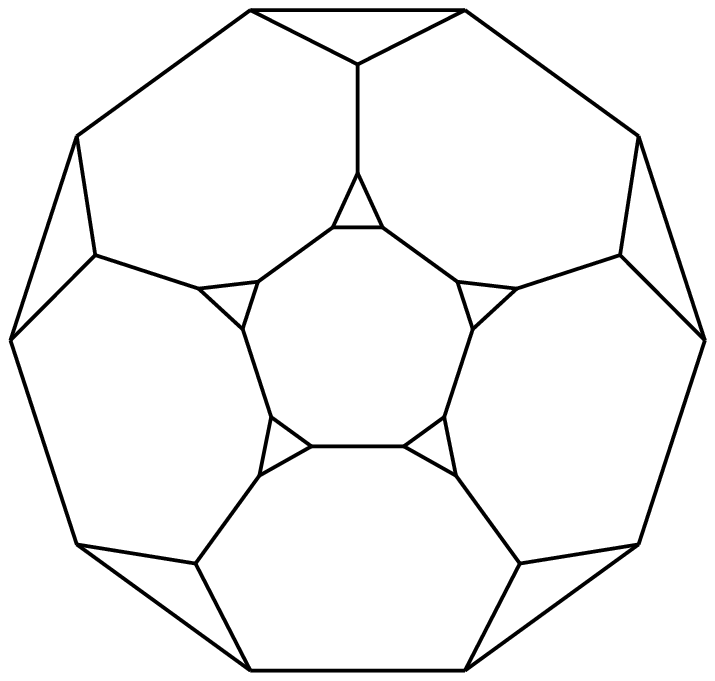}\\$truncated$\\$prism$ $over$\\$a$ $prism$} &
\specialcell[t]{$N=$55\\$F$=16, $S$=28\\\includegraphics[height=1.9cm]{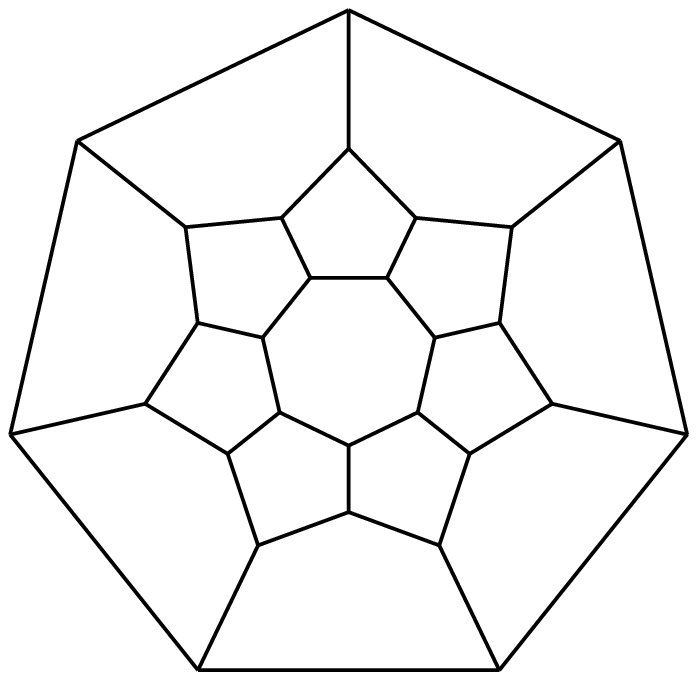}\\ $pentagonal$\\$antiprism$ $over$\\$heptagon$} & 
\specialcell[t]{$N=623$\\$F$=14, $S$=48\\\includegraphics[height=1.9cm]{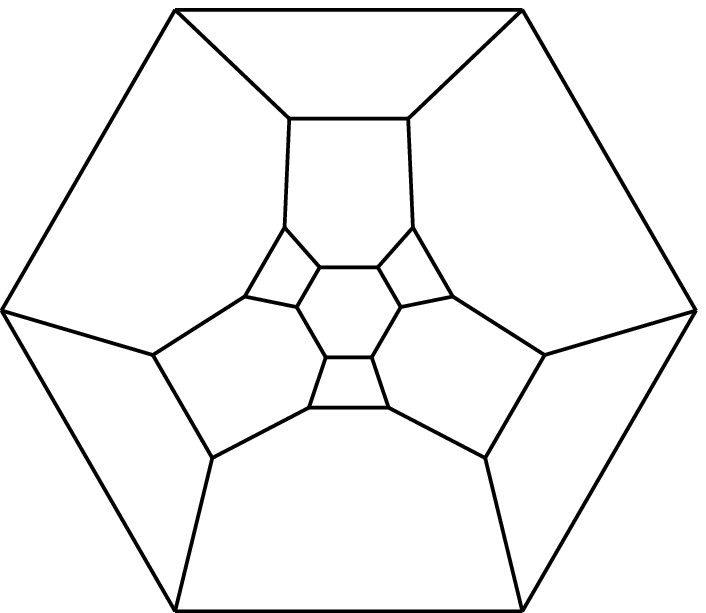}\\ \\$Kelvin$ $cell$} \\
\hline
\end{tabular}
\caption{Highly symmetric polyhedra.  For each type, we include the number of times $N$ it appears in the 250,000,000 cell data set, its number of faces $F$, and the order $S$ of its symmetry group.}
\label{famous-diagrams}
\end{figure*}

Figure \ref{famous-diagrams} illustrates Schlegel diagrams of a number of highly symmetric polyhedra: the tetrahedron, truncated tetrahedron,  cube, truncated cube,  pentagonal dodecahedron,  truncated pentagonal prism,  pentagonal antiprism over a heptagon, and  truncated octahedron.  The first, third, and fifth of these are Platonic solids that occur with non-zero probability in the Poisson-Voronoi structures.  The last of these shapes, often referred to as the {\it Kelvin tetrakaidecaheron} or {\it Kelvin cell}, was conjectured by Lord Kelvin \cite{sir1887division, kelvin1894homogeneous} to tile three-dimensional space with a minimal surface area, very much like the regular hexagon tiles the plane with minimal perimeter \cite{hales2001honeycomb}.  
\setlength{\tabcolsep}{3pt}

It can be shown that every topological type appears in the Poisson-Voronoi tessellation with a non-zero frequency, and so the appearance of these highly symmetric shapes is not surprising.  However, their relative frequencies warrant  attention.  The truncated cube and the Kelvin cell both have 14 faces and a symmetry group of order $S=48$, and yet they occur with substantially different frequencies.  It is clear that frequencies are not entirely determined by the number of faces of a cell nor by the order of its associated symmetry group.  It is unclear how these topological features impact frequency.

The 24 most common topological types in Poisson-Voronoi structures account for less than 2.5\% of all cells.  By contrast, the distribution of topological types in grain growth structures is substantially more concentrated \cite{2012lazar}.  There, the 24 most common types account for over 25\% of all cells \cite{2012lazar}.  While space-filling constraints in both the Poisson-Voronoi and grain growth structures create a bias towards certain topological types, the curvature flow process that governs the evolution of grain growth structures leads to a secondary bias towards cells that exhibit a low surface area to volume ratio.  Distributions of topological types have not been collected, to the best of our knowledge, for other cellular structures.

\subsection{Order of symmetry groups}
As noted earlier, the algorithm which determines the Weinberg vector of a cell also determines the order of its associated symmetry group.
\begin{figure}
\centering
\includegraphics[width=1.\columnwidth]{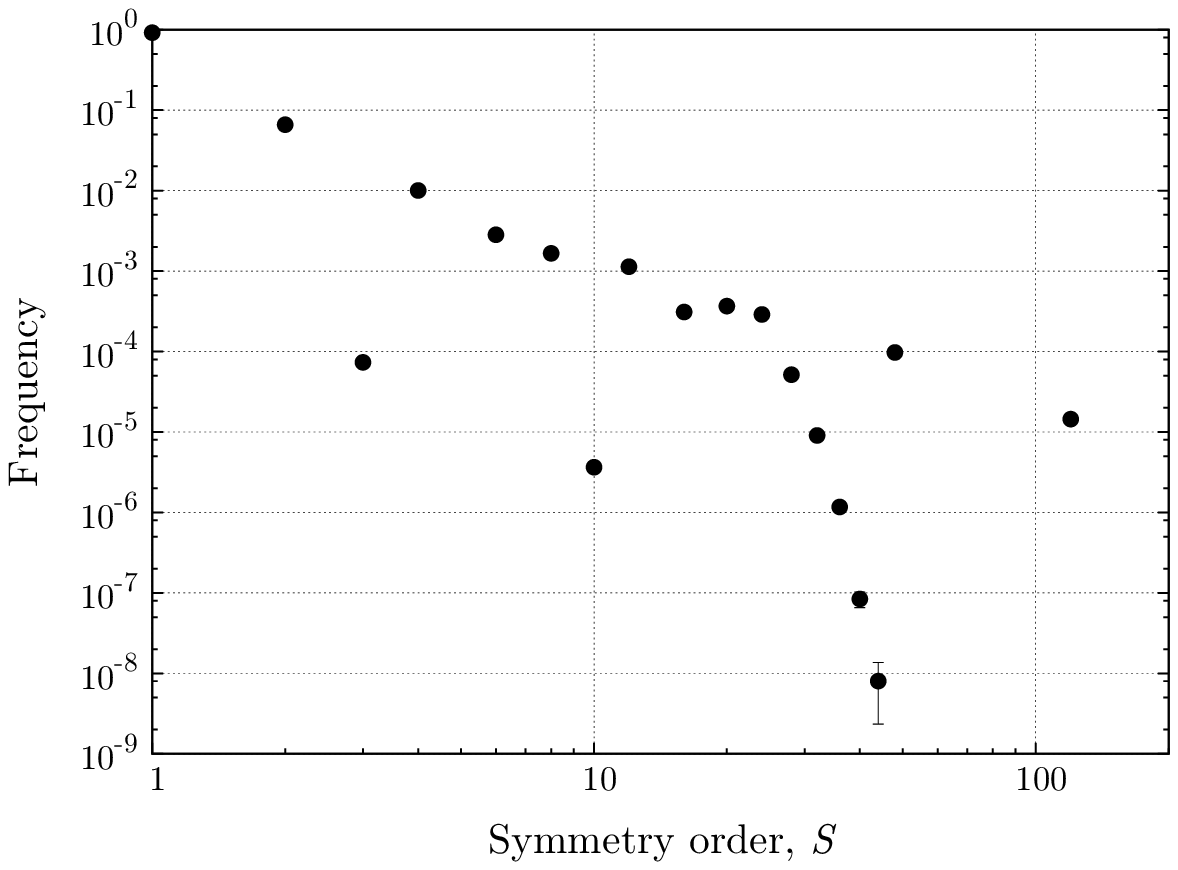}
\caption{Distribution of orders of cell symmetry groups.  Error bars indicate standard error from the mean; in many cases the error bars are not visible because they are smaller than the points.}
\label{symmetries}
\end{figure}
Figure \ref{symmetries} shows the distribution of symmetry orders among all cells.  Roughly 91.71\% of cells have only the trivial symmetry (order 1), 6.61\% have a symmetry of order 2, and 1.00\% have a symmetry of order 4.  The remaining 0.68\% have symmetries of order 3 or higher than 4.  The probability of finding a cell with a particular symmetry order generally decreases quickly with the order, subject to certain secondary rules.  More specifically, odd numbers and numbers whose prime factors are all large appear highly infrequently.  Odd orders appear in topological types with rotational symmetries but without mirror or inversion symmetries.  Therefore, we find no cells with symmetry order 13, for example, even though we find many with symmetry orders 16, 24 and 48.

The average symmetry order of Poisson-Voronoi cells is 1.16.  This might be contrasted with the case of grain growth structures \cite{2011lazar}, where the average observed symmetry order is 3.09 \cite{2012lazar}. This discrepancy may be due to the tendency of mean curvature flow to minimize surface area, although how this correlates with topological symmetry is unclear since curvature is a geometric quantity.

We note that although some symmetries of a cell can be observed in its Schlegel diagram, the diagram can often obscure other symmetries.  Entries 5 and 21 in Fig.~\ref{schlegel-diagrams}, for example, might appear at first sight more symmetric than entry 6, and yet the latter has the highest symmetry order of the three.  To understand this apparent inconsistency between the diagram and the data, we note that entries 5 and 21 both have only one hexagonal face.  Therefore, aside from rotations or reflections, there is no way to redraw identical graphs using a different face as the outside polygon.  Entry 6, in contrast, has four pentagonal faces, and the graph can be redrawn with each of those faces as the outside polygon.  These contribute additional symmetries which might be initially overlooked when considering the Schlegel diagrams.

\subsection{Cloths and swatches}
\label{subsec_cloths_and_swatches}

The types of topological information considered up to this point concern the configuration of faces and edges on cell surfaces, but not the topology of the network of cells extending throughout the tessellation. This is more difficult to address for at least two reasons. First, much more information is involved in characterizing the topology of the cell network than a single cell.  Second, collecting statistics relating to the topological features of the boundary network requires a much larger computational effort. 

One approach \cite{2012mason} is to construct and collect statistics of {\it swatches}, where a swatch is roughly a collection of labels for the vertices (intersection of four cells) in a portion of the tessellation. The labeling procedure is performed as follows. Let one of the vertices of the structure be designated as the root, and assign a label to this vertex. The swatch is expanded by a canonical procedure that assigns labels to any vertices connected by a single edge to one of the most recently labeled vertices. While performing this procedure on a quadrivalent Cayley tree would give a single, unique label for every vertex, in practice the network of edges contains loops around every face.  The result is that vertices are often assigned multiple labels; the labels of such a vertex are considered to be equivalent, and define an equivalence relation. After $r$ iterations of assigning labels, the set of equivalence relations is known as a swatch of order $2r$.  A swatch contains all of the topological information about the network of cells in the region around the root; as evidence of this, consider that applying the equivalence relations to a labeled quadrivalent Cayley tree exactly reproduces the network of edges. A swatch therefore classifies the topology of the locale, analogous to the way a Weinberg vector classifies a single cell.

For a positive integer $k$, vertices may be randomly selected from the tessellation to serve as root vertices for the construction of swatches of order $k$. The frequencies at which the different types of swatches appear during this sampling gives a probability distribution that effectively describes the distribution of local topological environments. Allowing $k$ to vary over the positive integers gives an infinite set of probability distributions, collectively known as the {\it cloth} of the cell network.

The probability distribution for a given value of $k$ further defines a {\it $k$-entropy} via the Shannon entropy formula \cite{1948shannon}. The $k$-entropy indicates the variability of the local topological environment, and is a well-defined property of an infinite and statistically homogenous \cite{2012mason} cellular structure. The $k$-entropies are reported here as a function of $k$.

\begin{figure}[h]
\centering
\vspace{3mm}
\includegraphics[width=1.\columnwidth]{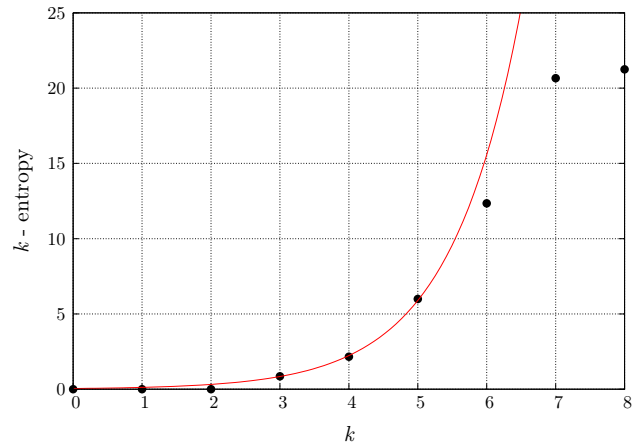}
\caption{The $k$-entropy calculated using all $V$ = 1,691,911,665 vertices in the three-dimensional Poisson-Voronoi structures, containing 250,000,000 cells.  The red line is $0.0457e^{0.972k}$.}
\label{kentropy}
\end{figure}

We constructed swatches for all $V$ = 1,691,911,665 vertices in the data set, and report the $k$-entropies for $k = 0$ to $8$ in Fig.~\ref{kentropy}.
The $k$-entropy of the system is 0 for $k = {0,1,2}$ since a sufficiently small neighborhood around any vertex is topologically trivial (e.g., an isolated vertex, or a vertex connected to four edges). On the other hand, for $k = {7,8}$ there are so many possible local environments that the number of swatch types is much larger than the number of vertices in the system. As a result, no swatch type is sampled more than once during the sampling procedure, apparently bounding the $k$-entropy from above due to the finite system size. This probably affects the $k$-entropy for $k = 6$ as well, where the probability distribution of swatch types appears to be insufficiently sampled. The $k$-entropy for the remaining three values of $k = {3,4,5}$ is adequately fit by a least-squares procedure to an exponential function.

Suppose that the $k$-entropy is roughly proportional to the natural logarithm of the number of swatch types (this is precisely true in the case of a uniform distribution), and that the exponential form suggested above holds, i.e.,
\begin{equation}
ln(N_k) \sim c_0 \exp( c_1 \cdot k ),
\end{equation}
where $N_k$ is the number of swatch types of order $k$. This implies that $N_k$ grows roughly as a double exponential,
\begin{equation}
N_k \sim \exp( c_0 \exp( c_1 \cdot k ) ).
\end{equation}
Although more data points would certainly help to validate this suggestion, the apparent growth rate means that adequately sampling the $k$-entropy for even $k = 6$ is extremely computationally demanding.

Comparing our results with the $k$-entropies for grain growth structures \cite{2012mason}, we find that the $k$-entropies of the Poisson-Voronoi tessellation are slightly higher. The slightly higher values are consistent with the greater variability of cell types in the Poisson-Voronoi tessellation, as is evidenced by the differences in the distributions of the $p$-vectors or of the Weinberg vectors for the two structures. That said, the similarity of the $k$-entropies does not imply the similarity of the local topological environments, but only that the amount of variability in the two structures is similar.

\section{Geometrical data}

One of the most frequently studied geometrical-topological relations is that between the number of faces of a cell and its expected volume.  Before considering that relationship, we look at the distribution of volumes over all cells and at the partial distributions of volumes limited to cells with fixed numbers of faces.  Likewise, we consider the distribution of surface areas of cells, as well as areas and perimeters of faces.

\subsection{Distribution of volumes}
\label{Distribution of volumes}

\begin{figure}[h]
\centering
\vspace{3mm}
\includegraphics[width=1.\columnwidth, trim = 4mm 0 2mm 0mm]{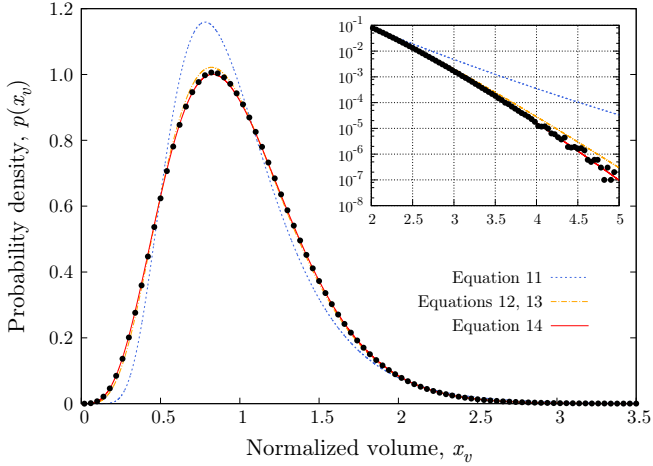}
\caption{(Color online) Distribution of normalized cell volumes, $x_v = v/\langle v \rangle$.  The standard deviation of the data set is 0.4231, consistent with analytical results to within numerical accuracy.  The curves represent suggested forms of the distribution, as described in the text.  The inset shows a subset of the data on a semilogarithmic plot.}
\label{volumefits}
\end{figure}

Despite much interest in understanding the distribution of volumes among three-dimensional Poisson-Voronoi cells, few rigorous results are available.  Throughout this section we use $x_v = v/\langle v \rangle$ to denote normalized cell volumes, where $v$ is the volume of a particular cell and $\langle v \rangle$ is the average volume per cell.  Throughout the paper we use $p(x)$ to denote the probability  distribution of a variable $x$.  Gilbert \cite{1962gilbert} and Brakke \cite{1985brakke} obtained exact integral expressions for the variance of this distribution; numerical integration yields a variance of 0.1790.  Figure \ref{volumefits} shows the distribution of cell volumes in our data set.  The distribution exhibits a maximum at roughy $x_v = 0.831$  with a probability density of $p(x_v)=1.006$.  Several suggestions have been made for the form of this distribution.  

Hanson \cite{hanson1983voronoi} suggested that the volume is distributed according to a Maxwell distribution:
\begin{equation}
p(x) = \frac{32}{\pi^2}x^2e^{-4x^2/\pi}.
\label{maxwelleq}
\end{equation}
Hanson acknowledged the lack of physical motivation to substantiate this suggestion and realized that this form does not provide a  particularly good fit to the data for all $x$. In addition, the variance of this distribution, $3\pi/8-1$, is not consistent with the exact results of Gilbert \cite{1962gilbert} and Brakke \cite{1985brakke}.  

Ferenc and N{\'e}da \cite{2007ferenc}, motivated by a known result of one-dimensional Poisson-Voronoi structures, and based on the study of 18,000,000 three-dimensional Poisson-Voronoi cells, proposed 
\begin{equation}
p(x) = \frac{3125}{24}x^4e^{-5x}.
\label{ferenceq}
\end{equation}
The variance of this distribution is 1/5 which, again, is inconsistent with the known exact result.  Ferenc and N{\'e}da acknowledged that this form is empirical and  not completely consistent with the true distribution.  Because Eqs.~(\ref{maxwelleq}) and (\ref{ferenceq}) are inconsistent with the exact, known properties of this distribution,  we do not consider them further.

Kumar {\it et al.}~\cite{1992kumar} considered a lognormal distribution as an approximation of the Poisson-Voronoi volumes distribution:
\begin{equation}
p(x) = \frac{1}{x\sqrt{2\pi}\sigma}\exp\left[-\frac{(\ln x - \mu)^2}{2\sigma^2}\right],
\label{lognormaleq}
\end{equation}
where $\mu$ and $\sigma$ are determined by fitting.  Using simulation data, Kumar {\it et al.}~\cite{1992kumar} obtained $\sigma = 0.4332$ and $\mu = -0.0735$.  However, since we know the mean and variance exactly \cite{1962gilbert,1985brakke}, these two parameters are completely determined: $\sigma=0.4058$ and $\mu=-0.0823$.  As noted by Kumar {\it et al.}~\cite{1992kumar} and others \cite{fatima1988grain}, a lognormal distribution appears to have little physical justification, and given its weakness in fitting the data, can serve only as a rough guide to the actual distribution.

Another suggested form for the Poisson-Voronoi volume distribution is a $\Gamma$ distribution function with one, two, or three fitting parameters.  Kiang \cite{kiang1966random} attempted to extend results known for one-dimensional systems and limited simulation data to suggest a volume distribution of the form:
\begin{equation}
p(x) = \frac{\gamma}{\Gamma(\gamma)}(\gamma x)^{\gamma-1}e^{-\gamma x},
\label{gammaeqA}
\end{equation}
where $\gamma$ is a constant which Kiang believed to be 6.  Andrade and Fortes \cite{andrade1988distribution}, using a larger data set, concluded that $\gamma\approx 5.56$.  Kumar {\it et al.}~\cite{1992kumar} found $\gamma = 5.7869$.  All these fits should only be considered approximations, since the variance $\sigma^2$ is known exactly.  This, then, determines $\gamma = 1/\sigma^2 = 5.586$.

Kumar {\it et al.}~\cite{1992kumar} also suggested a two-parameter version of this distribution,
\begin{equation}
p(x) = \frac{x^{\gamma-1}}{\beta^\gamma\Gamma(\gamma)}e^{-x/\beta}.
\label{gammaeqB}
\end{equation}
Using simulation data, Kumar {\it et al.}~obtained best fit values of the constants, $\beta=0.1782$ and $\gamma=5.6333$.  However, the exact variance results require $\beta=\sigma^2=0.1790$ and $\gamma=1/\sigma^2 = 5.586$. With these values, Eq.~(\ref{gammaeqB}) reduces to Eq.~(\ref{gammaeqA}). 

Tanemura \cite{2003tanemura} suggested a three-parameter version of the distribution,
\begin{equation}
p(x) = \frac{\alpha\beta^{\gamma/\alpha}}{\Gamma(\gamma/\alpha)}x^{\gamma-1}e^{-\beta x^\alpha}.
\label{gammaeqC}
\end{equation}
Fitting to simulation data, Tanemura  found $\alpha=1.409$, $\beta=2.813$, and $\gamma=4.120$.  However, this can be simplified using the exact values for  the mean and variance; hence, there is  only one free parameter.  Fitting to our own data and using these exact results yields $\alpha=1.1580$, which fixes $\beta=4.0681$ and $\gamma=4.7868$.

Figure \ref{volumefits} shows a comparison of the volume distribution for our large data set and the various suggested fits.  
The parameters in Eqs.~(\ref{lognormaleq}), (\ref{gammaeqA}), and (\ref{gammaeqB}) are determined using the known exact results, with the single free parameter  in Eq.~(\ref{gammaeqC}) determined via a least squares fit to our data set.

Inspection of Fig.~\ref{volumefits} shows that Eq.~(\ref{lognormaleq}) does a poor job reproducing the simulation data.   Equations \ref{gammaeqA} and \ref{gammaeqB}  exhibit systematic errors compared with the simulation data (see both the peak position and the large $x$ behavior), although they are far superior  to Eq.~(\ref{lognormaleq}).  The adjustable three-parameter $\Gamma$ distribution function [Eq.~\ref{gammaeqC})] provide a best fit to the data. 

\begin{figure}
\centering
\vspace{3mm}
\includegraphics[width=1.\columnwidth]{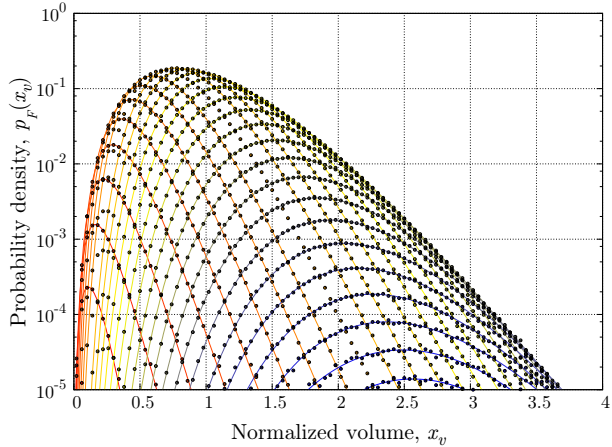}
\caption{(Color online) Partial distributions $p_{_F}(x_v)$ of cell volumes for each number of faces $F$.  The red-most curve, at the bottom left of the plot, corresponds to $F=5$; the blue-most curve, at the bottom right of the plot, corresponds to $F=32$.  Data were binned in intervals of width 0.04.} 
\label{faces_volumes_distributions}
\end{figure}

We next consider the partial distributions $p_{_F}(x_v)$ of cell volumes for each number of faces $F$; the partial distributions are normalized so $p(x) = \sum_{F=1}^{\infty} p_{_F}(x_v)$.  Data for $4\leq F \leq 32$ are shown in Fig.~\ref{faces_volumes_distributions}; a semi-log scale is used to help differentiate the data for very small and very large volumes.  
Tanemura \cite{2003tanemura} used a relatively large data set (5 million cells) and suggested that each of these partial distributions could be accurately described by the three-parameter $\Gamma$ function considered earlier [Eq.~(\ref{gammaeqC})], where $\alpha$, $\beta$, and $\gamma$ for each curve are parameters that depend on $F$.   We test this suggestion using a least squares fit to obtain parameters $\alpha$, $\beta$, and $\gamma$ for each $F$.  Figure \ref{faces_volumes_distributions} shows least squares fits of Eq.~(\ref{gammaeqC}) for each $F$.  Obtained parameters are provided in Table IX of the Supplemental Material.  While we know of no theoretical reason to expect this form, it appears to match the data very well.

\subsection{Distribution of surface areas}
\label{Distribution of surface areas}

\begin{figure}
\centering
\vspace{3mm}
\includegraphics[width=1.\columnwidth]{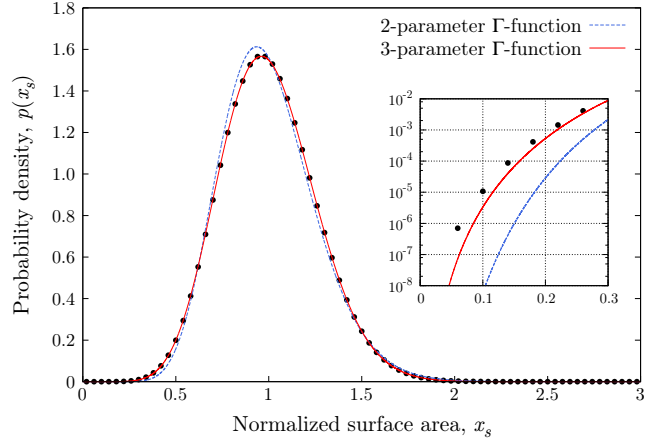}
\caption{(Color online) Distribution of surface areas among all cells.  Data were binned in intervals of width 0.04.  The standard deviation is 0.254, shown to three decimal places. The inset shows a subset of the data on a semilogarithmic plot.} 
\label{surface_areas_distribution}
\end{figure}

We next consider the distribution of surface areas over all cells.  In this section we use $x_s = s/\langle s \rangle$ to denote normalized surface area, where $s$ is the surface area of a particular cell and $\langle s \rangle$ is the average surface area per cell.  Brakke \cite{1985brakke} provided an integral equation for  the variance of this distribution, and numerically evaluated it to be 0.064679.  Our data reproduce this exact result to within 0.0001\%.  Figure \ref{surface_areas_distribution} plots the distribution of surface areas in our data set.  The curve appears to peak at roughly $x_s=0.96$ with a probability density of $p(x_s)= 1.57$. 

Kumar {\it et al.}~suggested that this distribution can be described by a two-parameter $\Gamma$ function [Eq.~(\ref{gammaeqB})] with fitted parameters $\alpha=15.4847$ and $\beta=0.06490$.  However, since both the mean and variance are known, there are no degrees of freedom in fitting two parameters.  The analytic constraints yield $\alpha=15.461$ and $\beta=0.06468$. 

If we consider a three-parameter $\Gamma$ function [Eq.~(\ref{gammaeqC})], then we are left with one degree of freedom in choosing the parameters.  A least squares fit finds that $\alpha=1.845$, $\beta=4.416$, and $\gamma=8.557$ fit the data most closely, while satisfying the known analytic constraints.
%
Figure \ref{surface_areas_distribution} shows both fits and the collected data.  Although the three-parameter version slightly underestimates $p(x)$ for small $x$, as can be seen on the inset plot, overall it provides excellent agreement with the data.

\begin{figure}
\centering
\vspace{3mm}
\includegraphics[width=1.\columnwidth]{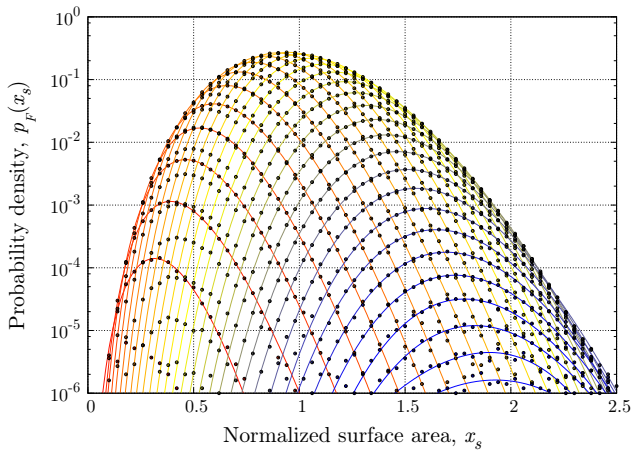}
\caption{(Color online) Partial distributions $p_{_F}(x_s)$ of cell surface areas for each number of faces $F$.  The red-most curve, at the bottom left of the plot, corresponds to $F=4$; the blue-most curve, at the bottom right of the plot, corresponds to $F=33$.  Data were binned in intervals of width 0.04.} 
\label{faces_surface_areas_distributions}
\end{figure}

Figure \ref{faces_surface_areas_distributions} shows the partial distributions $p_{_F}(x_s)$ of surface areas for cells with fixed numbers of faces. 
We show the data on a semi-log plot to focus attention on data of very large and small surface areas.  Tanemura \cite{2003tanemura} suggested that these distributions could also be accurately described by the three-parameter $\Gamma$ function considered earlier [Eq.~(\ref{gammaeqC})], where $\alpha$, $\beta$, and $\gamma$ for each curve are parameters that depend on $F$.  We test this suggestion using a least squares fit to obtain parameters $\alpha$, $\beta$, and $\gamma$ for each $F$.  Figure \ref{faces_volumes_distributions} shows least squares fits of Eq.~(\ref{gammaeqC}) for each $F$; the parameters are provided in Table X of the Supplemental Material.  While we cannot provide justification to expect this form, it appears to match the data very well.

\subsection{Distribution of face areas}
\begin{figure}
\centering
\vspace{3mm}
\includegraphics[width=1.\columnwidth]{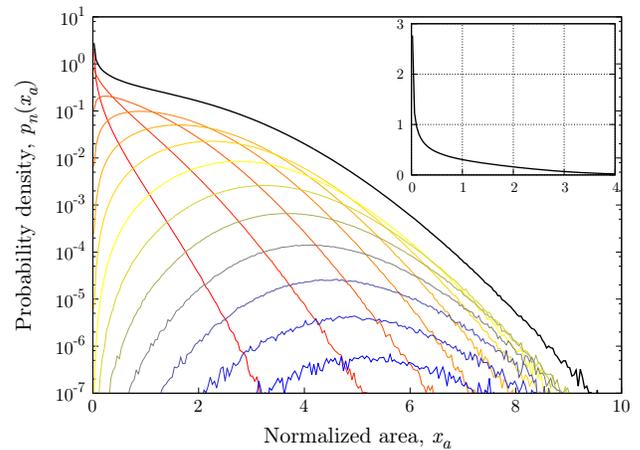}
\caption{(Color online) Partial distributions $p_n(x_a)$ of face areas for each number of edges $n$.  The red-most curve, located to the left of the other curves, corresponds to $n=3$; the blue-most curve, located at the bottom-center of the plot, corresponds to $n=15$.  The black curves show the distribution of areas summed over all $n$.  Data were binned in intervals of width 0.04.} 
\label{faces_areas_distributions}
\end{figure}

We next consider the distribution of areas of faces.  In this section we use $x_a = a/\langle a \rangle$ to denote normalized areas, where $a$ is the area of a particular face and $\langle a \rangle$ is the average area over all faces.  Brakke \cite{1985brakke} provided an integral expression for the variance of this distribution; numerical evaluation shows that it is equal to 1.01426.  Our data reproduce this exact result to within 0.005\%.  The black curves in Fig.~\ref{faces_areas_distributions} show the distribution of areas among all faces in our data set.  Unlike the distributions considered earlier, this one is far from symmetric; instead, it is strongly biased towards faces with very small areas.

We also consider the partial distributions $p_n(x_a)$ of areas limited to faces with fixed numbers of edges $n$.  The colored curves in Fig.~\ref{faces_areas_distributions} show these distributions.
It appears from the figure that $p_n(0)>0$ for $n=3$ and $4$. Hence, these curves cannot be fitted using a $\Gamma$ function [Eq.~(\ref{gammaeqC})], for which $p(0)$ always evaluates to $0$.

\subsection{Distribution of face perimeters}

\begin{figure}
\centering
\vspace{3mm}
\includegraphics[width=1.\columnwidth]{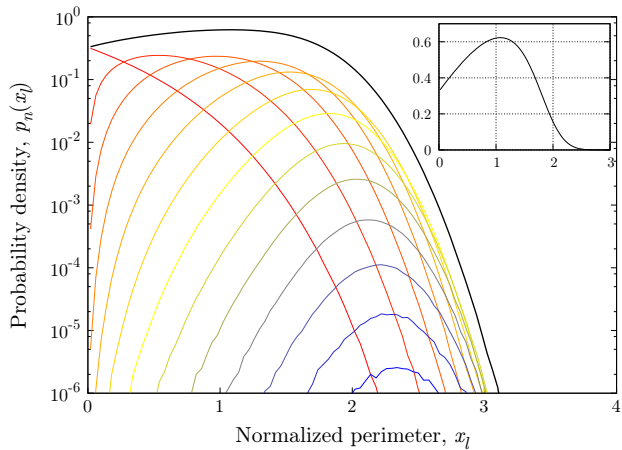}
\caption{(Color online) Partial distributions $p_n(x_l)$ of face perimeters for each number of edges $n$.  The red-most curve, located to the left of the other curves, corresponds to $n=3$; the blue-most curve, located at the bottom-center of the plot, corresponds to $n=15$.  The black curves show the distribution of perimeters summed over all $n$.  Data were binned in intervals of width 0.04.} 
\label{faces_perim_distributions}
\end{figure}

Last, we consider the distribution of perimeters of faces.  In this section we use $x_l = l/\langle l \rangle$ to denote a normalized perimeter, where $l$ is the perimeter of a particular face and $\langle l \rangle$ is the average perimeter over all faces.  Again, Brakke \cite{1985brakke} derived an exact analytical expression for the variance that evaluates to 0.2898.  Our data reproduce this  result to within 0.004\%.  The black curves in Fig.~\ref{faces_perim_distributions} show the distribution of perimeters among all faces in our data set.  The shape of this figure is similar to that calculated analytically by Brakke \cite{brakke1987statistics} for the distribution of edge lengths in two-dimensional Poisson-Voronoi structures.  

We also consider the partial distributions $p_n(x_l)$ of perimeters limited to faces with fixed numbers of edges $n$.  The colored curves in Fig.~\ref{faces_perim_distributions} show these distributions.
It appears from the data that $p_n(0)>0$ for $n=3$; this implies that the partial perimeter distributions cannot be fitted to a $\Gamma$ function.

\section{Correlations between geometry and topology}

We now consider how the average volume and surface area of a cell depend on its number of faces $F$, and how the average area and perimeter of a face depend on its number of edges $n$.  In two-dimensional systems, the study of this type of relationship was pioneered by Lewis \cite{1928lewis}, who observed in some natural structures that the area of a cell was proportional to its number of edges.

\begin{figure}
\centering
\vspace{3mm}
\includegraphics[width=1.\columnwidth, trim=5mm 0 0 0]{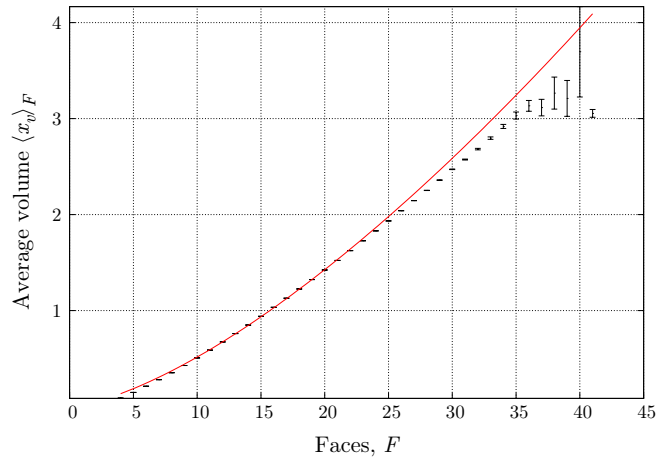}
\caption{Average normalized volume $\langle x_v \rangle_{_F}$ as a function of number of faces $F$; error bars indicate standard error from the mean.  The red curve is a least squares fit to $AF^b$, as explained in the text.} 
\label{nvolumes}
\end{figure}

Figure \ref{nvolumes} shows the average volume of a cell as a function of its number of faces; we use $\langle x_v \rangle_F$ to denote the average volume of cells with $F$ faces.  Based on a data set with 102,000 cells, Kumar {\it et al.}~\cite{1992kumar} suggested that $\langle x_v \rangle_F = AF^b$, where $A$ and $b$ are fitting parameters.  Kumar {\it et al.}~found $A=0.0164$ and $b=1.498$; the more extensive data collected here yield similar values, $A=0.0176$ and $b=1.468$.  
The curve appears to fit the data well for $10\leq F \leq 20$, though not for large or small $F$.  

\begin{figure}
\centering
\vspace{3mm}
\includegraphics[width=1.\columnwidth, trim=5mm 0 0 0]{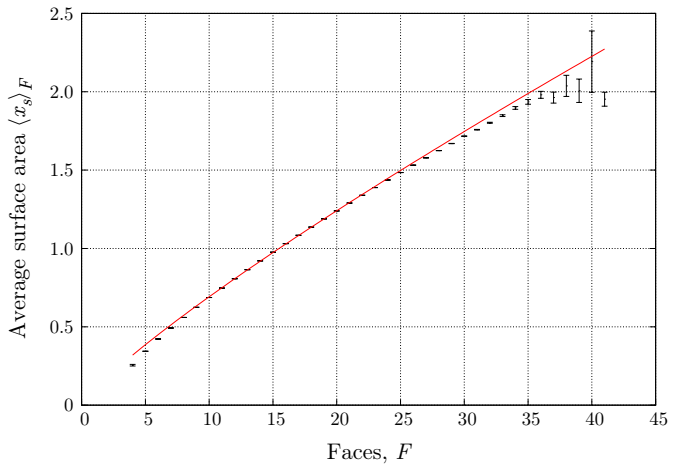}
\caption{Average normalized surface area $\langle x_s \rangle_{_F}$ as a function of number of faces $F$; error bars indicate standard error from the mean.  The red curve is a least squares fit to $AF^b$, as explained in the text.} 
\label{nareas}
\end{figure}

A similar relation might be considered for the average surface area of a cell.  Based on simulation results, Kumar {\it et al.}~\cite{1992kumar} suggested that $\langle x_s \rangle_F = AF^b$, where $\langle x_s \rangle_F$ is the average surface area of cells with $F$ faces and $A$ and $b$ are fitting parameters (Fig.~\ref{nareas}).  Kumar {\it et al.}~found $A=0.09645$ and $b=0.8526$; our data yield similar values, $A=0.0993$ and $b=0.843$.  
This curve too appears to fit the data well for $10\leq F \leq 20$, though also fails for large and small $F$.  

\begin{figure}
\centering
\vspace{3mm}
\includegraphics[width=1.\columnwidth, trim=5mm 0 0 0]{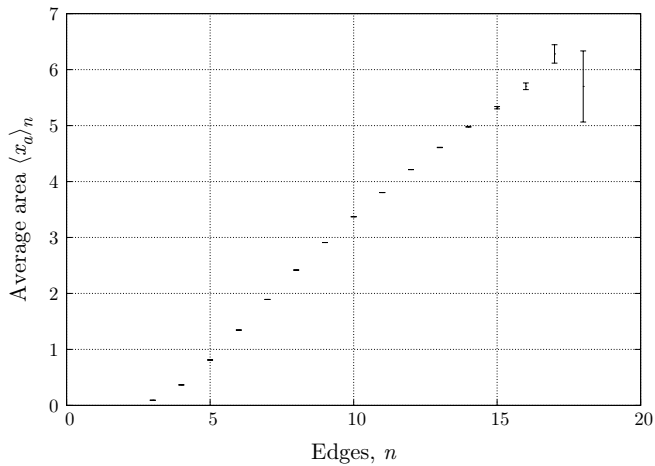}
\caption{Average normalized area $\langle x_a \rangle_n$ as a function of number of edges $n$; error bars indicate standard error from the mean.} 
\label{eareas}
\end{figure}

\begin{figure}
\centering
\vspace{3mm}
\includegraphics[width=1.\columnwidth, trim=5mm 0 0 0]{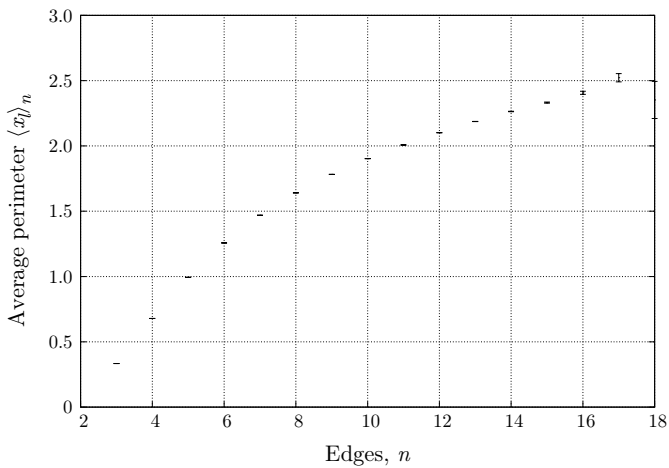}
\caption{Average normalized perimeter $\langle x_l \rangle_n$ as a function of number of edges $n$; error bars indicate standard error from the mean.} 
\label{eperims}
\end{figure}

Finally we turn to the dependance of the expected area and perimeter of a face on its number of edges $n$, illustrated in Figs.~\ref{eareas} and \ref{eperims}.
Although the exact forms of these relationships cannot be determined, it is clear that neither the average area nor perimeter of a face increase linearly with $n$.

\section{Conclusions}

Poisson-Voronoi networks are widely used across the physical and biological sciences as canonical cell structures.  While two-dimensional Poisson-Voronoi networks have been widely studied and often used as surrogates for three-dimensional applications, such three-dimensional networks have been much less widely examined.  In this report, we have provided a much more complete characterization of three-dimensional Poisson-Voronoi networks than exists in the literature.

In particular, we report a wide range of statistical properties of three-dimensional Poisson-Voronoi structures containing a combined total of 250,000,000 cells.  The data demonstrate that although the Poisson-Voronoi structure is generated using a random distribution of points, it exhibits a rich topological and geometrical structure.  

The size of the data set considered here has enabled us to resolve properties of such structures that have been impossible to investigate previously.  While some of the results corroborate earlier work at much higher precision, the results also clearly contradict other conjectures.

In particular, we  found that the natural extension of the Aboav-Weaire relation to three dimensions is {\bf not} consistent with our very large data set, contrary to what was previously reported \cite{1992kumar, fortes1993applicability}.  In particular, for $F < 12$ faces, the average number of faces of a cell's neighbors increases with the number of faces $F$ of a central cell.  This is consistent with  recent theoretical results \cite{hilhorst2009heuristic, 2012masonB}.

Considering more refined  topological data, we observed that some $p$-vectors appear significantly more frequently than others.  We also observed that even when considering a fixed $p$-vector, not all topological types appear with equal frequencies.  Understanding such topological distributions may provide new insight into the topological structure of other natural cellular structures and the forces under which those systems evolve.  One particularly interesting set of results shows that the order of the symmetry groups of the three-dimensional Poisson-Voronoi cells shows clear trends that can be used to distinguish it from other types of cellular networks.  

Our data set supports the conjecture of Tanemura \cite{2003tanemura} regarding the distribution of cell volumes and surface areas when restricted to cells with fixed numbers of faces.  In particular, a three-parameter $\Gamma$ function [Eq.~(\ref{gammaeqC})] appears to fit these data precisely.  This equation also appears to fit the distribution of volumes over all cells.  However, this functional form does not  accurately describe the distribution of cell surface areas or cell face areas and perimeters.

We considered the dependence of the expected volume and surface area of a cell on its number of faces.  The data presented here counters  conjectures of Kumar {\it et al.}~\cite{1992kumar} regarding the form of this relationship.  Unfortunately, we were unable to provide a well-founded alternative.

Extensive geometrical and topological statistics from our data structures are included in the Supplemental Material and an extensive set of measures of the cells in the entire 250,000,000 cell data set is available online at \cite{webpage2}.

{\bf Acknowledgments.} We thank Ken Brakke for providing computer programs to generate Poisson-Voronoi structures, and for ongoing support of his Surface Evolver program.  Most of the computations reported herein were performed using the computational resources of the Institute for Advanced Study.

\bibliographystyle{apsrev4-1.bst}
\bibliography{refs}

\setcounter{table}{1}
\setcounter{figure}{19}

\onecolumngrid
\newpage

\begin{center}
{\Large \bf Supplemental Material}
\end{center}

In the following pages, we report  a wide range of data describing the topology of faces, cells, and cell edge networks of three-dimensional Poisson-Voronoi structures.  We include the distribution of face edges, cell faces, $p$-vectors, topological types, and symmetry orders.  We also report the distribution of cell areas and perimeters, and of cell volumes and surface areas.  Then, we break these distributions down further into distributions for faces with a fixed number of sides, and to cells with a fixed number of faces.  The complete data are available online at http://web.math.princeton.edu/\textasciitilde lazar/voronoi.html.  

\begin{center}
    \begin{tabular}{ | p{7cm} | p{10.5cm} |}
    \hline
    {\bf File} & {\bf Description}  \\ \hline\hline
    {\tt all\_pvectors.data} & Distribution of all 375,410 observed $p$-vectors. \\\hline
    {\tt top\_10000\_wvectors.data}& Distribution of 10,000 most common Weinberg vectors. \\\hline
    {\tt edges\_dist\_areas\_perims.data} & Distribution of edges, and average area and perimeter. \\\hline
    {\tt faces\_dist\_vols\_areas.data} & Distribution of faces, and average volume and surface area. \\\hline
    {\tt p\_vectors\_and\_types.data} & Counts of possible and observed $p$-vectors and topological types. \\\hline
    {\tt symmetries.data} & Distribution of symmetry orders. \\\hline
    {\tt areas\_distribution.data} & Distribution of areas of all faces. \\\hline
    {\tt faces\_areas\_distributions.data} & Distribution of areas of all faces with fixed number of edges. \\\hline
    {\tt perims\_distribution.data} & Distribution of perimeters of all faces. \\\hline
    {\tt faces\_perim\_distributions.data} & Distribution of perimeters of all faces with fixed number of edges. \\\hline
    {\tt volumes\_distribution.data} & Distribution of volumes of all cells. \\\hline
    {\tt faces\_volumes\_distributions.data} & Distribution of volumes of all cells with fixed number of faces. \\\hline
    {\tt surface\_areas\_distribution.data} & Distribution of surface areas of all cells. \\\hline
    {\tt faces\_surface\_areas\_distributions.data} & Distribution of surface areas of all cells with fixed number of faces. \\
    \hline
    \end{tabular}
\end{center}

\begin{table}
\centering
\makebox[0pt][c]{\parbox{1.\columnwidth}{
\begin{minipage}[b]{0.32\hsize}\centering
\begin{tabular}{|r|c|r|c|c|}
\hline
& $p$-vector &  \multicolumn{1}{c|}{$F$} & $N$ & {\it f} \\
\hline
1 & $(001343100...)$ & 12 & 970356 & 0.00388 \\
2 & $(001342100...)$ & 11 & 854284 & 0.00342 \\
3 & $(001433200...)$ & 13 & 744186 & 0.00298 \\
4 & $(001344100...)$ & 13 & 722099 & 0.00289 \\
5 & $(001423100...)$ & 11 & 718819 & 0.00288 \\
6 & $(002333110...)$ & 13 & 710824 & 0.00284 \\
7 & $(001332000...)$ & 9 & 686096 & 0.00274 \\
8 & $(000442000...)$ & 10 & 662391 & 0.00265 \\
9 & $(001352200...)$ & 13 & 657484 & 0.00263 \\
10 & $(002233100...)$ & 11 & 652330 & 0.00261 \\
11 & $(001432200...)$ & 12 & 645966 & 0.00258 \\
12 & $(001353200...)$ & 14 & 644724 & 0.00258 \\
13 & $(002332110...)$ & 12 & 640657 & 0.00256 \\
14 & $(001422100...)$ & 10 & 633915 & 0.00254 \\
15 & $(002322200...)$ & 11 & 629166 & 0.00252 \\
16 & $(002242200...)$ & 12 & 619653 & 0.00248 \\
17 & $(002342210...)$ & 14 & 617595 & 0.00247 \\
18 & $(001443110...)$ & 14 & 608920 & 0.00244 \\
19 & $(000443000...)$ & 11 & 597989 & 0.00239 \\
20 & $(002343210...)$ & 15 & 582316 & 0.00233 \\
21 & $(001442110...)$ & 13 & 579534 & 0.00232 \\
22 & $(001424100...)$ & 12 & 578374 & 0.00231 \\
23 & $(001434200...)$ & 14 & 557186 & 0.00223 \\
24 & $(002243200...)$ & 13 & 543534 & 0.00217 \\
25 & $(002323200...)$ & 12 & 531980 & 0.00213 \\
26 & $(002232100...)$ & 10 & 524974 & 0.00210 \\
27 & $(002423210...)$ & 14 & 508054 & 0.00203 \\
28 & $(002334110...)$ & 14 & 504469 & 0.00202 \\
29 & $(001252000...)$ & 10 & 502450 & 0.00201 \\
30 & $(000533100...)$ & 12 & 496934 & 0.00199 \\
31 & $(001263100...)$ & 13 & 494245 & 0.00198 \\
32 & $(001341100...)$ & 10 & 488878 & 0.00196 \\
33 & $(002234100...)$ & 12 & 482640 & 0.00193 \\
34 & $(001354200...)$ & 15 & 479810 & 0.00192 \\
35 & $(001345100...)$ & 14 & 477182 & 0.00191 \\
36 & $(001334000...)$ & 11 & 475735 & 0.00190 \\
37 & $(002422210...)$ & 13 & 473982 & 0.00190 \\
38 & $(002333300...)$ & 14 & 471010 & 0.00188 \\
39 & $(002324200...)$ & 13 & 468805 & 0.00188 \\
40 & $(001442300...)$ & 14 & 463908 & 0.00186 \\
41 & $(001444110...)$ & 15 & 462925 & 0.00185 \\
42 & $(001443300...)$ & 15 & 461700 & 0.00185 \\
43 & $(002332300...)$ & 13 & 449667 & 0.00180 \\
44 & $(001351200...)$ & 12 & 448514 & 0.00179 \\
45 & $(000453100...)$ & 13 & 445787 & 0.00178 \\
46 & $(001533210...)$ & 15 & 437472 & 0.00175 \\
47 & $(001333000...)$ & 10 & 433409 & 0.00173 \\
48 & $(001453210...)$ & 16 & 429016 & 0.00172 \\
49 & $(002344210...)$ & 16 & 424623 & 0.00170 \\
50 & $(001452210...)$ & 15 & 417387 & 0.00167 \\
51 & $(001331000...)$ & 8 & 414904 & 0.00166 \\
52 & $(002244200...)$ & 14 & 413495 & 0.00165 \\
53 & $(001262100...)$ & 12 & 413322 & 0.00165 \\
54 & $(000454100...)$ & 14 & 407166 & 0.00163 \\
55 & $(000452100...)$ & 12 & 400062 & 0.00160 \\
56 & $(003232210...)$ & 13 & 389187 & 0.00156 \\
57 & $(002432310...)$ & 15 & 383600 & 0.00153 \\
58 & $(003233210...)$ & 14 & 383450 & 0.00153 \\
59 & $(000534100...)$ & 13 & 382470 & 0.00153 \\
60 & $(002253110...)$ & 14 & 380894 & 0.00152 \\
\hline
\end{tabular}
\end{minipage}
\hfill
\begin{minipage}[b]{0.32\hsize}\centering
\begin{tabular}{|r|c|r|c|c|}
\hline
& $p$-vector &  \multicolumn{1}{c|}{$F$} & $N$ & {\it f} \\
\hline
61 & $(001532210...)$ & 14 & 380290 & 0.00152 \\
62 & $(000363000...)$ & 12 & 378639 & 0.00152 \\
63 & $(000444000...)$ & 12 & 369036 & 0.00148 \\
64 & $(002341210...)$ & 13 & 365352 & 0.00146 \\
65 & $(002424210...)$ & 15 & 363294 & 0.00145 \\
66 & $(003223110...)$ & 12 & 360507 & 0.00144 \\
67 & $(002331300...)$ & 12 & 360474 & 0.00144 \\
68 & $(002433310...)$ & 16 & 359007 & 0.00144 \\
69 & $(001253000...)$ & 11 & 343540 & 0.00137 \\
70 & $(001363110...)$ & 15 & 342759 & 0.00137 \\
71 & $(002224000...)$ & 10 & 342223 & 0.00137 \\
72 & $(002252110...)$ & 13 & 338813 & 0.00136 \\
73 & $(001261100...)$ & 11 & 337744 & 0.00135 \\
74 & $(001435200...)$ & 15 & 337609 & 0.00135 \\
75 & $(000532100...)$ & 11 & 335062 & 0.00134 \\
76 & $(001421100...)$ & 9 & 326573 & 0.00131 \\
77 & $(001441300...)$ & 13 & 324983 & 0.00130 \\
78 & $(002334300...)$ & 15 & 324192 & 0.00130 \\
79 & $(000542200...)$ & 13 & 321311 & 0.00129 \\
80 & $(001254000...)$ & 12 & 320972 & 0.00128 \\
81 & $(001264100...)$ & 14 & 317806 & 0.00127 \\
82 & $(001534210...)$ & 16 & 317202 & 0.00127 \\
83 & $(001444300...)$ & 16 & 316502 & 0.00127 \\
84 & $(000364000...)$ & 13 & 316407 & 0.00127 \\
85 & $(000543200...)$ & 14 & 315810 & 0.00126 \\
86 & $(000362000...)$ & 11 & 315678 & 0.00126 \\
87 & $(001425100...)$ & 13 & 315177 & 0.00126 \\
88 & $(002433120...)$ & 15 & 311076 & 0.00124 \\
89 & $(001454210...)$ & 17 & 308362 & 0.00123 \\
90 & $(001523110...)$ & 13 & 308320 & 0.00123 \\
91 & $(002254110...)$ & 15 & 300175 & 0.00120 \\
92 & $(002335110...)$ & 15 & 297450 & 0.00119 \\
93 & $(001524110...)$ & 14 & 297275 & 0.00119 \\
94 & $(000462200...)$ & 14 & 296941 & 0.00119 \\
95 & $(001441110...)$ & 12 & 296622 & 0.00119 \\
96 & $(001362110...)$ & 14 & 295553 & 0.00118 \\
97 & $(000441000...)$ & 9 & 291934 & 0.00117 \\
98 & $(002331110...)$ & 11 & 285885 & 0.00114 \\
99 & $(002432120...)$ & 14 & 284740 & 0.00114 \\
100 & $(002352310...)$ & 16 & 281706 & 0.00113 \\
101 & $(002241200...)$ & 11 & 280656 & 0.00112 \\
102 & $(001445110...)$ & 16 & 280426 & 0.00112 \\
103 & $(001355200...)$ & 16 & 279018 & 0.00112 \\
104 & $(001522110...)$ & 12 & 278044 & 0.00111 \\
105 & $(001451210...)$ & 14 & 275887 & 0.00110 \\
106 & $(002421210...)$ & 12 & 275389 & 0.00110 \\
107 & $(002235100...)$ & 13 & 275247 & 0.00110 \\
108 & $(003312210...)$ & 12 & 272966 & 0.00109 \\
109 & $(000463200...)$ & 15 & 272633 & 0.00109 \\
110 & $(002231100...)$ & 9 & 269118 & 0.00108 \\
111 & $(000544200...)$ & 15 & 266645 & 0.00107 \\
112 & $(001433010...)$ & 12 & 265340 & 0.00106 \\
113 & $(003234210...)$ & 15 & 264604 & 0.00106 \\
114 & $(002242010...)$ & 11 & 264147 & 0.00106 \\
115 & $(002442220...)$ & 16 & 264107 & 0.00106 \\
116 & $(001512200...)$ & 11 & 263778 & 0.00106 \\
117 & $(001431200...)$ & 11 & 261858 & 0.00105 \\
118 & $(002353310...)$ & 17 & 257245 & 0.00103 \\
119 & $(003332220...)$ & 15 & 256849 & 0.00103 \\
120 & $(001542310...)$ & 16 & 255888 & 0.00102 \\
\hline
\end{tabular}
\end{minipage}
\hfill
\begin{minipage}[b]{0.32\hsize}\centering
\begin{tabular}{|r|c|r|c|c|}
\hline
& $p$-vector &  \multicolumn{1}{c|}{$F$} & $N$ & {\it f} \\
\hline
121 & $(001364110...)$ & 16 & 255752 & 0.00102 \\
122 & $(000451100...)$ & 11 & 255748 & 0.00102 \\
123 & $(001523300...)$ & 14 & 255714 & 0.00102 \\
124 & $(001353010...)$ & 13 & 253371 & 0.00101 \\
125 & $(002434310...)$ & 17 & 253316 & 0.00101 \\
126 & $(000440000...)$ & 8 & 252807 & 0.00101 \\
127 & $(003313210...)$ & 13 & 252543 & 0.00101 \\
128 & $(003322310...)$ & 14 & 252509 & 0.00101 \\
129 & $(001434010...)$ & 13 & 252496 & 0.00101 \\
130 & $(002325200...)$ & 14 & 251827 & 0.00101 \\
131 & $(000372100...)$ & 13 & 251247 & 0.00101 \\
132 & $(002252300...)$ & 14 & 248444 & 0.00099 \\
133 & $(002431310...)$ & 14 & 246295 & 0.00099 \\
134 & $(002253300...)$ & 15 & 245427 & 0.00098 \\
135 & $(002443220...)$ & 17 & 244899 & 0.00098 \\
136 & $(003123100...)$ & 10 & 244466 & 0.00098 \\
137 & $(001513200...)$ & 12 & 244121 & 0.00098 \\
138 & $(001354010...)$ & 14 & 242450 & 0.00097 \\
139 & $(001543310...)$ & 17 & 242275 & 0.00097 \\
140 & $(001335000...)$ & 12 & 240198 & 0.00096 \\
141 & $(001352010...)$ & 12 & 239410 & 0.00096 \\
142 & $(002243010...)$ & 12 & 238926 & 0.00096 \\
143 & $(003323310...)$ & 15 & 238132 & 0.00095 \\
144 & $(002345210...)$ & 17 & 235822 & 0.00094 \\
145 & $(003223300...)$ & 13 & 235583 & 0.00094 \\
146 & $(000373100...)$ & 14 & 235216 & 0.00094 \\
147 & $(003222110...)$ & 11 & 234520 & 0.00094 \\
148 & $(001362300...)$ & 15 & 233130 & 0.00093 \\
149 & $(000455100...)$ & 15 & 232498 & 0.00093 \\
150 & $(003242310...)$ & 15 & 231928 & 0.00093 \\
151 & $(002262210...)$ & 15 & 231609 & 0.00093 \\
152 & $(003224110...)$ & 13 & 230872 & 0.00092 \\
153 & $(003212200...)$ & 10 & 230605 & 0.00092 \\
154 & $(003333220...)$ & 16 & 229635 & 0.00092 \\
155 & $(003231210...)$ & 12 & 228407 & 0.00091 \\
156 & $(001363300...)$ & 16 & 227814 & 0.00091 \\
157 & $(001251000...)$ & 9 & 227684 & 0.00091 \\
158 & $(002245200...)$ & 15 & 226914 & 0.00091 \\
159 & $(001514200...)$ & 13 & 226510 & 0.00091 \\
160 & $(002314100...)$ & 11 & 226362 & 0.00091 \\
161 & $(000522000...)$ & 9 & 223821 & 0.00090 \\
162 & $(002352120...)$ & 15 & 222637 & 0.00089 \\
163 & $(002313100...)$ & 10 & 222542 & 0.00089 \\
164 & $(002263210...)$ & 16 & 222046 & 0.00089 \\
165 & $(003323120...)$ & 14 & 221973 & 0.00089 \\
166 & $(003132200...)$ & 11 & 219674 & 0.00088 \\
167 & $(000445000...)$ & 13 & 218836 & 0.00088 \\
168 & $(003222300...)$ & 12 & 218196 & 0.00087 \\
169 & $(002353120...)$ & 16 & 217689 & 0.00087 \\
170 & $(002433201...)$ & 15 & 217182 & 0.00087 \\
171 & $(002244010...)$ & 13 & 212408 & 0.00085 \\
172 & $(001531210...)$ & 13 & 212259 & 0.00085 \\
173 & $(002432201...)$ & 14 & 211664 & 0.00085 \\
174 & $(002434120...)$ & 16 & 210983 & 0.00084 \\
175 & $(001346100...)$ & 15 & 210485 & 0.00084 \\
176 & $(003143110...)$ & 13 & 210362 & 0.00084 \\
177 & $(002324010...)$ & 12 & 210247 & 0.00084 \\
178 & $(002251300...)$ & 13 & 210141 & 0.00084 \\
179 & $(000464200...)$ & 16 & 209003 & 0.00084 \\
180 & $(001522300...)$ & 13 & 208541 & 0.00083 \\
\hline
\end{tabular}
\end{minipage}
}}
\caption{The 180 most frequent $p$-vectors in the Poisson-Voronoi and grain growth microstructures.  $F$ indicates the number of faces, $N$ indicates the absolute frequency, and $f$ indicates the relative frequency.  A list of all $p$-vectors found in the data set along with frequencies can be found in the file {\tt all\_pvectors.data}.}
\label{pvectortableextended}
\end{table}

\begin{table}
\begin{tabular}{|r|l|r|c|r|r|c|}
\hline
{\hspace*{5mm}}& {\bf  Weinberg vector} &  \multicolumn{1}{c|}{$F$} & {\bf $p$-vector} &  \multicolumn{1}{c|}{$S$} & \multicolumn{1}{c|}{$N$} & $f$ \\
\hline
\hline
1 & ABCACDEFAFGHIBIJKDKLELMGMNHNJNMLKJIHGFEDCB & 9 & (00133200...) & 1 & 686096 & 0.00274 \\
2 & ABCACDEFAFGHBHIJDJKEKLGLILKJIHGFEDCB & 8 & (00133100...) & 2 & 414904 & 0.00166 \\
3 & ABCDADEFAFGHIBIJKCKLMEMNGNOHOPJPLPONMLKJIHGFEDCB & 10 & (00044200...) & 2 & 394197 & 0.00158 \\
4 & ABCACDEFAFGHIBIJKLDLMEMNGNOPHPJPOKONMLKJIHGFEDCB & 10 & (00134110...) & 1 & 300751 & 0.00120 \\
5 & ABCDADEFAFGHBHIJCJKLELMGMNINKNMLKJIHGFEDCB & 9 & (00044100...) & 4 & 291934 & 0.00117 \\
6 & ABCDADEFAFGHBHIJCJKEKLGLILKJIHGFEDCB & 8 & (00044000...) & 8 & 252807 & 0.00101 \\
7 & ABCACDEFAFGHIBIJKLDLMEMNOGOPHPJPONKNMLKJIHGFEDCB & 10 & (00142210...) & 1 & 239579 & 0.00096 \\
8 & ABCDADEFGAGHIBIJKCKLMEMNFNOPHPQJQRLRORQPONMLKJIH & 11 & (00036200...) & 2 & 237965 & 0.00095 \\
9 & ABCACDEFAFGHIBIJKDKLELMNGNOHOPJPMPONMLKJIHGFEDCB & 10 & (00133300...) & 1 & 236692 & 0.00095 \\
10 & ABCDADEFAFGHBHIJKCKLMEMNOGOPIPQJQRLRNRQPONMLKJIH & 11 & (00044300...) & 1 & 234533 & 0.00094 \\
11 & ABCACDEFAFGHIBIJKLDLMNENGNMOPHPJPOKOMLKJIHGFEDCB & 10 & (00142210...) & 1 & 234180 & 0.00094 \\
12 & ABCACDEFGAGHIJBJKLMDMNENOPFPHPOQRIRKRQLQONMLKJIH & 11 & (00142310...) & 1 & 233523 & 0.00093 \\
13 & ABCDADEFAFGHBHIJCJKLELMNGNOIOPKPMPONMLKJIHGFEDCB & 10 & (00044200...) & 2 & 227853 & 0.00091 \\
14 & ABCACDEFAFGHBHIJDJKLELMGMNINKNMLKJIHGFEDCB & 9 & (00125100...) & 2 & 227684 & 0.00091 \\
15 & ABCDADEFAFGHBHIJKCKLELMNGNINMJMLKJIHGFEDCB & 9 & (00052200...) & 4 & 223821 & 0.00090 \\
16 & ABCACDEFAFGHIBIJKDKLMEMNGNOPHPJPOLONMLKJIHGFEDCB & 10 & (00125200...) & 2 & 219647 & 0.00088 \\
17 & ABCACDEFAFGHIBIJKDKLMEMNGNOHOPJPLPONMLKJIHGFEDCB & 10 & (00125200...) & 2 & 204714 & 0.00082 \\
18 & ABCACDEFAFGHIBIJKLDLMNENOGOPQHQJQPRKRMRPONMLKJIH & 11 & (00126110...) & 1 & 204666 & 0.00082 \\
19 & ABCACDEFAFGHIBIJKLDLMNENOGOPKPMPONMLKJHJIHGFEDCB & 10 & (00223210...) & 1 & 202792 & 0.00081 \\
20 & ABCACDEFAFGHIBIJKLDLMEMNGNKNMLKJHJIHGFEDCB & 9 & (00223110...) & 2 & 200420 & 0.00080 \\
21 & ABCDADEFGAGHIBIJKCKLMEMNFNOHOPJPLPONMLKJIHGFEDCB & 10 & (00036100...) & 6 & 196335 & 0.00079 \\
22 & ABCACDEFGAGHIJBJKLDLMEMNFNOHOPIPKPONMLKJIHGFEDCB & 10 & (00141400...) & 2 & 192647 & 0.00077 \\
23 & ABCDADEFAFGHIBIJKCKLMEMNOGOPHPQJQRLRNRQPONMLKJIH & 11 & (00044300...) & 2 & 188196 & 0.00075 \\
24 & ABCACDEFAFGHIJBJKLMDMNENOPGPHPOLONMLKIKJIHGFEDCB & 10 & (00321220...) & 1 & 182319 & 0.00073 \\
25 & ABCDADEFGAGHIBIJKCKLELMFMNHNJNMLKJIHGFEDCB & 9 & (00036000...) & 12 & 175908 & 0.00070 \\
26 & ABCACDEFAFGHIBIJKDKLELGLKJHJIHGFEDCB & 8 & (00222200...) & 4 & 172020 & 0.00069 \\
27 & ABCACDEFAFGHIBIJKLDLMNENGNMKMLKJHJIHGFEDCB & 9 & (00312210...) & 2 & 167986 & 0.00067 \\
28 & ABCDADEFGAGHIBIJKCKLMEMNOFOPQHQRJRSTLTNTSPSRQPON & 12 & (00036300...) & 1 & 167251 & 0.00067 \\
29 & ABCACDEFAFGHIBIJKLDLMEMNGNOHOPJPKPONMLKJIHGFEDCB & 10 & (00223210...) & 1 & 167141 & 0.00067 \\
30 & ABCACDEFAFGHBHIJKDKLELMNGNINMJMLKJIHGFEDCB & 9 & (00142110...) & 2 & 166543 & 0.00067 \\
31 & ABCDADEFAFGHBHIJCJKLMEMNGNOPIPKPOLONMLKJIHGFEDCB & 10 & (00053110...) & 2 & 165628 & 0.00066 \\
32 & ABCACDEFAFGHIBIJKDKLMEMNOGOPHPQJQRLRNRQPONMLKJIH & 11 & (00125300...) & 1 & 159324 & 0.00064 \\
33 & ABCACDEFGAGHIJBJKLMDMEMLNOFOHONPIPKPNLKJIHGFEDCB & 10 & (00231310...) & 2 & 158213 & 0.00063 \\
34 & ABCACDEFAFGHBHIDIJEJGJIHGFEDCB & 7 & (00133000...) & 6 & 154332 & 0.00062 \\
35 & ABCACDEAEFGHBHIJKDKLFLJLKJIGIHGFEDCB & 8 & (00312110...) & 1 & 150240 & 0.00060 \\
36 & ABCDADEFAFGHIBIJKCKLMNENOGOPQHQRJRLRQPMPONMLKJIH & 11 & (00045110...) & 2 & 150040 & 0.00060 \\
37 & ABCACDEFAFGHBHIJKDKLELMNGNOIOPJPMPONMLKJIHGFEDCB & 10 & (00134110...) & 1 & 148303 & 0.00059 \\
38 & ABCACDEFAFGHIBIJKLDLMNENOGOPQRHRJRQKQPMPONMLKJIH & 11 & (00134210...) & 1 & 145835 & 0.00058 \\
39 & ABCDADEFGAGHIBIJKCKLMEMNOFOPHPQJQRLRNRQPONMLKJIH & 11 & (00028100...) & 4 & 144758 & 0.00058 \\
40 & ABCACDEFAFGHIBIJKLDLMEMNOGOPQHQJQPRKRNRPONMLKJIH & 11 & (00134210...) & 1 & 137500 & 0.00055 \\
41 & ABCACDEFAFGHIBIJKDKLELMNGNHNMJMLKJIHGFEDCB & 9 & (00222300...) & 2 & 135758 & 0.00054 \\
42 & ABCACDEFGAGHIJBJKLMDMNENOFOPHPQRIRKRQLQPONMLKJIH & 11 & (00142310...) & 1 & 133849 & 0.00054 \\
43 & ABCACDEAEFGHBHIJKDKLFLMNGNINMJMLKJIHGFEDCB & 9 & (00142110...) & 1 & 132858 & 0.00053 \\
44 & ABCACDEFGAGHIJBJKLDLMEMNOFOPHPQRIRKRQNQPONMLKJIH & 11 & (00133400...) & 2 & 132080 & 0.00053 \\
45 & ABCACDEFGAGHIJBJKLDLMEMNOFOPHPQIQRKRNRQPONMLKJIH & 11 & (00125300...) & 2 & 131980 & 0.00053 \\
46 & ABCACDEFAFGHIBIJKLDLMNENOGOPHPQJQRKRMRQPONMLKJIH & 11 & (00134210...) & 1 & 130636 & 0.00052 \\
47 & ABCACDEFAFGHIBIJKDKLMEMNOGOHONPJPLPNMLKJIHGFEDCB & 10 & (00222400...) & 1 & 130506 & 0.00052 \\
48 & ABCDADEFAFGHBHIJCJKLMEMNOGOPQIQKQPRLRNRPONMLKJIH & 11 & (00053210...) & 1 & 129216 & 0.00052 \\
49 & ABCACDEFAFGHIBIJKLDLMEMNOGOPHPQJQRKRNRQPONMLKJIH & 11 & (00134210...) & 1 & 127830 & 0.00051 \\
50 & ABCACDEFGAGHIJBJKLMDMEMLNOFOPHPQIQRKRNRQPONLKJIH & 11 & (00223310...) & 1 & 125216 & 0.00050 \\
\hline
\end{tabular}
\caption{The 100 most common Weinberg vectors and their number of faces $F$,  $p$-vector, the order $S$ of the associated symmetry group, and frequencies $f$ in the Poisson-Voronoi microstructure.  To consolidate the data, integers are replaced with letters (e.g., 1 is replaced with A, 2 with B, etc.).  Long Weinberg vectors are truncated in a manner that leaves the truncated part unambiguous.  The 10,000 most common Weinberg vectors can be found in the file {\tt top\_10000\_wvectors.data}; the entire list of all 72,101,233 observed Weinberg vectors is too large to include.}
\label{wvectortable-pv}
\end{table}

\begin{table}
\begin{tabular}{|r|l|r|c|r|r|c|}
\hline
{\hspace*{5mm}}& {\bf  Weinberg vector} &  \multicolumn{1}{c|}{$F$} & {\bf $p$-vector} &  \multicolumn{1}{c|}{$S$} & \multicolumn{1}{c|}{$N$} & $f$ \\
\hline
\hline
51 & ABCACDEFGAGHIJBJKLDLELKMNFNHNMIMKJIHGFEDCB & 9 & (00230400...) & 4 & 124995 & 0.00050 \\
52 & ABCACDEFAFGHIJBJKLMDMNENOGOPHPLPONMLKIKJIHGFEDCB & 10 & (00224020...) & 2 & 124137 & 0.00050 \\
53 & ABCDADEFAFGHBHIJKCKLMEMNOGOIONPJPLPNMLKJIHGFEDCB & 10 & (00052300...) & 4 & 124069 & 0.00050 \\
54 & ABCACDEAEFGHBHIJDJKFKLGLILKJIHGFEDCB & 8 & (00141200...) & 2 & 123938 & 0.00050 \\
55 & ABCACDEAEFGHBHIJKDKLMFMNGNINMLJLKJIHGFEDCB & 9 & (00231210...) & 1 & 122663 & 0.00049 \\
56 & ABCDADEFGAGHIBIJKLCLMNENOFOPQHQJQPRKRMRPONMLKJIH & 11 & (00044300...) & 4 & 122348 & 0.00049 \\
57 & ABCACDEFAFGHIBIJKLDLMNOEOPGPQRHRJRQNQPONMKMLKJIH & 11 & (00223310...) & 1 & 120472 & 0.00048 \\
58 & ABCDADEFAFGHIBIJKCKLMEMNOGOPHPQRJRLRQNQPONMLKJIH & 11 & (00052400...) & 2 & 119443 & 0.00048 \\
59 & ABCDADEFAFGHIBIJKCKLMNENOPGPQHQRSJSLSRTMTOTRQPO & 12 & (00053310...) & 1 & 117048 & 0.00047 \\
60 & ABCACDEFAFGHBHIJKDKLELMGMNINJNMLKJIHGFEDCB & 9 & (00215010...) & 2 & 114530 & 0.00046 \\
61 & ABCACDEFGAGHIJBJKLDLMEMNOFOHONPIPKPNMLKJIHGFEDCB & 10 & (00133300...) & 6 & 114298 & 0.00046 \\
62 & ABCACDEFAFGHIBIJKDKLMEMNOGOPQHQJQPRLRNRPONMLKJIH & 11 & (00133400...) & 1 & 113239 & 0.00045 \\
63 & ABCACDEFAFGHBHIJKDKLMEMNGNOPIPJPOLONMLKJIHGFEDCB & 10 & (00215110...) & 1 & 111236 & 0.00044 \\
64 & ABCACDEFAFGHIJBJKLMDMNENOGOPHPQRIRKRQLQPONMLKJIH & 11 & (00143120...) & 1 & 110830 & 0.00044 \\
65 & ABCDADEFAFGHIBIJKCKLMNENOGOPHPQRJRLRQMQPONMLKJIH & 11 & (00053210...) & 2 & 109614 & 0.00044 \\
66 & ABCACDEFAFGHIBIJKDKLMEMNOGOPHPQRJRLRQNQPONMLKJIH & 11 & (00133400...) & 1 & 104620 & 0.00042 \\
67 & ABCDADEFAFGHIBIJKCKLMEMNOGOPHPQRJRSLSTNTQTSRQPO & 12 & (00044400...) & 1 & 104459 & 0.00042 \\
68 & ABCACDEFAFGHBHIJDJKLELGLKIKJIHGFEDCB & 8 & (00214100...) & 4 & 103424 & 0.00041 \\
69 & ABCDADEFGAGHIBIJKLCLMNENOPFPQHQRJRSKSTMTOTSRQPO & 12 & (00036300...) & 2 & 103145 & 0.00041 \\
70 & ABCACDEFAFGHBHIJDJKEKLMGMNINLNMLKJIHGFEDCB & 9 & (00214200...) & 2 & 101630 & 0.00041 \\
71 & ABCACDEFGAGHIJBJKLDLMEMNFNOPHPQIQRKRORQPONMLKJIH & 11 & (00141500...) & 1 & 100197 & 0.00040 \\
72 & ABCACDEFAFGHIJBJKLMDMNENOGOPQHQRIRKRQPLPONMLKJIH & 11 & (00135020...) & 1 & 99613 & 0.00040 \\
73 & ABCACDEFGAGHIJBJKLMDMNENOPQRFRHRQIQPKPOLONMLKJIH & 11 & (00151220...) & 1 & 98188 & 0.00039 \\
74 & ABCACDEFGAGHIJBJKLMDMNENOPQFQHQPRIRSKSTLTOTSRPO & 12 & (00134310...) & 1 & 96027 & 0.00038 \\
75 & ABCACDEFAFGHIJBJKLMDMNENOPGPQHQRLRORQPONMLKIKJIH & 11 & (00232220...) & 1 & 95947 & 0.00038 \\
76 & ABCDADEFGAGHIBIJKLCLMNENOFOPQHQRJRSTKTMTSPSRQPO & 12 & (00044400...) & 2 & 95423 & 0.00038 \\
77 & ABCACDEFGAGHIJBJKLMDMNENOPFPQHQRIRSKSTLTOTSRQPO & 12 & (00134310...) & 1 & 94607 & 0.00038 \\
78 & ABCACDEFAFGHBHIJDJKLELMNGNOPIPKPOMONMLKJIHGFEDCB & 10 & (00214300...) & 1 & 94178 & 0.00038 \\
79 & ABCACDEFAFGHIBIJKLDLMNENOPGPQHQJQPORKRMRONMLKJIH & 11 & (00142310...) & 1 & 92974 & 0.00037 \\
80 & ABCACDEFGAGHIJBJKLMDMEMLNOPFPHPOQIQRKRNRQONLKJIH & 11 & (00232220...) & 1 & 92593 & 0.00037 \\
81 & ABCACDEFGAGHIJBJKLDLMEMNOFOPHPQRIRSKSTNTQTSRQPO & 12 & (00125400...) & 1 & 92511 & 0.00037 \\
82 & ABCACDEFAFGHIBIJKLDLMNENOGOPHPJPONMKMLKJIHGFEDCB & 10 & (00223210...) & 2 & 92455 & 0.00037 \\
83 & ABCDADEFAFGHBHIJCJKLELMNOGOPIPQRKRMRQNQPONMLKJIH & 11 & (00053210...) & 1 & 92066 & 0.00037 \\
84 & ABCACDEFAFGHBHIJKDKLMEMNGNOIOPJPLPONMLKJIHGFEDCB & 10 & (00126010...) & 2 & 91831 & 0.00037 \\
85 & ABCDADEFAFGHBHICIJEJGJIHGFEDCB & 7 & (00052000...) & 20 & 91654 & 0.00037 \\
86 & ABCDADEFAFGHIBIJKCKLMNENOGOPQHQRSJSLSRTMTPTRQPO & 12 & (00045210...) & 1 & 91577 & 0.00037 \\
87 & ABCDADEFGAGHIBIJKCKLMEMNOFOPHPQRJRSLSTNTQTSRQPO & 12 & (00028200...) & 2 & 90317 & 0.00036 \\
88 & ABCACDEFAFGHBHIJKDKLELMNGNOPIPJPOMONMLKJIHGFEDCB & 10 & (00312310...) & 1 & 86892 & 0.00035 \\
89 & ABCACDEFAFGHBHIJDJKLMEMNGNLNMLKIKJIHGFEDCB & 9 & (00303300...) & 2 & 86215 & 0.00034 \\
90 & ABCACDEFAFGHIJBJKLMDMNOEOPGPQHQRLRNRQPONMLKIKJIH & 11 & (00224120...) & 2 & 85589 & 0.00034 \\
91 & ABCACDEFAFGHBHIJKDKLMEMNOGOPIPQJQRLRNRQPONMLKJIH & 11 & (00126110...) & 1 & 85417 & 0.00034 \\
92 & ABCACDEFAFGHIBIJKLDLMNOEOGONPQHQJQPRKRMRPNMLKJIH & 11 & (00142310...) & 1 & 84838 & 0.00034 \\
93 & ABCACDEFAFGHIJBJKLMDMNENOPGPHPOQRIRKRQLQONMLKJIH & 11 & (00232220...) & 1 & 84609 & 0.00034 \\
94 & ABCACDEFGAGHIJBJKLMDMNENOPFPQHQRSTITKTSLSRORQPO & 12 & (00142410...) & 1 & 83475 & 0.00033 \\
95 & ABCACDEFAFGHIBIJKDKLMEMNGNOPHPQJQRLRORQPONMLKJIH & 11 & (00117200...) & 2 & 82555 & 0.00033 \\
96 & ABCACDEFAFGHIJBJKLMDMNOEOPQGQHQPRLRNRPONMLKIKJIH & 11 & (00322130...) & 1 & 81394 & 0.00033 \\
97 & ABCACDEAEFGHIBIJKLDLMFMNGNOPHPJPOKONMLKJIHGFEDCB & 10 & (00143020...) & 1 & 81141 & 0.00032 \\
98 & ABCACDEFAFGHIBIJKDKLELMNOGOHONPJPMPNMLKJIHGFEDCB & 10 & (00312310...) & 1 & 81045 & 0.00032 \\
99 & ABCACDEFAFGHIJBJKLMNDNOEOPGPQHQRLRMRQPONMLKIKJIH & 11 & (00322211...) & 1 & 81022 & 0.00032 \\
100 & ABCDADEFAFGHIJBJKCKLMNENOGOPHPQRIRLRQMQPONMLKJIH & 11 & (00054020...) & 2 & 80250 & 0.00032 \\
\hline
\end{tabular}
\caption{The 100 most common Weinberg vectors and their number of faces $F$,  $p$-vector, the order $S$ of the associated symmetry group, and frequencies $f$ in the Poisson-Voronoi microstructure.  To consolidate the data, integers are replaced with letters (e.g., 1 is replaced with A, 2 with B, etc.).  Long Weinberg vectors are truncated in a manner that leaves the truncated part unambiguous.  The 10,000 most common Weinberg vectors can be found in the file {\tt top\_10000\_wvectors.data}; the entire list of all 72,101,233 observed Weinberg vectors is too large to include.}
\label{wvectortable-pv2}
\end{table}

\begin{table}
\begin{tabular}{|c|r|r|r|r|}
\hline
\multicolumn{1}{|c}{$n$} & \multicolumn{1}{|c}{$N$} & \multicolumn{1}{|c}{$f$} & \multicolumn{1}{|c}{$\langle x_a \rangle_n$} & \multicolumn{1}{|c|}{$\langle x_l \rangle_n$}\\
\hline
1 & 0 & 0.0000 & 0.0000 $\pm$ 0.0000 & 0.0000 $\pm$ 0.0000 \\
2 & 0 & 0.0000 & 0.0000 $\pm$ 0.0000 & 0.0000 $\pm$ 0.0000 \\
3 & 261661169 & 0.1347 & 0.0928 $\pm$ 0.1595 & 1.7435 $\pm$ 1.4534\\ 
4 & 446276118 & 0.2298 & 0.3651 $\pm$ 0.3952 & 3.5491 $\pm$ 1.8703\\ 
5 & 469146641 & 0.2416 & 0.8115 $\pm$ 0.6274 & 5.1951 $\pm$ 1.9856\\ 
6 & 369632159 & 0.1903 & 1.3458 $\pm$ 0.7981 & 6.5723 $\pm$ 1.9352\\ 
7 & 226308795 & 0.1165 & 1.8932 $\pm$ 0.9062 & 7.6811 $\pm$ 1.8219\\ 
8 & 109221807 & 0.0562 & 2.4178 $\pm$ 0.9727 & 8.5766 $\pm$ 1.7028\\ 
9 & 42138921 & 0.0217 & 2.9102 $\pm$ 1.0157 & 9.3172 $\pm$ 1.5987\\ 
10 & 13206845 & 0.0068 & 3.3708 $\pm$ 1.0463 & 9.9465 $\pm$ 1.5134\\ 
11 & 3419199 & 0.0018 & 3.8028 $\pm$ 1.0698 & 10.4938 $\pm$ 1.4443\\ 
12 & 738992 & 0.0004 & 4.2134 $\pm$ 1.0887 & 10.9834 $\pm$ 1.3875\\ 
13 & 136155 & 0.0001 & 4.6086 $\pm$ 1.1078 & 11.4328 $\pm$ 1.3433\\ 
14 & 21470 & 0.0000 & 4.9779 $\pm$ 1.1227 & 11.8333 $\pm$ 1.3061\\ 
15 & 2962 & 0.0000 & 5.3197 $\pm$ 1.1298 & 12.1899 $\pm$ 1.2624\\ 
16 & 375 & 0.0000 & 5.7022 $\pm$ 1.1513 & 12.5823 $\pm$ 1.2147\\ 
17 & 52 & 0.0000 & 6.2794 $\pm$ 1.1861 & 13.1853 $\pm$ 1.1916\\ 
18 & 5 & 0.0000 & 5.6993 $\pm$ 1.4205 & 12.2969 $\pm$ 1.6639\\
19 & 0 & 0.0000 & 0.0000 $\pm$ 0.0000 & 0.0000 $\pm$ 0.0000 \\
20 & 0 & 0.0000 & 0.0000 $\pm$ 0.0000 & 0.0000 $\pm$ 0.0000 \\
\hline
\end{tabular}
\caption{Faces of different topological types; a total of 1,941,911,665 faces.  Data included are number of edges $n$, absolute frequency $N$, relative frequency $f$, average normalized area $\langle x_a \rangle_n$, and average normalized perimeter $\langle x_l \rangle_n$.  Included for area and perimeter is standard deviation from the mean.  Areas are normalized so their mean is 1; perimeters are measured in units of average edge lengths. All data is included in the file {\tt edges\_dist\_areas\_perims.data}.}
\label{wvectortable}
\end{table}

\begin{table}
\centering
\makebox[0pt][c]{\parbox{1.\columnwidth}{
\begin{minipage}[b]{0.48\hsize}\centering
\begin{tabular}{|r|r|r|r|r|r|}
\hline
\multicolumn{1}{|c|}{$F$} & \multicolumn{1}{c|}{$N$} & \multicolumn{1}{c|}{$f$} & \multicolumn{1}{c|}{$\langle x_v \rangle_{_F}$} & \multicolumn{1}{c|}{$\langle x_s \rangle_{_F}$} & \multicolumn{1}{c|}{$m(F)$}\\
\hline
1 & 0 & 0.000 & 0.000 $\pm$ 0.000 & 0.000 $\pm$ 0.000 & 0.000 \\
2 & 0 & 0.000 & 0.000 $\pm$ 0.000 & 0.000 $\pm$ 0.000 & 0.000 \\
3 & 0 & 0.000 & 0.000 $\pm$ 0.000 & 0.000 $\pm$ 0.000 & 0.000 \\
4 & 325 & 0.000 & 0.088 $\pm$ 0.049 & 0.255 $\pm$ 0.089 & 15.187 \\
5 & 9517 & 0.000 & 0.147 $\pm$ 0.076 & 0.344 $\pm$ 0.111 & 15.701 \\
6 & 86847 & 0.000 & 0.212 $\pm$ 0.100 & 0.422 $\pm$ 0.124 & 15.981 \\
7 & 434884 & 0.002 & 0.279 $\pm$ 0.122 & 0.493 $\pm$ 0.133 & 16.125 \\
8 & 1474073 & 0.006 & 0.352 $\pm$ 0.144 & 0.560 $\pm$ 0.141 & 16.218 \\
9 & 3757407 & 0.015 & 0.428 $\pm$ 0.164 & 0.625 $\pm$ 0.148 & 16.279 \\
10 & 7673585 & 0.031 & 0.506 $\pm$ 0.184 & 0.687 $\pm$ 0.154 & 16.311 \\
11 & 13105393 & 0.052 & 0.588 $\pm$ 0.203 & 0.748 $\pm$ 0.160 & 16.327 \\
12 & 19278192 & 0.077 & 0.673 $\pm$ 0.221 & 0.806 $\pm$ 0.165 & 16.329 \\
13 & 24913588 & 0.100 & 0.760 $\pm$ 0.239 & 0.864 $\pm$ 0.169 & 16.323 \\
14 & 28773718 & 0.115 & 0.849 $\pm$ 0.257 & 0.921 $\pm$ 0.173 & 16.310 \\
15 & 30069917 & 0.120 & 0.941 $\pm$ 0.274 & 0.976 $\pm$ 0.177 & 16.293 \\
16 & 28709465 & 0.115 & 1.034 $\pm$ 0.291 & 1.031 $\pm$ 0.181 & 16.272 \\
17 & 25258283 & 0.101 & 1.129 $\pm$ 0.307 & 1.084 $\pm$ 0.185 & 16.249 \\
18 & 20627759 & 0.083 & 1.225 $\pm$ 0.323 & 1.137 $\pm$ 0.188 & 16.224 \\
19 & 15733136 & 0.063 & 1.323 $\pm$ 0.339 & 1.188 $\pm$ 0.191 & 16.198 \\
20 & 11240712 & 0.045 & 1.422 $\pm$ 0.354 & 1.240 $\pm$ 0.194 & 16.171 \\
21 & 7575444 & 0.030 & 1.523 $\pm$ 0.369 & 1.290 $\pm$ 0.197 & 16.142 \\
\hline
\end{tabular}
\end{minipage}
\hfill
\begin{minipage}[b]{0.48\hsize}\centering
\begin{tabular}{|r|r|r|r|r|r|}
\hline
\multicolumn{1}{|c|}{$F$} & \multicolumn{1}{c|}{$N$} & \multicolumn{1}{c|}{$f$} & \multicolumn{1}{c|}{$\langle x_v \rangle_{_F}$} & \multicolumn{1}{c|}{$\langle x_s \rangle_{_F}$} & \multicolumn{1}{c|}{$m(F)$}\\
\hline
22 & 4821143 & 0.019 & 1.624 $\pm$ 0.383 & 1.340 $\pm$ 0.200 & 16.115 \\
23 & 2919932 & 0.012 & 1.727 $\pm$ 0.397 & 1.389 $\pm$ 0.202 & 16.086 \\
24 & 1678180 & 0.007 & 1.830 $\pm$ 0.412 & 1.437 $\pm$ 0.205 & 16.058 \\
25 & 919317 & 0.004 & 1.934 $\pm$ 0.425 & 1.484 $\pm$ 0.207 & 16.030 \\
26 & 482492 & 0.002 & 2.040 $\pm$ 0.438 & 1.532 $\pm$ 0.209 & 16.002 \\
27 & 243551 & 0.001 & 2.145 $\pm$ 0.452 & 1.578 $\pm$ 0.212 & 15.976 \\
28 & 116870 & 0.001 & 2.252 $\pm$ 0.465 & 1.624 $\pm$ 0.213 & 15.948 \\
29 & 54391 & 0.000 & 2.359 $\pm$ 0.477 & 1.669 $\pm$ 0.215 & 15.920 \\
30 & 24236 & 0.000 & 2.471 $\pm$ 0.493 & 1.717 $\pm$ 0.219 & 15.899 \\
31 & 10467 & 0.000 & 2.573 $\pm$ 0.503 & 1.758 $\pm$ 0.219 & 15.862 \\
32 & 4393 & 0.000 & 2.681 $\pm$ 0.522 & 1.802 $\pm$ 0.224 & 15.845 \\
33 & 1713 & 0.000 & 2.796 $\pm$ 0.536 & 1.848 $\pm$ 0.228 & 15.824 \\
34 & 643 & 0.000 & 2.918 $\pm$ 0.537 & 1.897 $\pm$ 0.222 & 15.796 \\
35 & 262 & 0.000 & 3.032 $\pm$ 0.593 & 1.935 $\pm$ 0.244 & 15.725 \\
36 & 104 & 0.000 & 3.133 $\pm$ 0.569 & 1.980 $\pm$ 0.227 & 15.702 \\
37 & 33 & 0.000 & 3.115 $\pm$ 0.488 & 1.963 $\pm$ 0.198 & 15.555 \\
38 & 14 & 0.000 & 3.266 $\pm$ 0.601 & 2.038 $\pm$ 0.244 & 15.455 \\
39 & 9 & 0.000 & 3.210 $\pm$ 0.528 & 2.006 $\pm$ 0.211 & 15.664 \\
40 & 3 & 0.000 & 3.696 $\pm$ 0.666 & 2.192 $\pm$ 0.277 & 15.692 \\
41 & 2 & 0.000 & 3.054 $\pm$ 0.041 & 1.953 $\pm$ 0.045 & 15.634 \\
42 & 0 & 0.000 & 0.000 $\pm$ 0.000 & 0.000 $\pm$ 0.000 & 0.000 \\
\hline
\end{tabular}
\end{minipage}
}}
\caption{Cells with different numbers of faces; a total of 250,000,000 cells.  Data included are the absolute frequency $N$, the relative frequency $f$, average normalized volume $\langle x_v \rangle_{_F}$, average normalized surface area $\langle x_s \rangle_{_F}$, and average number of neighbors' neighbors $m(F)$.  Included for volume and area is standard deviation from the mean.  Volumes and surface areas normalized so means are 1.  All data is included in the file {\tt faces\_dist\_vols\_areas.data}.}
\label{facestable}
\end{table}

\begin{table}
\centering
\makebox[0pt][c]{\parbox{1.\columnwidth}{
\begin{minipage}[b]{0.48\hsize}\centering
\begin{tabular}{|r|r|r|r|r|}
\hline
 & \multicolumn{2}{c|}{$p$-vectors} & \multicolumn{2}{c|}{Topological types} \\ 
\hline
\multicolumn{1}{|c|}{F} & \multicolumn{1}{c|}{observed} & \multicolumn{1}{c|}{possible} & \multicolumn{1}{c|}{observed} & \multicolumn{1}{c|}{possible} \\
\hline    
1 & 0 & 0 & 0 & 0 \\
2 & 0 & 0 & 0 & 0 \\
3 & 0 & 0 & 0 & 0 \\
4 & 1 & 1 & 1 & 1 \\
5 & 1 & 1 & 1 & 1 \\
6 & 2 & 2 & 2 & 2 \\
7 & 5 & 5 & 5 & 5 \\
8 & 13 & 13 & 14 & 14 \\
9 & 33 & 33 & 50 & 50 \\
10 & 85 & 85 & 233 & 233 \\
11 & 199 & 199 & 1249 & 1249 \\
12 & 440 & 445 & 7370 & 7595 \\
13 & 917 & 947 & 41,337 & 49,566 \\
14 & 1779 & 1909 & 204,367 & 339,722 \\
15 & 3254 & --\hspace{2.2mm} & 864,084 & 2,406,841 \\
16 & 5458 & --\hspace{2.2mm} & 3,735,112 & 17,490,241 \\
17 & 8455 & --\hspace{2.2mm} & 9,771,618 & 129,664,753 \\
18 & 12,417 & --\hspace{2.2mm} & 13,858,582 & 977,526,957 \\
19 & 16,998 & --\hspace{2.2mm} & 13,852,480 & 7,475,907,149 \\
20 & 22,267 & --\hspace{2.2mm} & 10,944,266 & 57,896,349,553 \\
21 & 27,539 & --\hspace{2.2mm} & 7,545,337 & 453,382,272,049 \\
\hline    
\end{tabular}    
\end{minipage}    
\hfill    
\begin{minipage}[b]{0.48\hsize}\centering    
\begin{tabular}{|r|r|r|r|r|}    
\hline    
 & \multicolumn{2}{c|}{$p$-vectors} & \multicolumn{2}{c|}{Topological types} \\ 
\hline
\multicolumn{1}{|c|}{F} & \multicolumn{1}{c|}{observed} & \multicolumn{1}{c|}{possible} & \multicolumn{1}{c|}{observed} & \multicolumn{1}{c|}{possible} \\
\hline    
22 & 32,295 & --\hspace{2.2mm} & 4,818,687 & 3,585,853,662,949 \\
23 & 35,847 & --\hspace{2.2mm} & 2,919,767 & 28,615,703,421,545 \\
24 & 37,821 & --\hspace{2.2mm} & 1,678,171 & --\hspace{2.2mm} \\
25 & 37,327 & --\hspace{2.2mm} & 919,317 & --\hspace{2.2mm} \\
26 & 34,619 & --\hspace{2.2mm} & 482,492 & --\hspace{2.2mm} \\
27 & 30,142 & --\hspace{2.2mm} & 243,551 & --\hspace{2.2mm} \\
28 & 24,280 & --\hspace{2.2mm} & 116,870 & --\hspace{2.2mm} \\
29 & 17,879 & --\hspace{2.2mm} & 54,391 & --\hspace{2.2mm} \\
30 & 11,971 & --\hspace{2.2mm} & 24,236 & --\hspace{2.2mm} \\
31 & 7062 & --\hspace{2.2mm} & 10,467 & --\hspace{2.2mm} \\
32 & 3654 & --\hspace{2.2mm} & 4393 & --\hspace{2.2mm} \\
33 & 1590 & --\hspace{2.2mm} & 1713 & --\hspace{2.2mm} \\
34 & 634 & --\hspace{2.2mm} & 643 & --\hspace{2.2mm} \\
35 & 262 & --\hspace{2.2mm} & 262 & --\hspace{2.2mm} \\
36 & 103 & --\hspace{2.2mm} & 104 & --\hspace{2.2mm} \\
37 & 33 & --\hspace{2.2mm} & 33 & --\hspace{2.2mm} \\
38 & 14 & --\hspace{2.2mm} & 14 & --\hspace{2.2mm} \\
39 & 9 & --\hspace{2.2mm} & 9 & --\hspace{2.2mm} \\
40 & 3 & --\hspace{2.2mm} & 3 & --\hspace{2.2mm} \\
41 & 2 & --\hspace{2.2mm} & 2 & --\hspace{2.2mm} \\
42 & 0 & --\hspace{2.2mm} & 0 & --\hspace{2.2mm} \\
\hline
\end{tabular}
\end{minipage}
}}
\caption{This table lists the number of distinct $p$-vectors and Weinberg vectors we observed in our data sets of 250,000,000 cells.  For each fixed number of faces $F$, we list both the number of observed $p$-vectors and Weinberg vectors, and also the total possible number of these.  Data are obtained from [63].  There are a total of 375,410 distinct $p$-vectors in our data set; there are a total of 72,101,233 topological types. All data is included in the file {\tt p\_vectors\_and\_types.data}.}
\label{facestable2}
\end{table}

\begin{table}
\begin{tabular}{|r|r|r|}
\hline
 \multicolumn{1}{|c|}{$S$} & \multicolumn{1}{c|}{$N$} & \multicolumn{1}{c|}{$f$} \\
\hline
1 & 229260326 & 0.9170 \\
2 & 16524698 & 0.0661 \\
3 & 18385 & 0.0001 \\
4 & 2506544 & 0.0100 \\
6 & 705748 & 0.0028 \\
8 & 414776 & 0.0017 \\
10 & 914 & 0.0000 \\
12 & 283733 & 0.0011 \\
16 & 77449 & 0.0003 \\
20 & 91681 & 0.0004 \\
24 & 72273 & 0.0003 \\
28 & 12877 & 0.0000 \\
32 & 2261 & 0.0000 \\
36 & 292 & 0.0000 \\
40 & 21 & 0.0000 \\
44 & 2 & 0.0000 \\
48 & 24408 & 0.0000 \\
120 & 3612 & 0.0000 \\
\hline
\end{tabular}
\caption{Absolute and relative frequencies of grains with various symmetry group orders $S$.  All data is included in the file {\tt symmetries.data}.}
\label{symmetriestable}
\end{table}

\setlength{\tabcolsep}{3pt}
\begin{figure*}[ht]
\centering
\begin{tabular}{|c|c|c|c|c|c|c|c|}
\hline
\specialcell[t]{ 1. $N = 96027$\\\includegraphics[scale=0.275]{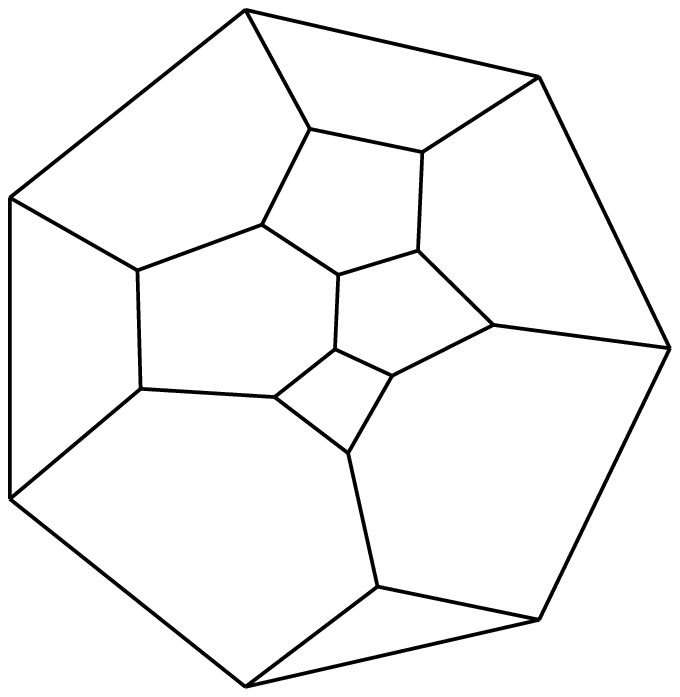}\\ {\it f} = 9.90\%} &
\specialcell[t]{ 2. $N = 94607$\\\includegraphics[scale=0.275]{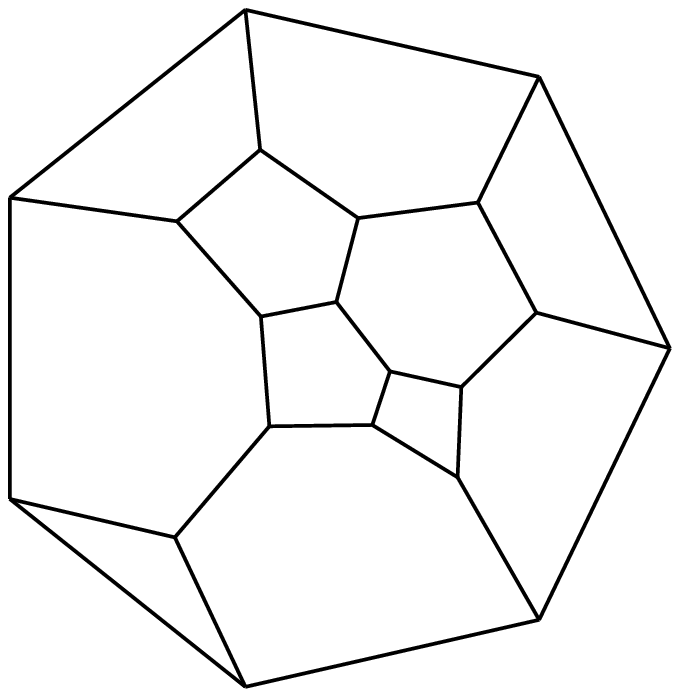}\\ {\it f} = 9.75\%} &
\specialcell[t]{ 3. $N = 64120$\\\includegraphics[scale=0.275]{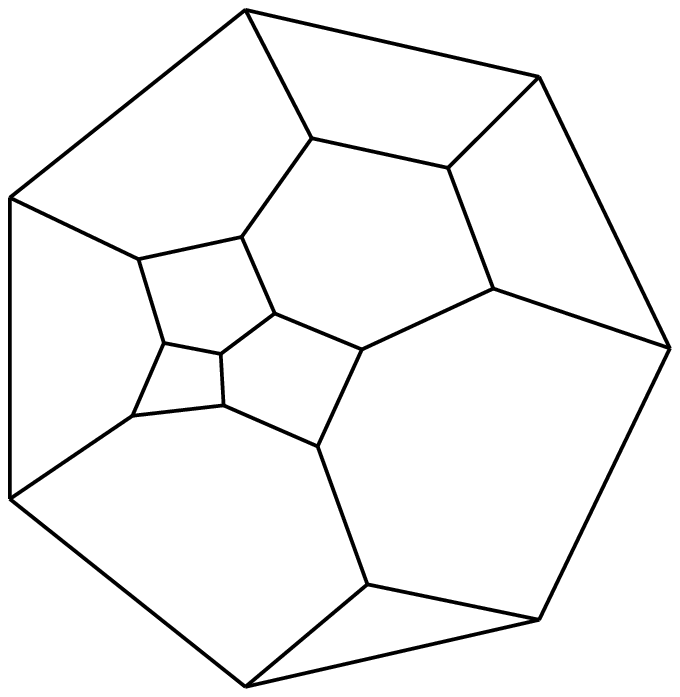}\\ {\it f} = 6.61\%} &
\specialcell[t]{ 4. $N = 63883$\\\includegraphics[scale=0.275]{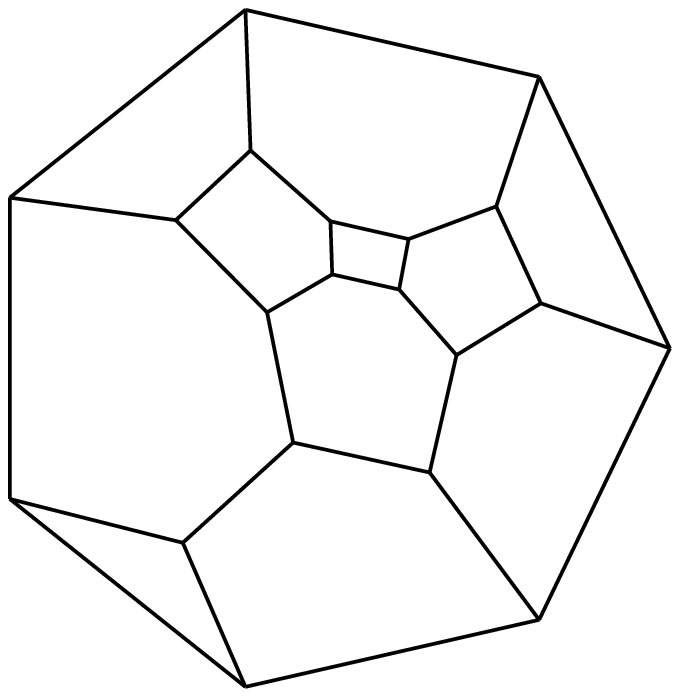}\\ {\it f} = 6.58\%} &
\specialcell[t]{ 5. $N = 63648$\\\includegraphics[scale=0.275]{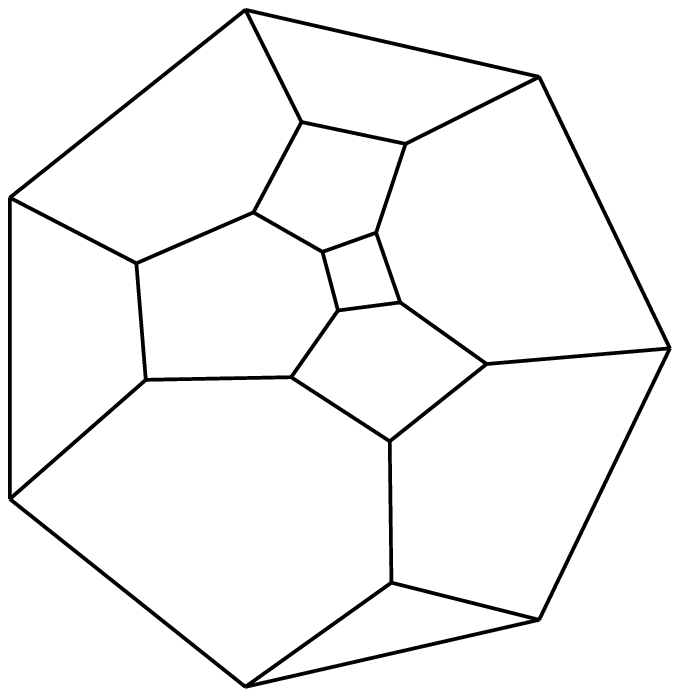}\\ {\it f} = 6.56\%} &
\specialcell[t]{ 6. $N = 60283$\\\includegraphics[scale=0.275]{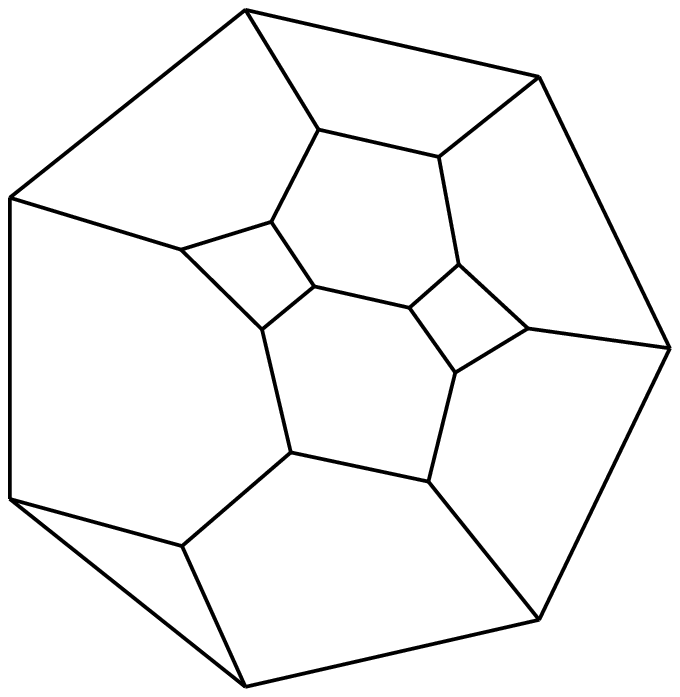}\\ {\it f} = 6.21\%} &
\specialcell[t]{ 7. $N = 48475$\\\includegraphics[scale=0.275]{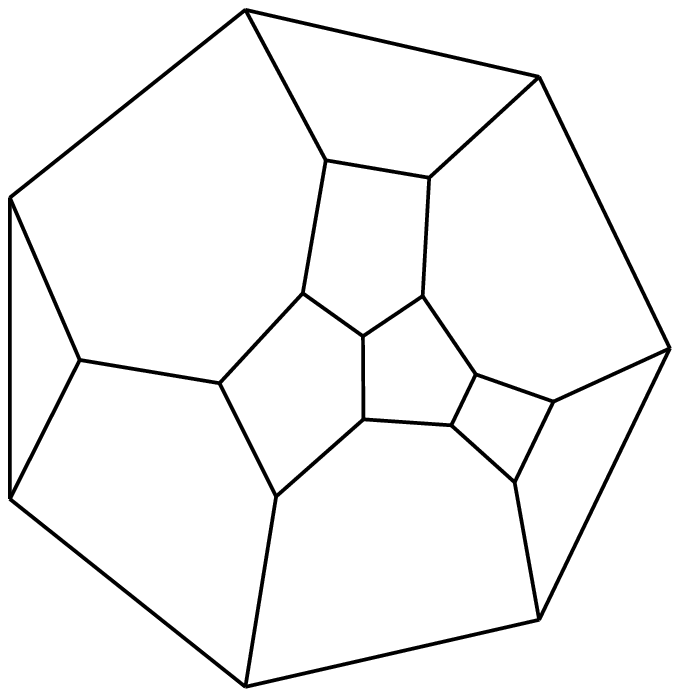}\\ {\it f} = 5.00\%} &
\specialcell[t]{ 8. $N = 46864$\\\includegraphics[scale=0.275]{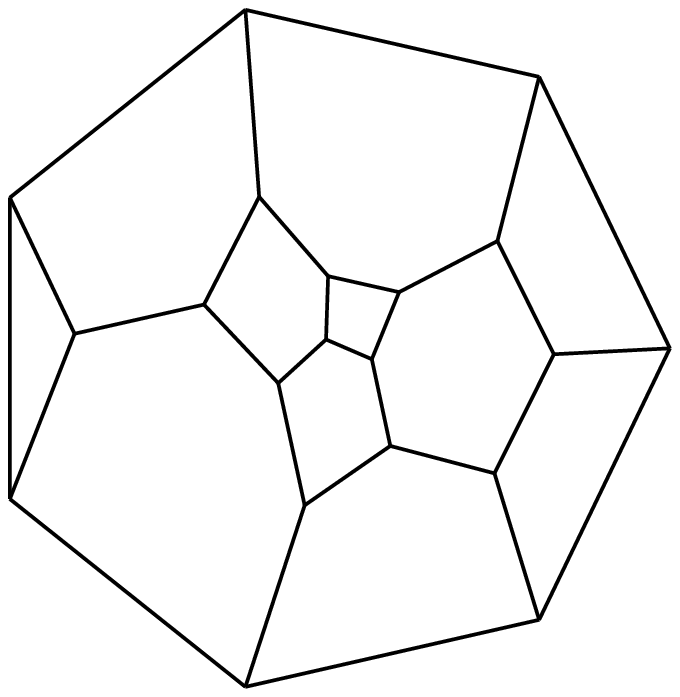}\\ {\it f} = 4.83\%} \\
\hline
\hline
\specialcell[t]{ 9. $N = 44690$\\\includegraphics[scale=0.275]{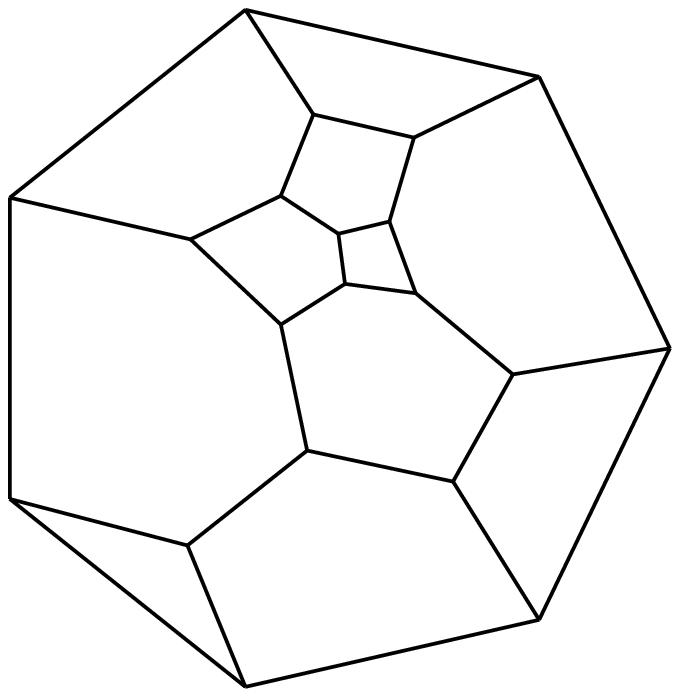}\\ {\it f} = 4.61\%} &
\specialcell[t]{ 10. $N = 30594$\\\includegraphics[scale=0.275]{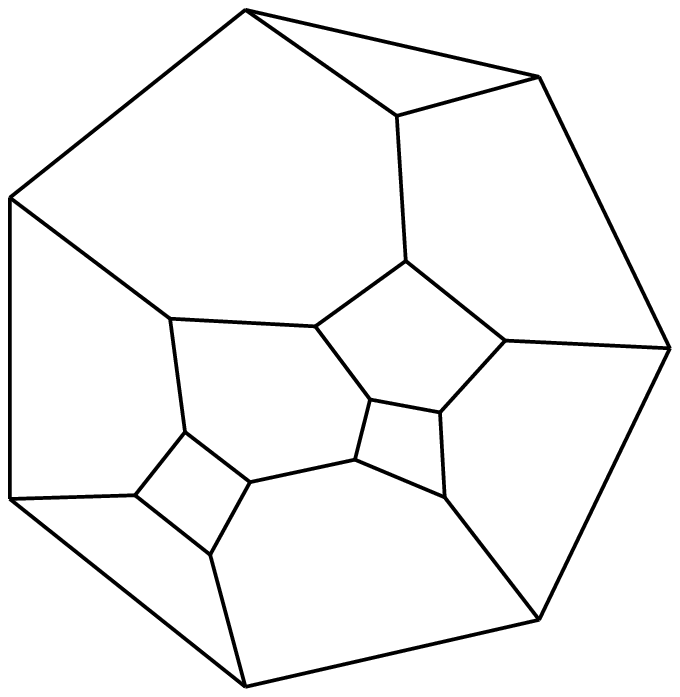}\\ {\it f} = 3.15\%} &
\specialcell[t]{ 11. $N = 30102$\\\includegraphics[scale=0.275]{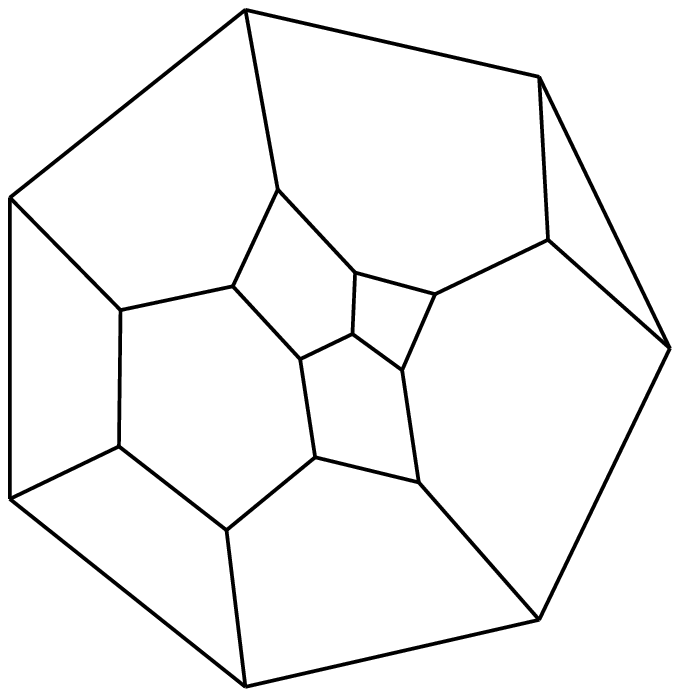}\\ {\it f} = 3.10\%} &
\specialcell[t]{ 12. $N = 29884$\\\includegraphics[scale=0.275]{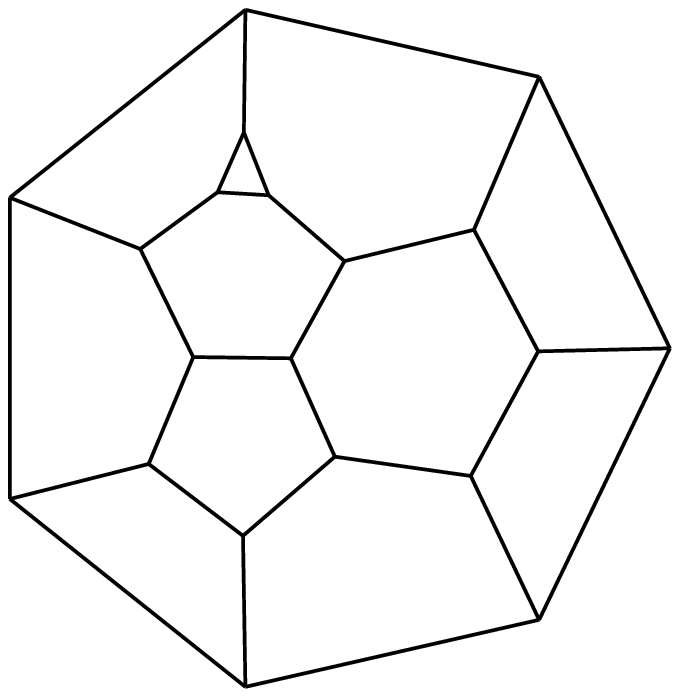}\\ {\it f} = 3.08\%} &
\specialcell[t]{ 13. $N = 29632$\\\includegraphics[scale=0.275]{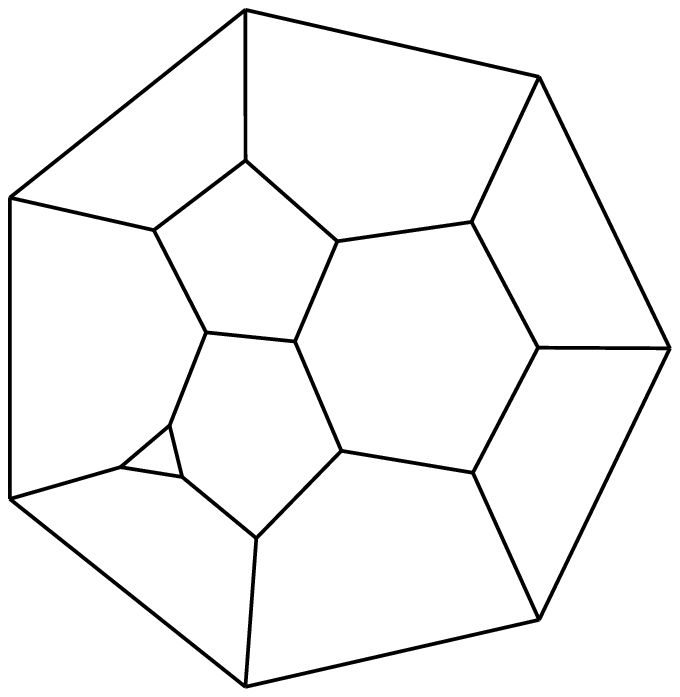}\\ {\it f} = 3.05\%} &
\specialcell[t]{ 14. $N = 28481$\\\includegraphics[scale=0.275]{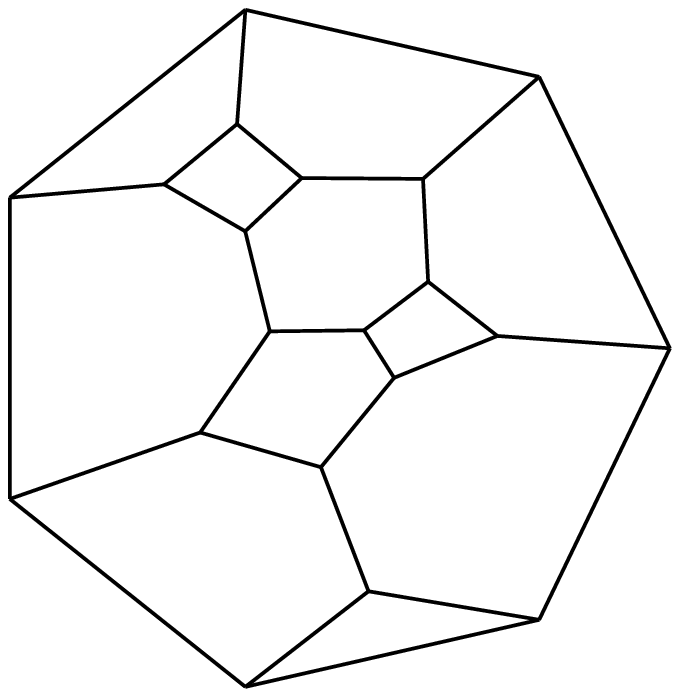}\\ {\it f} = 2.94\%} &
\specialcell[t]{ 15. $N = 27457$\\\includegraphics[scale=0.275]{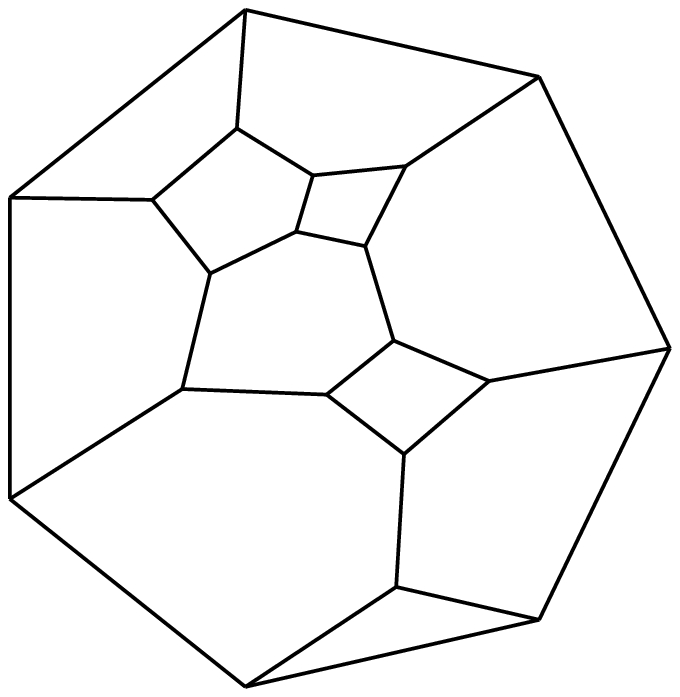}\\ {\it f} = 2.83\%} &
\specialcell[t]{ 16. $N = 27283$\\\includegraphics[scale=0.275]{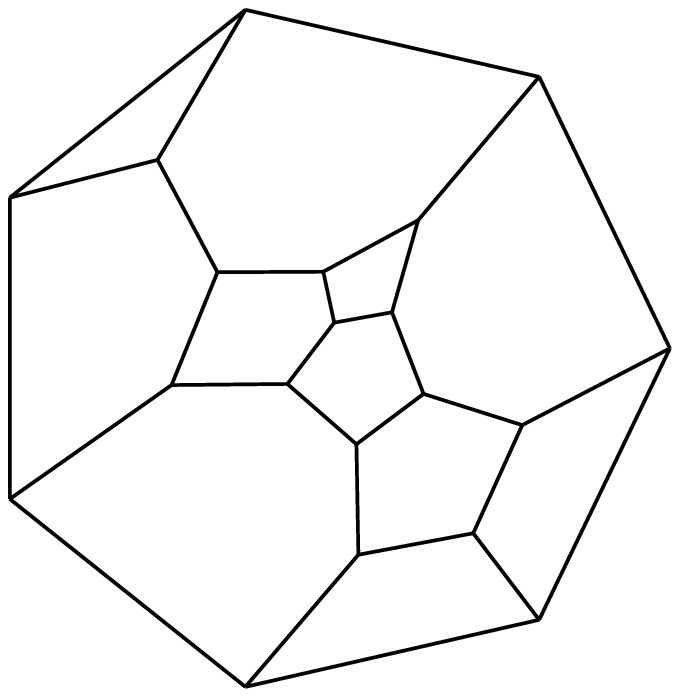}\\ {\it f} = 2.81\%} \\
\hline
\hline
\specialcell[t]{ 17. $N = 27241$\\\includegraphics[scale=0.275]{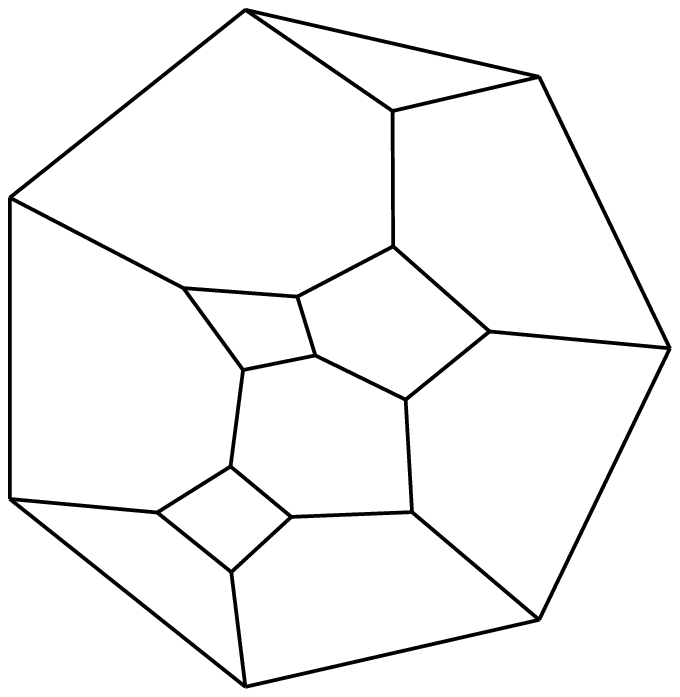}\\ {\it f} = 2.81\%} &
\specialcell[t]{ 18. $N = 27204$\\\includegraphics[scale=0.275]{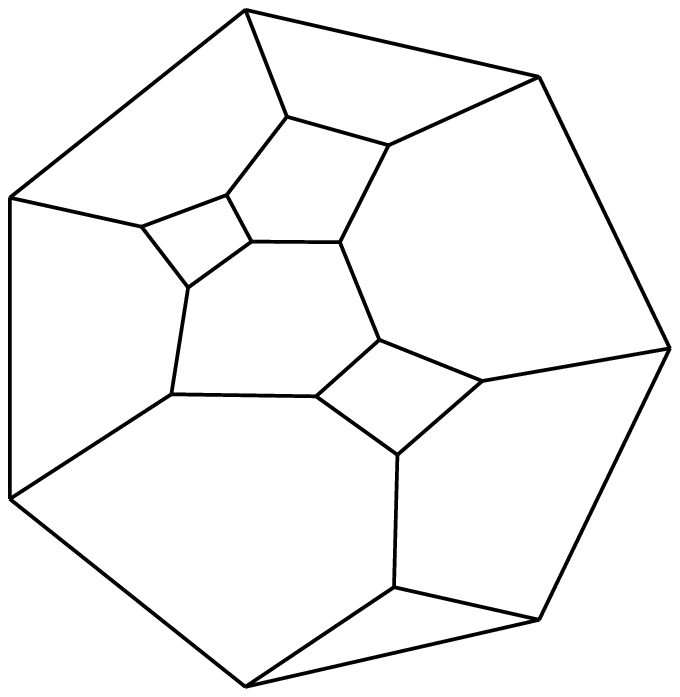}\\ {\it f} = 2.80\%} &
\specialcell[t]{ 19. $N = 18969$\\\includegraphics[scale=0.275]{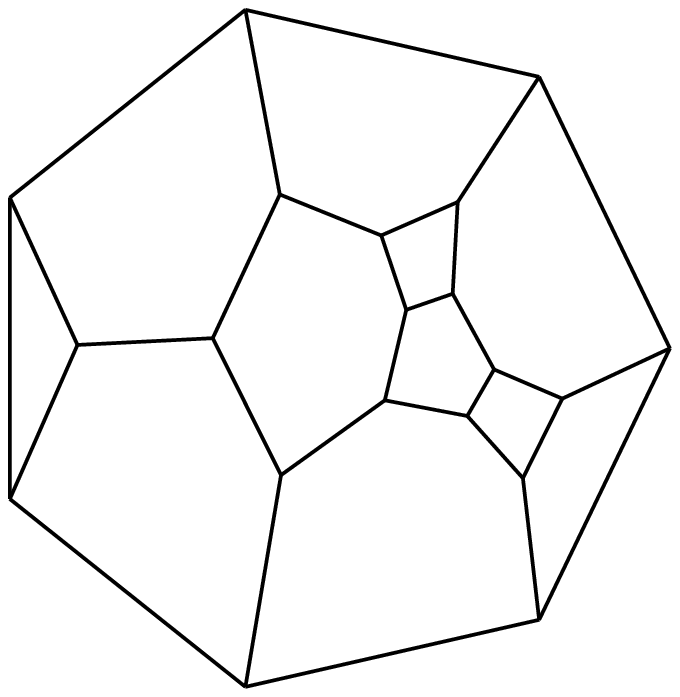}\\ {\it f} = 1.95\%} &
\specialcell[t]{ 20. $N = 18718$\\\includegraphics[scale=0.275]{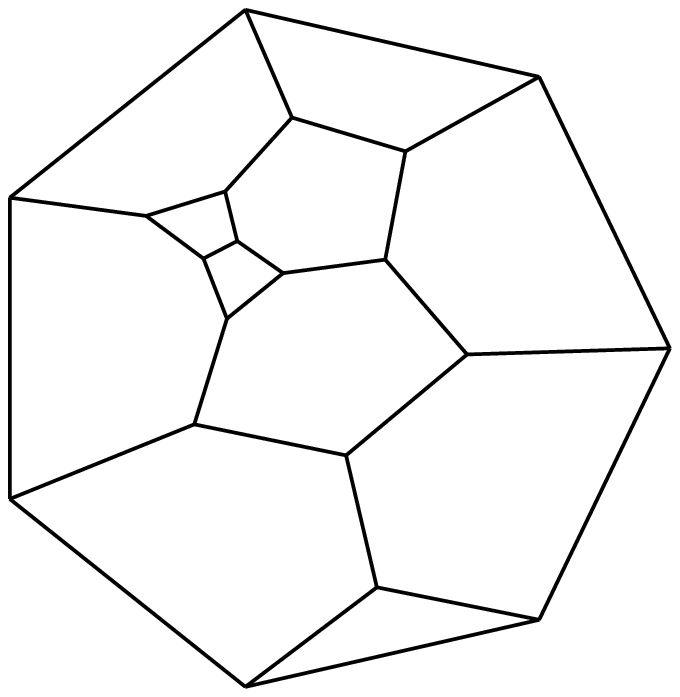}\\ {\it f} = 1.93\%} &
\specialcell[t]{ 21. $N = 12240$\\\includegraphics[scale=0.275]{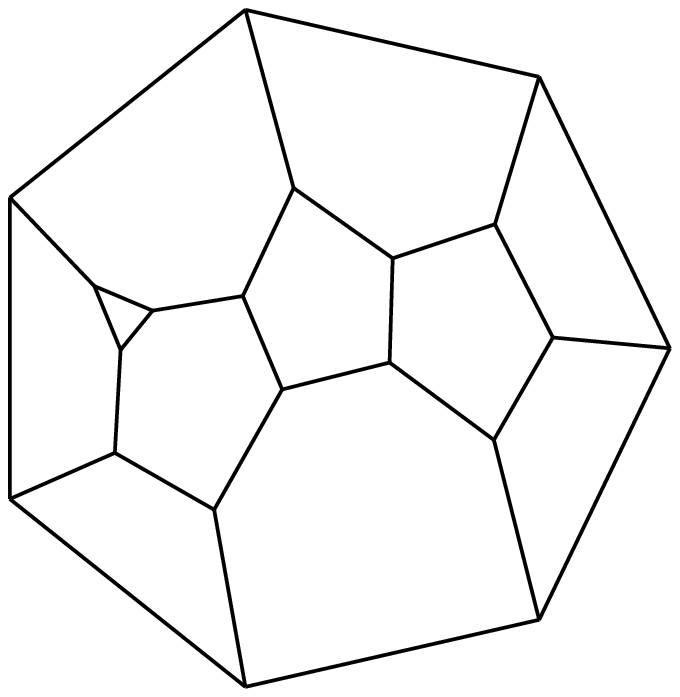}\\ {\it f} = 1.26\%} &
\specialcell[t]{ 22. $N = 11918$\\\includegraphics[scale=0.275]{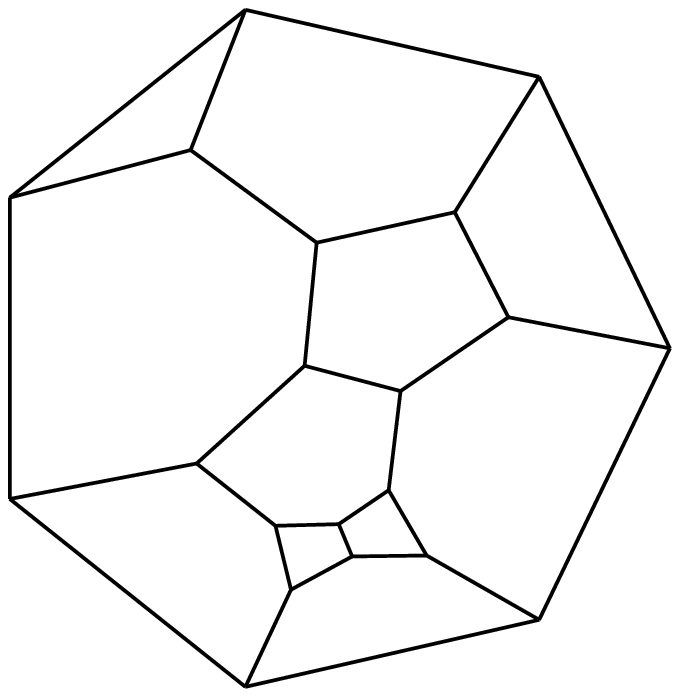}\\ {\it f} = 1.23\%} &
\specialcell[t]{ 23. $N = 11881$\\\includegraphics[scale=0.275]{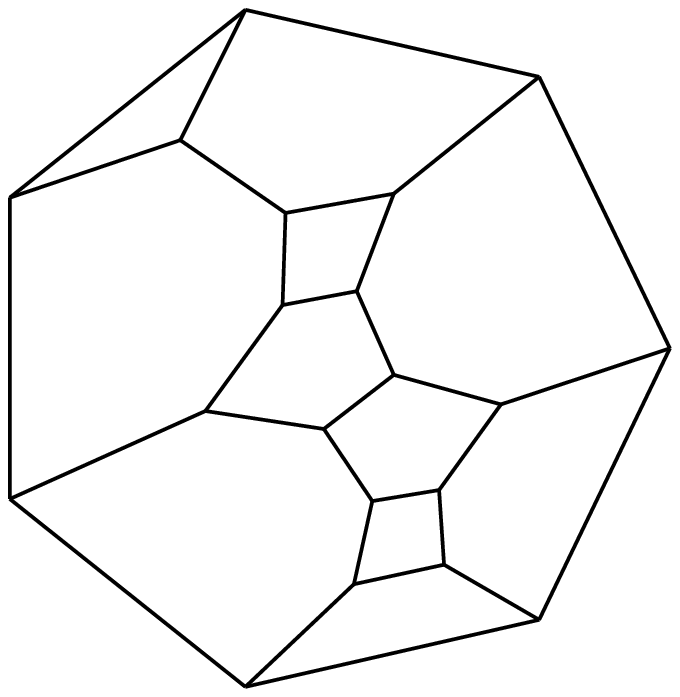}\\ {\it f} = 1.22\%} &
\specialcell[t]{ 24. $N = 7733$\\\includegraphics[scale=0.275]{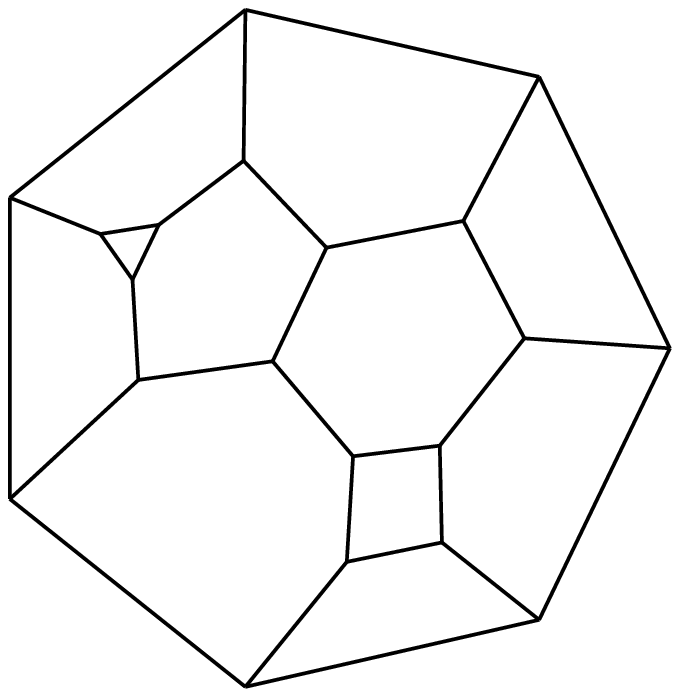}\\ {\it f} = 0.80\%} \\
\hline
\hline
\specialcell[t]{ 25. $N = 7606$\\\includegraphics[scale=0.275]{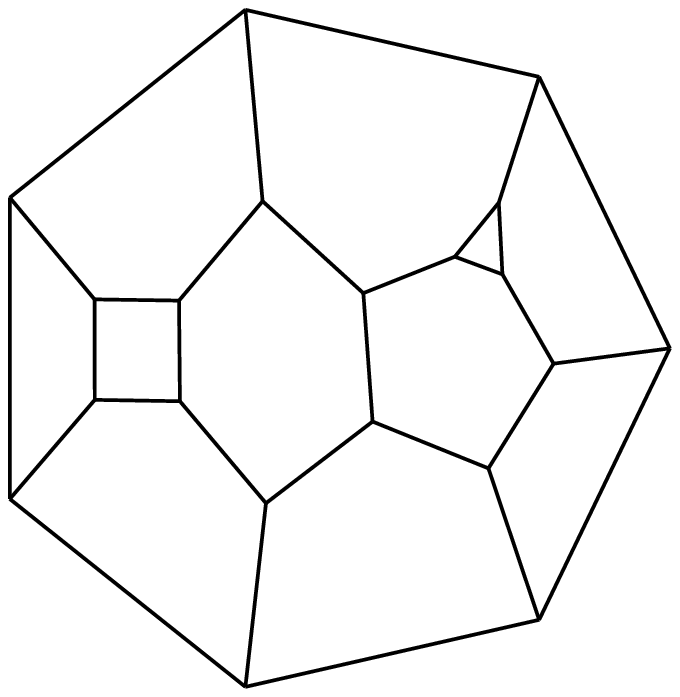}\\ {\it f} = 0.78\%} &
\specialcell[t]{ 26. $N = 7298$\\\includegraphics[scale=0.275]{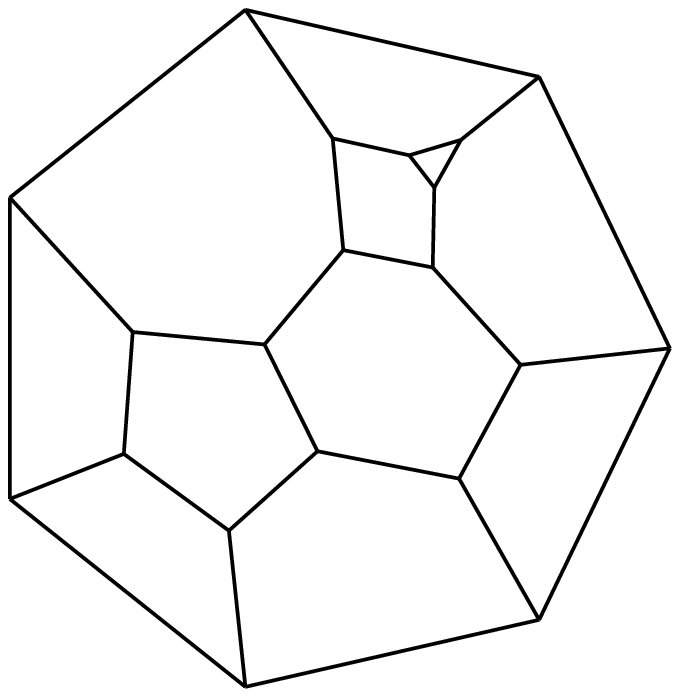}\\ {\it f} = 0.75\%} &
\specialcell[t]{ 27. $N = 7291$\\\includegraphics[scale=0.275]{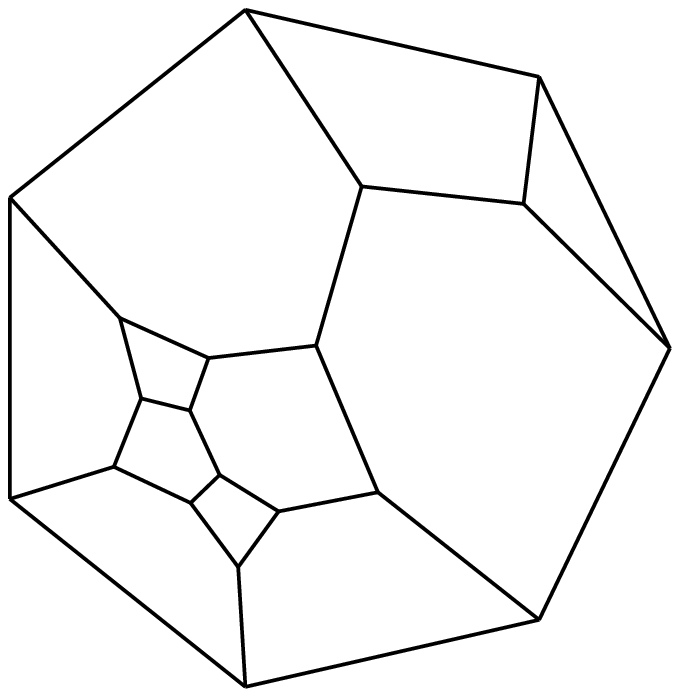}\\ {\it f} = 0.75\%} &
\specialcell[t]{ 28. $N = 6897$\\\includegraphics[scale=0.275]{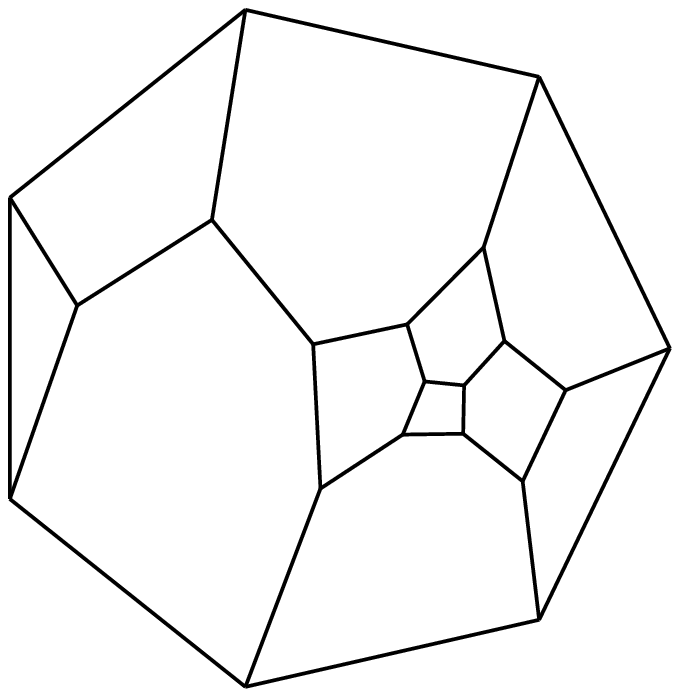}\\ {\it f} = 0.71\%} &
\specialcell[t]{ 29. $N = 5281$\\\includegraphics[scale=0.275]{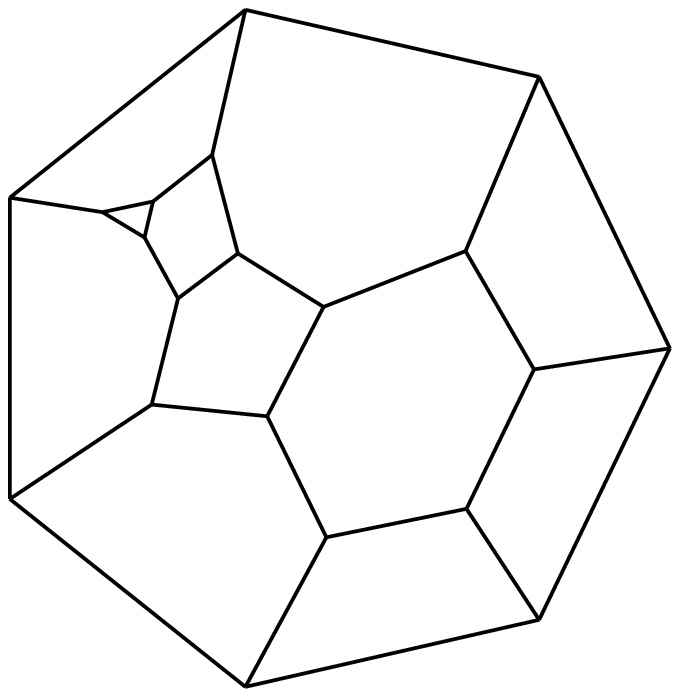}\\ {\it f} = 0.54\%} &
\specialcell[t]{ 30. $N = 4572$\\\includegraphics[scale=0.275]{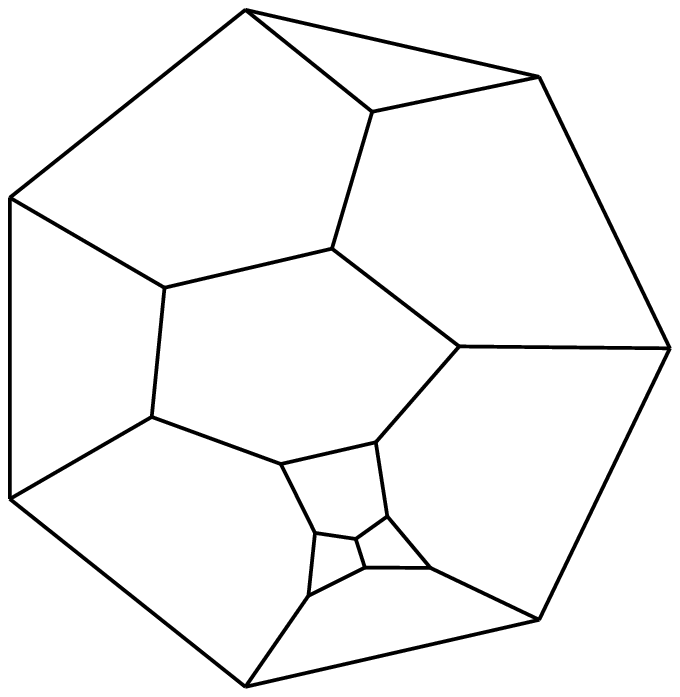}\\ {\it f} = 0.47\%} &
\specialcell[t]{ 31. $N = 3227$\\\includegraphics[scale=0.275]{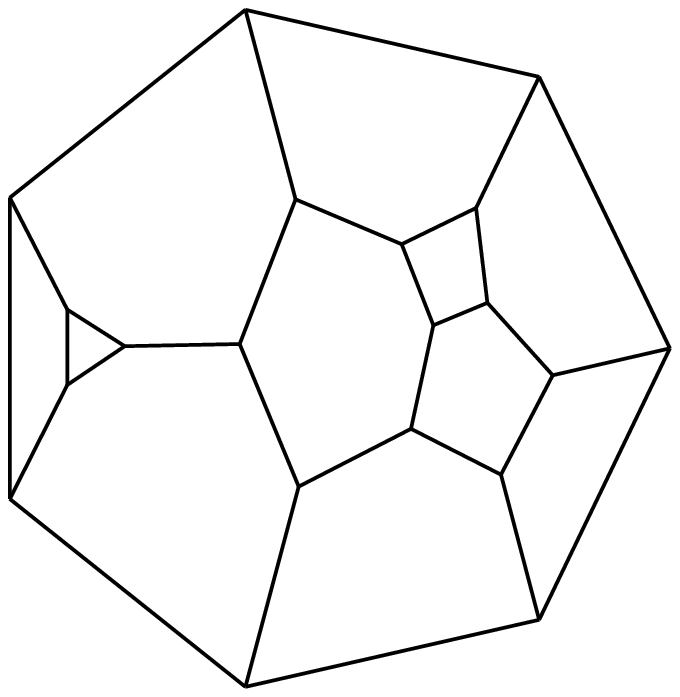}\\ {\it f} = 0.33\%} &
\specialcell[t]{ 32. $N = 1812$\\\includegraphics[scale=0.275]{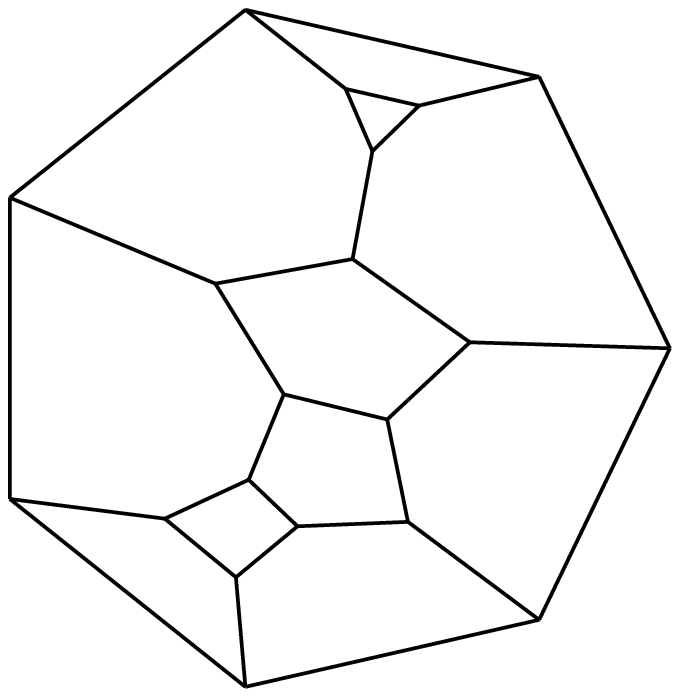}\\ {\it f} = 0.19\%} \\
\hline
\hline
\specialcell[t]{ 33. $N = 1425$\\\includegraphics[scale=0.275]{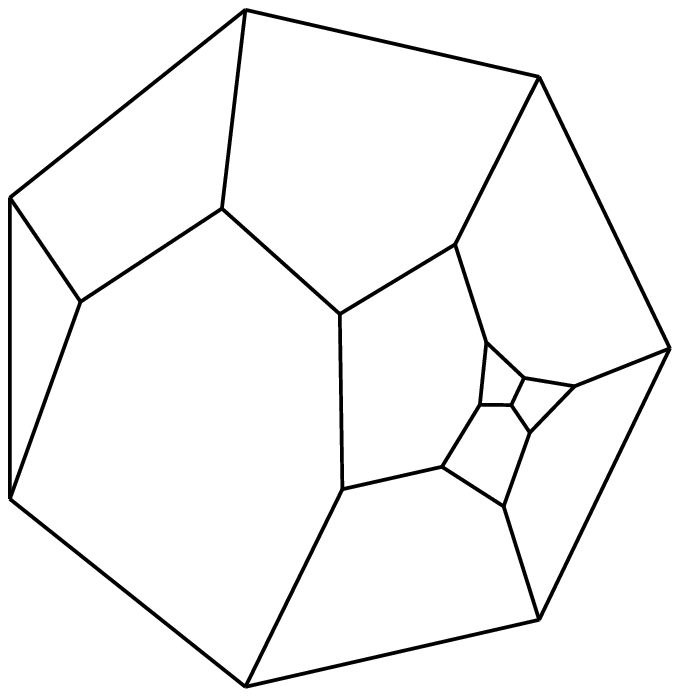}\\ {\it f} = 0.15\%} &
\specialcell[t]{ 34. $N = 1323$\\\includegraphics[scale=0.275]{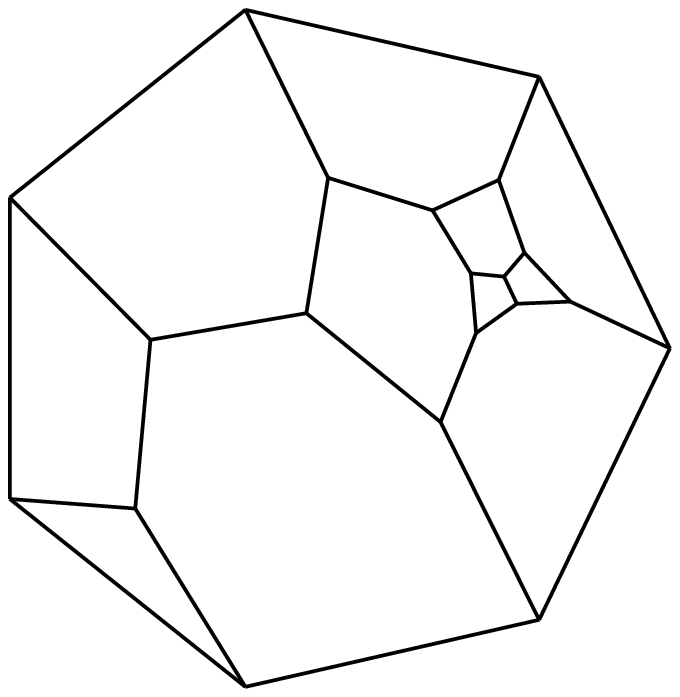}\\ {\it f} = 0.14\%} &
\specialcell[t]{ 35. $N = 1013$\\\includegraphics[scale=0.275]{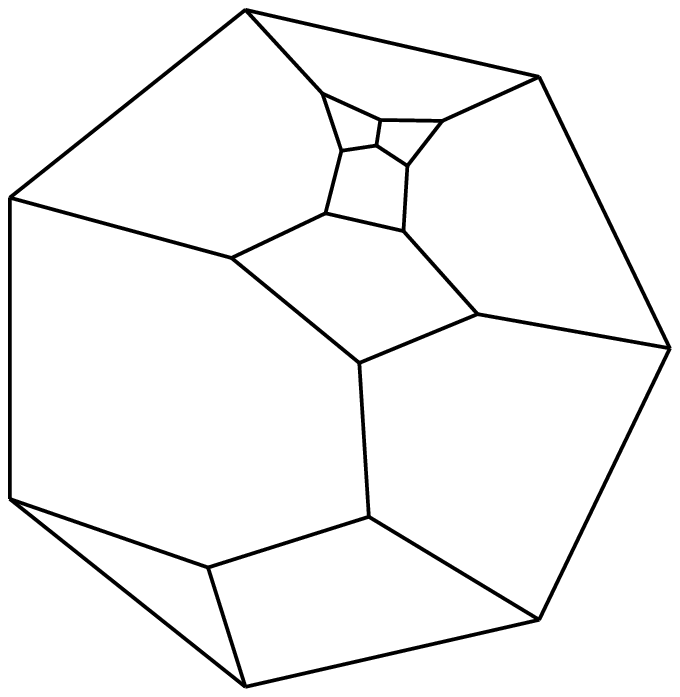}\\ {\it f} = 0.10\%} &
\specialcell[t]{ 36. $N = 460$\\\includegraphics[scale=0.275]{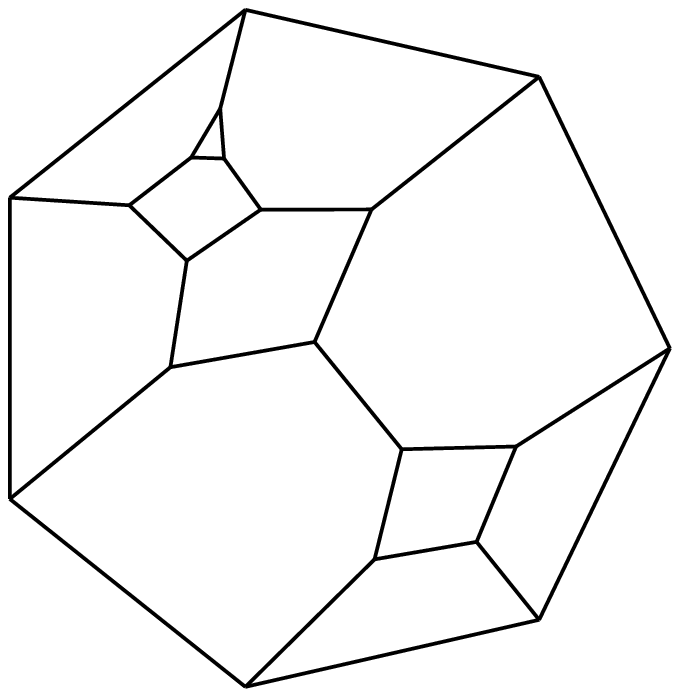}\\ {\it f} = 0.05\%} &
\specialcell[t]{ 37. $N = 192$\\\includegraphics[scale=0.275]{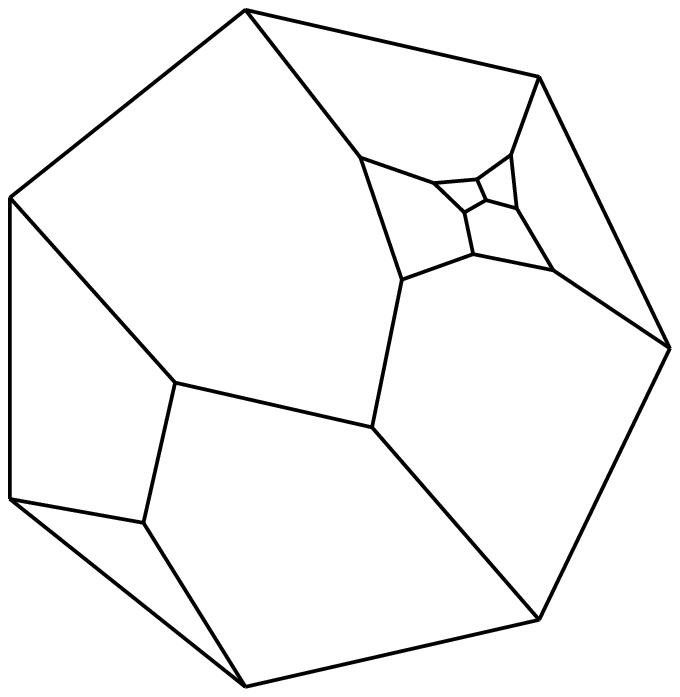}\\ {\it f} = 0.02\%} &
\specialcell[t]{ 38. $N = 25$\\\includegraphics[scale=0.275]{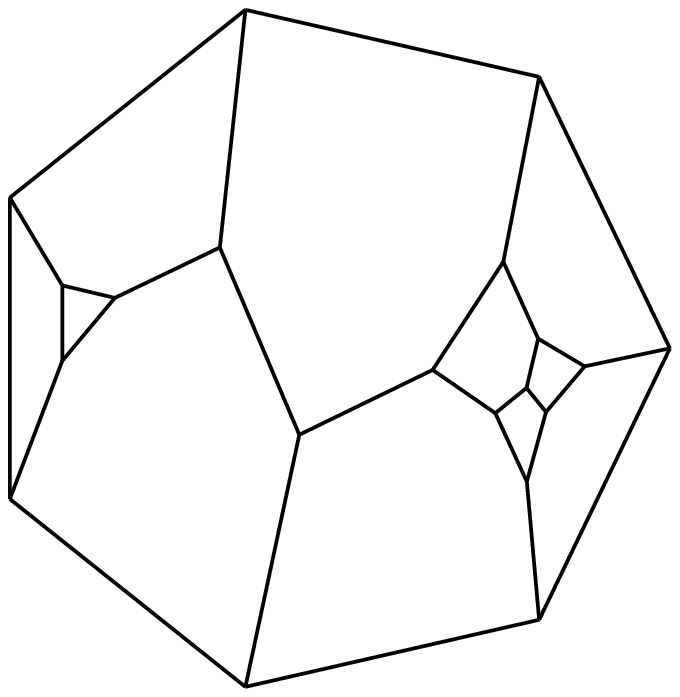}\\ {\it f} = 0.00\%} &&\\
\hline
\end{tabular}
\caption{Schlegel diagrams of all 38 topological types sharing $p$-vector $(001343100...)$.  Listed for each type is its absolute frequency $N$ and the relative frequency $f$ among all types with the given $p$-vector.  Entry 11 is the only one of these types which has a non-trivial symmetry ($S=2$).}
\label{schlegel-diagrams13431}
\end{figure*}

\begin{table}
\begin{center}
  \begin{tabular}{ | r | r | r | r |}
    \hline
$F$       & \multicolumn{1}{c|}{$\alpha$} & \multicolumn{1}{c|}{$\beta$} & \multicolumn{1}{c|}{$\gamma$} \\ \hline
5 & 1.0516 $\pm$ 0.1211 & 26.8590 $\pm$ 0.7646 & 3.7451 $\pm$ 0.3563 \\
6 & 0.8235 $\pm$ 0.0360 & 23.9845 $\pm$ 0.6768 & 5.4220 $\pm$ 0.2125 \\
7 & 0.8658 $\pm$ 0.0124 & 21.2568 $\pm$ 0.2521 & 6.0360 $\pm$ 0.0783 \\
8 & 0.8499 $\pm$ 0.0069 & 20.3902 $\pm$ 0.1754 & 7.0576 $\pm$ 0.0532 \\
9 & 0.8519 $\pm$ 0.0056 & 19.4911 $\pm$ 0.1576 & 7.9790 $\pm$ 0.0492 \\
10 & 0.8585 $\pm$ 0.0045 & 18.6500 $\pm$ 0.1362 & 8.8564 $\pm$ 0.0445 \\
11 & 0.8593 $\pm$ 0.0041 & 18.1090 $\pm$ 0.1306 & 9.7961 $\pm$ 0.0447 \\
12 & 0.8658 $\pm$ 0.0028 & 17.4918 $\pm$ 0.0910 & 10.6799 $\pm$ 0.0329 \\
13 & 0.8740 $\pm$ 0.0022 & 16.8727 $\pm$ 0.0731 & 11.5388 $\pm$ 0.0279 \\
14 & 0.8766 $\pm$ 0.0032 & 16.4873 $\pm$ 0.1106 & 12.4630 $\pm$ 0.0443 \\
15 & 0.8816 $\pm$ 0.0024 & 16.0797 $\pm$ 0.0856 & 13.3712 $\pm$ 0.0361 \\
16 & 0.8813 $\pm$ 0.0023 & 15.8695 $\pm$ 0.0826 & 14.3440 $\pm$ 0.0364 \\
17 & 0.8869 $\pm$ 0.0032 & 15.4893 $\pm$ 0.1186 & 15.2387 $\pm$ 0.0549 \\
18 & 0.8855 $\pm$ 0.0039 & 15.3826 $\pm$ 0.1474 & 16.2469 $\pm$ 0.0709 \\
19 & 0.8952 $\pm$ 0.0045 & 14.8780 $\pm$ 0.1683 & 17.0569 $\pm$ 0.0852 \\
20 & 0.9004 $\pm$ 0.0057 & 14.5516 $\pm$ 0.2130 & 17.9413 $\pm$ 0.1127 \\
21 & 0.9038 $\pm$ 0.0070 & 14.3131 $\pm$ 0.2653 & 18.8698 $\pm$ 0.1463 \\
22 & 0.8952 $\pm$ 0.0100 & 14.5412 $\pm$ 0.3981 & 20.0455 $\pm$ 0.2248 \\
23 & 0.9304 $\pm$ 0.0140 & 13.1567 $\pm$ 0.4973 & 20.3107 $\pm$ 0.3043 \\
24 & 0.9087 $\pm$ 0.0165 & 13.8693 $\pm$ 0.6431 & 21.7840 $\pm$ 0.3949 \\
25 & 0.9553 $\pm$ 0.0252 & 12.0889 $\pm$ 0.8403 & 21.6639 $\pm$ 0.5681 \\
26 & 0.9459 $\pm$ 0.0375 & 12.2951 $\pm$ 1.3040 & 22.8030 $\pm$ 0.8978 \\
27 & 0.8497 $\pm$ 0.0615 & 16.3169 $\pm$ 3.1440 & 26.4506 $\pm$ 1.9300 \\
28 & 0.8858 $\pm$ 0.0924 & 14.7106 $\pm$ 4.1820 & 26.6911 $\pm$ 2.7870 \\
29 & 1.0230 $\pm$ 0.1444 & 9.7563 $\pm$ 3.9500 & 23.9996 $\pm$ 3.3230 \\
30 & 0.9302 $\pm$ 0.2340 & 12.4305 $\pm$ 8.8750 & 26.7969 $\pm$ 6.6730 \\
    \hline
  \end{tabular}
\end{center}
\caption{Parameters for the partial distributions of cell volumes, as described in Section IV A.}
\label{gamma_params_volumes}
\end{table}

\begin{table}
\begin{center}
  \begin{tabular}{ | r | r | r | r |}
    \hline
$F$       & \multicolumn{1}{c|}{$\alpha$} & \multicolumn{1}{c|}{$\beta$} & \multicolumn{1}{c|}{$\gamma$} \\ \hline
5 & 1.5762 $\pm$ 0.3972 & 20.8402 $\pm$ 1.4590 & 6.3315 $\pm$ 1.3440 \\
6 & 1.2491 $\pm$ 0.0495 & 21.4514 $\pm$ 0.7675 & 9.2546 $\pm$ 0.3419 \\
7 & 1.3351 $\pm$ 0.0231 & 19.4808 $\pm$ 0.3514 & 10.2701 $\pm$ 0.1670 \\
8 & 1.3303 $\pm$ 0.0130 & 18.9765 $\pm$ 0.2260 & 11.8339 $\pm$ 0.1099 \\
9 & 1.3500 $\pm$ 0.0093 & 18.2157 $\pm$ 0.1716 & 13.1974 $\pm$ 0.0877 \\
10 & 1.3424 $\pm$ 0.0045 & 18.0385 $\pm$ 0.0900 & 14.7931 $\pm$ 0.0478 \\
11 & 1.3475 $\pm$ 0.0054 & 17.6702 $\pm$ 0.1138 & 16.2580 $\pm$ 0.0634 \\
12 & 1.3534 $\pm$ 0.0051 & 17.3403 $\pm$ 0.1124 & 17.7158 $\pm$ 0.0657 \\
13 & 1.3657 $\pm$ 0.0043 & 16.8931 $\pm$ 0.0965 & 19.0826 $\pm$ 0.0594 \\
14 & 1.3690 $\pm$ 0.0034 & 16.6523 $\pm$ 0.0780 & 20.5425 $\pm$ 0.0502 \\
15 & 1.3733 $\pm$ 0.0045 & 16.4274 $\pm$ 0.1064 & 22.0069 $\pm$ 0.0715 \\
16 & 1.3688 $\pm$ 0.0057 & 16.4174 $\pm$ 0.1405 & 23.6017 $\pm$ 0.0980 \\
17 & 1.3696 $\pm$ 0.0056 & 16.3018 $\pm$ 0.1413 & 25.1211 $\pm$ 0.1024 \\
18 & 1.3721 $\pm$ 0.0049 & 16.1507 $\pm$ 0.1271 & 26.6042 $\pm$ 0.0957 \\
19 & 1.3780 $\pm$ 0.0062 & 15.9226 $\pm$ 0.1626 & 28.0233 $\pm$ 0.1273 \\
20 & 1.3879 $\pm$ 0.0085 & 15.5909 $\pm$ 0.2223 & 29.3458 $\pm$ 0.1814 \\
21 & 1.4014 $\pm$ 0.0096 & 15.1868 $\pm$ 0.2482 & 30.6038 $\pm$ 0.2116 \\
22 & 1.3666 $\pm$ 0.0114 & 16.0624 $\pm$ 0.3245 & 32.9184 $\pm$ 0.2772 \\
23 & 1.4394 $\pm$ 0.0195 & 14.1189 $\pm$ 0.4768 & 32.8076 $\pm$ 0.4465 \\
24 & 1.3920 $\pm$ 0.0291 & 15.2752 $\pm$ 0.8076 & 35.4115 $\pm$ 0.7474 \\
25 & 1.4758 $\pm$ 0.0386 & 13.1021 $\pm$ 0.8904 & 34.8621 $\pm$ 0.9126 \\
26 & 1.4455 $\pm$ 0.0612 & 13.7373 $\pm$ 1.5320 & 37.0079 $\pm$ 1.5720 \\
27 & 1.2518 $\pm$ 0.1002 & 19.9525 $\pm$ 4.1610 & 44.3574 $\pm$ 3.6360 \\
28 & 1.3484 $\pm$ 0.1052 & 16.6244 $\pm$ 3.4740 & 43.2935 $\pm$ 3.4210 \\
29 & 1.4943 $\pm$ 0.2177 & 12.5015 $\pm$ 5.0580 & 40.4000 $\pm$ 5.8690 \\
30 & 1.4026 $\pm$ 0.3032 & 14.5465 $\pm$ 8.7210 & 43.7419 $\pm$ 9.4830 \\
    \hline
  \end{tabular}
\end{center}
\caption{Parameters for the partial distributions of cell surface areas, as described in Section IV B.}
\label{gamma_params_surface_areas}
\end{table}

\end{document}